\documentclass[reprint,amsmath,amssymb,aps]{revtex4-1}
\usepackage{array}
\usepackage{epsfig}
\usepackage{euscript}
\usepackage{comment}
\usepackage{xcolor}
\usepackage{hyperref}
\usepackage{subfigure}
\usepackage{graphicx}
\usepackage{dcolumn}
\usepackage{bm}
\usepackage{float}
\usepackage{url}
\allowdisplaybreaks
\usepackage{aas_macros}
\DeclareMathOperator{\sign}{sgn}

\begin{document}

\preprint{APS/123-QED}

\title{
Combining Post-Circular and  Pad\'e approximations to compute 
Fourier domain templates for eccentric inspirals }

\author{Srishti Tiwari}
\email{srishti.tiwari@tifr.res.in}

\author{Achamveedu Gopakumar}
 \affiliation{ Department of Astronomy and Astrophysics, Tata Institute of Fundamental Research, Mumbai 400005, India }

\date{\today}

\begin{abstract}
Observations of transient gravitational wave (GW) events with non-negligible orbital eccentricity can be highly rewarding from astrophysical considerations. Ready-to-use fully analytic
frequency domain inspiral GW templates are crucial ingredients to construct eccentric inspiral-merger-ringdown waveform families, required for the detection of such GW events.
It turns out that a fully analytic, post-Newtonian (PN) accurate frequency domain  inspiral template family, which uses certain post-circular approximation, may only be suitable to model events with initial eccentricities $e_0 \leq 0.2$.We here explore the possibility of combining Post-Circular and  Pad\'e approximations to obtain fully analytic frequency domain eccentric inspiral templates. The resulting 1PN-accurate approximant is capable of faithfully capturing eccentric inspirals having $e_0 \leq 0.6$ while employing our 1PN extension of a frequency domain template family that does not use post-circular approximation, detailed in 
Moore, B., et al.\ 2018, Classical and Quantum Gravity, 35, 235006.
We also discuss subtleties that arise while combining post-circular and Pad\'e approximations to obtain higher PN order templates for eccentric inspirals.
\end{abstract}

\maketitle

\section{\label{sec:level1} Introduction}
Gravitational wave events that involve compact binaries in non-circular orbits are of definite interest to the functional hecto-hertz GW observatories such as 
the Advanced LIGO (aLIGO), Advanced Virgo (aVirgo), and KAGRA \cite{ALIGO,AVirgo,KAGRA}.
This is despite the fact that all confirmed and recorded GW detections contain 
compact binaries inspiraling along quasi-circular orbits \cite{GWTC-1,O3_public,eBBH_19}.
In contrast, massive black hole (BH) binaries in eccentric orbits, like the one in bright blazar OJ~287 \cite{Laine_2020}, are promising nano-Hz GW sources for the 
rapidly maturing Pulsar Timing Array efforts\cite{ipta,AGGT}.
Orbital eccentricity is expected to be an important parameter for milli-hertz and deci-hertz GW astronomy that will be heralded by LISA and DECIGO, respectively \cite{mHZ-GW,Zwick_2020,DECIGO}.

   GW events that involve non-negligible orbital eccentricities are interesting to
 the LIGO-Virgo-Kagra consortium, because such events should allow us to 
 constrain possible formation scenarios for the observed binary BH coalescences
 and to test general relativity
 \cite{RLT_2019,MY_2020}.
 It turns out that formation scenarios for the observed O1, O2 and O3 
 binary BH events roughly fall in two distinct possibilities.
 The first scenario involves BH binaries formed in the galactic fields via isolated binary stellar evolution \cite{BKT_2002,Kruckow_2018}.
 These compact binaries are expected to have 
 orbital eccentricities $\sim 10^{-4}$ when their GWs enter aLIGO frequency window \cite{Kowalska_2011}.
 Such values are substantially below the levels at which we can constrain orbital eccentricities of aLIGO and aVirgo GW events \cite{GW150914,ENIGMA}.
 The second scenario involves formation of BH binaries at very close orbital 
 separations and this is astrophysically possible in globular clusters, young star 
 clusters,
 and active galactic nuclei \cite{Fragione_2019,Samsing_2018,KFT_2020,OKL_2009, Kremer_2019}. 
 These scenarios ensure that temporal evolution of BH binaries 
 are perturbed by other compact objects leading to the development 
 of orbital eccentricities. 
 The fact that GW emission reduces orbital eccentricity by a factor of 
 three when its semi-major axis shrinks by a factor of two 
 ensures that dynamically formed compact binaries
 with short orbital periods
 can display non-negligible 
 orbital eccentricities in the aLIGO frequency window \cite{Peters_64}.
 Additionally, BH binaries in such dense stellar environments can experience 
 Kozai-Lidov resonances due to gravitational perturbations of a third BH and 
 such a scenario can also provide eccentric BH binaries in the aLIGO frequency 
 window \cite{Kozai,Antonini_2014,Randall_2018}.
 It is important to note that the above two binary BH formation scenarios lead to 
 distinct distributions for the masses and spins of binary constituents \cite{Farr_2017,Arca_2020}.
  Unfortunately, GW observations from a few dozen BH binaries  can not provide 
 constraints on the most favorable formation scenarios for the so far recorded 
 GW events.
 
 There are on-going efforts to probe the presence of 
 eccentric compact binary mergers in the available interferometric data sets 
 \cite{Tiwari_2016,Abbott_2019,Nitz_2020,LNB_2020}.
 Measuring orbital eccentricity of a GW event should allow us to identify the prominent formation channel for aLIGO BH binaries. This is mainly because of few detailed and realistic evolution of compact binaries in globular clusters which suggest that $\sim 10 \%$ of such binaries can have eccentricities 
 $> 0.1$ when their GWs enter aLIGO frequency window \cite{Samsing_2018,Rodriguez_2018}.
 An efficient detection of such GW events and the accompanying accurate parameter estimation requires one to develop accurate and efficient 
 eccentric IMR template families both in the time and frequency domains,
 similar to template families developed for quasi-circular inspirals \cite{Hannam_2013,DN_2014}.
 
 
 There are few on-going efforts to compute such template families 
 for binary black hole systems,  merging along moderately eccentric orbits
 \cite{Hinder_2018,ENIGMA,DCAN_2020,Ramos_2020}. 
 These detailed investigations are being augmented by efforts that 
 explore the search sensitivity 
 of popular modeled and unmodeled LIGO-Virgo collaboration 
 search algorithms to capture eccentric 
 binary black hole coalescences, as pursued in Ref.~\cite{Antoni_2020}.
 An eccentric inspiral-merger-ringdown (IMR)
 family that extends the very popular \texttt{PhenomD} 
 templates for quasi-circular merger events \cite{Husa_2016,Khan_2016}  will be very helpful for extending efforts of Ref.~\cite{Antoni_2020} and 
 eventually searching for eccentric GW events.
 
 A crucial ingredient to such an eccentric IMR family will be a fully analytic 
  frequency domain GW response function for eccentric inspirals.
The post-circular (PC) scheme, developed and extended in Refs.~\cite{YABW,THG,Moore16,TGMH},
allowed one to compute fully analytic third post-Newtonian (3PN) accurate 
frequency domain $\tilde{h}(f)$ for eccentric inspirals by employing the 
method of stationary phase approximation \cite{BenOrz}.
Recall that PN approximation provides general relativity based corrections to the Newtonian dynamics of a compact binary system in terms of $(v/c)^2$, with $v$ denoting the orbital speed of the binary and $c$ being the speed of light in vacuum.
Therefore, $3$PN corrections provide  $(v/c)^{6}$ general relativity based 
contributions to relevant expressions and equations.

 The PN-accurate PC approach provides fully analytic 3PN-accurate expressions for the orbital eccentricity and the Fourier phases of $\tilde{h}(f)$ as functions of GW frequency and involves 
 series expansions in $e_0$, the value of orbital eccentricity at certain initial GW frequency, at every PN order.
 At present, Ref.~\cite{TGMH} provides the most PN-accurate $\tilde{h}(f)$ for compact binaries inspiraling along  eccentric orbits while 
employing the PC scheme. This fully analytic frequency domain GW response function incorporates 1PN-accurate amplitude, 3PN-accurate Fourier phase as well as 3PN-accurate evolution of orbital eccentricity $e_t$ in terms of orbital frequency $F$, while taking into account up to $\mathcal{O}(e_0^6)$ corrections at every PN order. An important feature of $\tilde{h}(f)$, given in Ref.~\cite{TGMH}, is the incorporation of general relativistic periastron advance in orbital motion of eccentric binaries. 
However, it was pointed out that PC scheme based $\tilde{h}(f)$ should be 
applicable to eccentric inspirals with $e_0 \leq 0.2$, especially 
if they incorporate only next-next-to leading order $e_0$ 
contributions \cite{MRLY18}. This prompted Ref.~\cite{MRLY18} to develop a semi-analytic frequency domain inspiral $\tilde{h}(f)$ that should be accurate to model eccentric inspirals 
with $e_0 > 0.2$. This initial investigation incorporated  the effects of dominant quadrupolar order GW emission while constructing their 
inspiral $\tilde{h}(f)$. Thereafter,  Ref.~\cite{MY19}
provided an inspiral eccentric $\tilde{h}(f)$ family that incorporated  GW emission 
effects to 3PN order and 
outlined a way to incorporate the effect of periastron advance in the Fourier phases.

The present effort explores the possibility of extending the ability of  PN-accurate PC approach to model eccentric inspirals with $e_0 \sim 0.6$.
This is influenced by the fact that the resulting $\tilde{h}(f)$ 
will be useful to extend the existing frequency domain \texttt{PhenomD}  
IMR templates with eccentric effects. 
And, there are on-going efforts to create such eccentric templates with inputs from 
Refs.~\cite{Antoni_2020} and \cite{TGMH}.
We  employ an elegant and simple  re-summation technique, namely  the {\it{Pad\'e approximation}} as detailed in Ref.~\cite{BenOrz}, on various Taylor expanded 
quantities of the PC scheme based $\tilde h(f)$.
We list below key findings of our investigations: 
\begin{itemize}
    \item We obtain Pad\'e approximation for the two crucial quantities,
     required to operationalise  Newtonian-order frequency-domain analytic inspiral waveform.
    This includes
    $e_t(F)$ that provides the frequency evolution of orbital eccentricity, and the associated Fourier phases $\Psi(F)$.
    The rational polynomials for these quantities 
    were computed from their post-circular scheme counterparts that included $\mathcal{O}(e_0^{19})$ and $\mathcal{O}(e_0^{20})$ corrections,
    respectively.
    \item 
    Our quadrupolar order Pad\'e approximation for $e_t$ 
    provides fractional relative errors that are $\leq 10^{-4}$ even 
    for $e_0$ values like  $0.6$. These estimates employ 
    numerically extracted $e_t(F)$ values from an exact 
     quadrupolar orbital 
    frequency $\omega(e_t, e_0, \omega_0)$ expression, present in
    Ref.~\cite{MRLY18} (hereafter referred to as the {\it MoRoLoYu} approach).
    \item We developed a fully analytic quadrupolar order Pad\'e approximation based 
     $\tilde h(f)$ (Pad\'e approximant $\tilde h(f)$).
     The usual {\it match} ($\mathcal{M}$) analysis reveals that 
     our approximant is {\it faithful} to the {\it MoRoLoYu} inspiral $\tilde{h}(f)$ 
     with $e_0 \sim 0.6$.
    \item
    We extended the above two inspiral template families,
    namely the Pad\'e and {\it MoRoLoYu} inspiral approximants,
    to 1PN order while restricting the amplitudes to the quadrupolar order.
    These two waveform families were also found to be {\it faithful} to each other for the classical aLIGO binaries with $e_0$ values $\sim 0.6$.
    \item 
     We discuss possible issues that need to be tackled to extend our Pad\'e 
     approximant to higher PN orders. This is influenced by the discussions of 
     Ref.~\cite{MY19} and the observed discrepancies between the analytically 
     and numerically extracted values of certain PN-accurate quantities.
\end{itemize}

We restricted our attention to $e_0 \lesssim 0.6$ values, influenced by the 
Laplace limit. This limit provides the maximum value for which the usual power series
in $e$ solution to the classical Kepler equation, namely $l= u-e\, \sin u$, converges \cite{KE_textbook}.
Note that we employ essentially such a solution to compute the starting point of 
our efforts, namely Eq.~(\ref{eq:hpct}), for the quadrupolar order GW polarization states. Interestingly, we may need to probe the existence of such a limit in PN-accurate Kepler Equation, given in Ref.~\cite{GMS}, as PN accurate version of
Eq.~(\ref{eq:hpct}) requires such a power series solution at PN orders \cite{Boetzel_17}.

Our paper is structured as follows: 
Sec.~\ref{sec:level2} provides brief descriptions of quadrupolar order
PC and MoRoLoYu  approaches to obtain eccentric $\tilde{h}(f)$ and introduces our 
Pad\'e approximant. Various comparisons between these approaches are presented 
in sub-sections of Sec.~\ref{sec:level2}. The 1PN extensions of these approaches 
are presented in Sec.~\ref{sec:level3} that includes data analysis relevant match computations. Sec.~\ref{sec:level3D} probes subtleties that we may face 
while extending our Pad\'e approximant to higher PN orders.
Appendices provide some underlying equations.

\section{\label{sec:level2} 
Analytic Fourier-domain eccentric GW waveform families at the quadrupolar order }

 We begin by summarizing how one formally obtains the frequency domain $\tilde{h}(f)$ 
 from its time-domain counterpart, influenced by Ref.~\cite{YABW}.
 How to operationalize the resulting $\tilde{h}(f)$ for two distinct approaches is described
 in the next two subsections.
 These two approaches are the fully analytic PC scheme of Ref.~\cite{YABW} and semi-analytic approach of 
 Ref.~\cite{MRLY18}  that should be valid essentially for arbitrary initial eccentricities. 
Thereafter, we present our fully analytic Pad\'e  approximant to model eccentric inspirals.
In what follows, we briefly summarize formulae that are required to compute FD GW response function for 
eccentric inspirals from its time domain counterpart. This is desirable as all the above three approaches employ these formulae.

The first step to obtain the FD GW response function $\tilde{h}(f)$ is to write down 
the time domain GW response (or strain) of a ground based GW detector as 
\begin{equation}
h(t) =  F_+ h_+(t) + F_\times h_\times(t), \label{eq:ht}     
\end{equation}
where $F_+$ and $F_\times$ are the antenna patterns of the interferometer that depend on certain angles, $\theta_S$, $\phi_S$ and $\psi_S$ that specify the declination and right ascension of the source as well as the polarisation angle ($\psi_S$), respectively. 
Further, $h_+(t)$ and $h_\times(t)$ represent the time-dependent GW polarization states at the Newtonian or quadrupolar order.
Following Ref.~\cite{YABW}, we write 
\begin{equation}
h_{+,\times}(t) = -\frac{G m \eta}{c^2 D_L}x \sum_{j=1}^{10} \left[C_{+,\times}^{(j)} \cos jl + S_{+,\times}^{(j)} \sin jl \right], \label{eq:hpct}
\end{equation}
and we have restricted  eccentricity contributions to $\mathcal{O}(e_t^8)$.
The additional symbols and variables that appear in the above equation are the luminosity distance to the source ($D_L$),
the usual PN expansion parameter
 $x = (G m \omega/c^3)^{2/3}$  while $\eta = m_1 m_2/m^2$ gives the symmetric mass ratio of a binary with component masses, $m_1$ and $m_2$ with total mass given as, $m = m_1 + m_2$. The secular orbital frequency of the binary is given by $\omega =2 \pi F$. 
 Note that $h_{+,\times}(t)$ expressions are given as a sum over harmonics ($j$) of $l$,
 the \textit{mean anomaly}, defined as 
 $l = n (t-t_0)$ where $ n = 2 \pi / P $ gives the mean motion of binary system having an orbital period of $P$ and $t_0$ is some initial epoch. 
 Further, the coefficients $C_{+,\times}^{(j)}$ of $\cos jl$ and $S_{+,\times}^{(j)}$ of $\sin jl$ in Eq.~(\ref{eq:hpct}) may be expressed as  power series in certain  \textit{time eccentricity} parameter $e_t$ that appear in the Keplerian type parametric solution 
whose coefficients are trigonometric functions of angles $\iota, \beta$ that describe the line of sight vector in certain inertial frame \cite{Boetzel_17}.
The explicit expressions for $C_{+,\times}^{(j)}$ and $S_{+,\times}^{(j)}$, accurate up to $\mathcal{O}(e_t^8)$, are given by Eqs.~(3.7-3.10) and (B1-B36) in Ref.~\cite{YABW}.
In general, the summation index $j$ goes to $\infty$ and the explicit 
expressions for $C_{+,\times}^{(j)}$ and $S_{+,\times}^{(j)}$ 
are written in terms of 
Bessel functions of first kind as given by Eqs.~(9) in Ref.~\cite{MRLY18}. In practice, only a finite number of harmonics {\it j} are  included while computing an eccentric inspiral waveform. Interestingly, the maximum number of harmonics depends on the highest order of eccentricity corrections included in the inspiral template \cite{YABW}. 
For example, a template that includes 
up to $\mathcal{O}(e_t^s)$ eccentricity corrections should have 
$s+2$ as the maximum 
$j$ value. This is why we include $10$ harmonics in our Eq.~(\ref{eq:ht}), which incorporates $\mathcal{O}(e_t^8)$ corrections in $e_t$.

It is fairly straightforward to obtain GW response function for eccentric inspirals 
by plugging  in the expressions for $h_{+,\times}$, namely Eq.~(\ref{eq:hpct}), into Eq.~(\ref{eq:ht}) and  this leads to 
\begin{equation}
h(t) = -\frac{G m \eta}{c^2 D_L}\left(\frac{G m \omega}{c^3}\right)^{2/3} \sum_{j=1}^{10} \alpha_j \cos (jl + \phi_j) \label{eq:ht1}.
\end{equation}
In above equation, $\alpha_j$ and $\phi_j$ are certain combination of  $F_{+,\times}$, $C_{+,\times}^{(j)}$ and $S_{+,\times}^{(j)}$ through $\Gamma_j$ and $\Sigma_j$ as ,
\begin{align} \label{eq:alpha_phi}
\alpha_j = &\, \sign(\Gamma_j)\sqrt{\Gamma_j^2 + \Sigma_j^2}, \\
\phi_j = &\, \arctan \left(-\frac{\Sigma_j}{\Gamma_j}\right),
\end{align}
where two new functions, $\Gamma_j = F_+C_+^{(j)} + F_\times C_\times^{(j)} $ and $\Sigma_j = F_+S_+^{(j)} + F_\times S_\times^{(j)}$ are introduced for simplicity \cite{YABW}. $\sign$ in Eq.~(\ref{eq:alpha_phi}) denotes the {\textit{Signum}} function such that $\sign(\Gamma_j) = 1$ if $\Gamma_j > 0$, $\sign(\Gamma_j) = -1$ if $\Gamma_j < 0$ and  $\sign(\Gamma_j) = 0$ if $\Gamma_j = 0$.

To model $h(t)$ from compact binaries that inspiral due to the emission of 
quadrupolar order GWs,
we introduce the following coupled differential equations for $\omega$ and $e_t$ 
\begin{align}
\frac{d\omega}{dt} = & \, \frac{\left(G\,m\,\omega\right)^{5/3}\,\omega^2\, \eta}{5\,c^5\,\left(1-e_t^2\right)^{7/2}}\left(96+292 e_t^2+37 e_t^4\right) \label{eq:domgdt} , \\
\frac{de_t}{dt} = & \, -\frac{\left(G\,m\,\omega\right)^{5/3}\,\omega\, \eta\,e_t}{15\,c^5\,\left(1-e_t^2\right)^{5/2}}\left(304+121 e_t^2\right).  \label{eq:detdt}
\end{align}
It is important to note that eccentricity contributions are fully incorporated in the above equations \cite{PM63}. The presence of these two coupled differential equations ensure that the prescription to compute $h(t)$ can be computationally expensive, especially for GW data analysis purposes.

 However, it is possible to obtain the Fourier transform of the resulting $h(t)$ 
 by employing the method of \textit{stationary phase approximation} (SPA) (see Chapter 6 in Ref.~\cite{BenOrz} for a nice description of the SPA method and Ref.~\cite{YABW} for its application to 
 eccentric $h(t)$ ). This leads to the following 
 expression for the Fourier domain GW response function
\begin{align}
\tilde{h}(f) = & \, \tilde{\mathcal{A}} \left(\frac{G m \pi f}{c^3}\right)^{-7/6} \sum_{j=1}^{10} \xi_j \left(\frac{j}{2}\right)^{2/3} e^{-i(\Psi_j+\pi/4)}, \label{eq:hf}
\end{align}
where the expressions for $\tilde{\mathcal{A}}$ and $\xi_j$ are given as,
\begin{align}
\tilde{\mathcal{A}} = & \, - \left(\frac{5 \pi \eta}{384}\right)^{1/2} \frac{G^2 m^2}{c^5 D_L}, \label{eq:Atilde} \\
\xi_j = & \, \frac{(1-e_t^2)^{7/4}}{\left(1+\frac{73}{24} e_t^2 + \frac{37}{96} e_t^4\right)^{1/2}} \alpha_j e^{-i \phi_j (f/j)}. \label{eq:xij}
\end{align}
Further, the crucial Fourier phase is given by 
\begin{align}
\Psi_j := j \phi(t^*_j) - 2 \pi f t^*_j\,. \label{eq:Psi0} 
\end{align}
and the use of the  stationary phase condition  demands the evaluation of 
 the Fourier phases only at the stationary points $t^*_j$( 
 $t^*_j$ represents those instances when $j F = f$).
 In other words,  $\Psi_j$ should only be computed  for those Fourier frequencies $f$ which are an integral multiple of orbital frequency $F$ as $j$ denotes the harmonic index in Eq.~(\ref{eq:hf}). Recall that $F = \omega/2 \pi$.

 Clearly, further efforts are required to operationalize the SPA based expression for 
 $\tilde{h}(f)$. Specifically, we need an accurate and efficient approach to specify the 
 way $e_t$ and $\Psi_j$ depends on $F$.  In what follows, we summarize 
 the existing two approaches, namely
  the \textit{post-circular} scheme of  Ref.~\cite{YABW} and the recent semi-analytical approach of Ref.~\cite{MRLY18} for  operationalizing the above prescription  for $\tilde{h}(f)$.
  Thereafter, we introduce our fully-analytic  Pad\'e approximation based approach to obtain  $e_t(f)$ and $\Psi_j(f)$ expressions in Section~\ref{sec:level2C}
and probe its preliminary data analysis implications.

\subsection{\label{sec:level2A}
Newtonian  Post-Circular scheme to compute 
$e_t(F)$ and $\Psi(F)$ }

 The starting point of the conventional PC scheme is a differential equation for $d \omega/de_t$ which arises from Eqs.~(\ref{eq:domgdt}) and (\ref{eq:detdt}).
 This leads to $d\omega/de_t = \omega\,\kappa_N(e_t)$  where
\begin{align}
\kappa_N = & \, -\frac{3}{e_t}\bigg[\frac{96 + 292 e_t^2 + 37 e_t^4}{(1-e_t^2)(304+121 e_t^2)}\bigg].  \label{eq:kappaN}    
\end{align}
It is rather straightforward to integrate a resulting expression,
namely  $d\omega/\omega  = \kappa_N(e_t)\, de_t$ and we get 
\begin{align}
\frac{\omega}{\omega_0} = &\, \frac{(1-e_t^2)^{3/2}e_0^{18/19}(304+121 e_0^2)^{1305/2299}}{(1-e_0^2)^{3/2} e_t^{18/19}(304+121 e_t^2)^{1305/2299}}. \label{eq:omegaofet}
\end{align}
where we have $\omega(e_0) = \omega_0$. This implies that 
$\omega_0$ and $e_0$ are the values of $\omega$ and $e_t$ at some initial epoch.

Clearly,  it is difficult to invert  Eq.~(\ref{eq:omegaofet})  analytically to obtain 
a closed form expression for $e_t(\omega,\omega_0,e_0)$. However, Eq.~(\ref{eq:omegaofet}) can be inverted \textit{numerically} to obtain frequency evolution of any arbitrary orbital eccentricity.

In contrast, the PC scheme which assumes $e_t\ll 1$, $e_0 \ll 1$  allows us to 
obtain analytical $e_t(\omega,\omega_0,e_0)$ expression from Eq.~(\ref{eq:omegaofet}).
This is possible as one can extract certain asymptotic eccentricity invariant 
from Eq.~(\ref{eq:omegaofet})  in the small eccentricity limit, as noted in Ref.~\cite{KKS},
which leads to the constancy of $e_t^2\, \omega^{19/9}$ in such a limit.
In practice, we Taylor expand 
Eq.~(\ref{eq:omegaofet}) around $e_t, e_0 = 0$ while keeping only the leading 
order terms in $e_t$ and $e_0$  to obtain 
\begin{align}\label{eq:ete0LO}
e_t \sim \frac{e_0}{\chi^{19/18}} + \mathcal{O}(e_0^3),     
\end{align}
where $\chi$ is defined as $\omega/\omega_0 = F/F_0$.

We are now in a position to implement analytically the equation for the  crucial Fourier phase,
given by Eq.~(\ref{eq:Psi0}). 
With the help of chain rule, 
 the time and phase variables that appear in the expression for $\Psi_j$  read
\begin{align}
t(F) = & \, \int_{}^{F} \frac{\tau'}{F'}dF', \label{eq:tofF} \\ 
\phi(F) = & \, 2 \pi \int_{}^{F} \tau' dF'\,. \label{eq:phiofF} 
\end{align}
This allows us to write Eq.~(\ref{eq:Psi0}) as 
\begin{align}
\Psi_j[F(t^*_j)] = & \, 2 \pi \int^{F(t^*_j)} \tau' \left(j-\frac{f}{F'}\right) dF', \label{eq:Psidef}   
\end{align}
where $\tau = F/\dot{F} = \omega/\dot{\omega}$. 
It is important to emphasize that the above integral for $\Psi_j[F(t^*_j)]$ 
should be evaluated at certain stationary points $t^*_j$ such that $F(t^*_j) = f/j$ as demanded by the {\it stationary phase condition} \cite{YABW}.
Therefore, one usually computes explicit expressions for the time and phase variables
of Eq.~(\ref{eq:Psi0}) with the help of Eqs.~(\ref{eq:tofF}) and (\ref{eq:phiofF}) in the PC scheme.
Clearly, we  require an expression for $\tau$ in terms of $e_0$, $F$ and $F_0$ to perform the integral in Eq.~(\ref{eq:Psidef}).  This is done by 
 taking the ratio of the orbital frequency $F$ and the orbital averaged time evolution
 equation for ${F}$ while using Eq.~(\ref{eq:domgdt}) as $\dot{F} = \dot{\omega}/2\pi$. 
 We employ Eq.~(\ref{eq:ete0LO}) for $e_t$ appearing 
  in the resulting expression of $\tau$ 
  and this leads to 
\begin{align}
\tau \sim & \, \frac{5}{96 \eta x^4}\left(\frac{G m}{c^3}\right)\left[1-\frac{157 e_0^2}{24 \chi^{19/9}}+\mathcal{O}(e_0^4)\right]\,. \label{eq:tau}     
\end{align}

 We now invoke Eq.~(\ref{eq:tau}) for $\tau$ in Eq.~(\ref{eq:Psidef}) for $\Psi_j(F)$
 and this results in 
\begin{widetext} 
\begin{align}
\Psi_j = &\, j \phi_c-2 \pi f t_c-\frac{3\,j}{256\,\eta\,x^{5/2}}\bigg[1-\frac{2355\,e_0^2}{1462}\chi^{-19/9}+\mathcal{O}(e_0^4)\bigg] \,, \label{eq:Psi_NLO}     
\end{align}
\end{widetext}
where 
$t_c$ and $\phi_c$ stand for the time and the corresponding phase at the coalescence
and arise as the constants of integration in Eq.~(\ref{eq:tofF}) and (\ref{eq:phiofF}). 
We note again that the above Fourier phase expression  should be 
 computed at the stationary points which will map the orbital frequency $F$ to the Fourier frequency $f$. In other words, we should replace $F$ and $F_0$ with $f/j$ and $f_0/j$ respectively,
 to operationalize the above expression.

It is fairly straightforward to extend the above computations to incorporate higher order
corrections in $e_0$. A crucial ingredient for that effort involves deriving 
an analytic expression for $e_t$ that extends Eq.~(\ref{eq:ete0LO}).
This requires us to Taylor expand  Eq.~(\ref{eq:omegaofet}) for
 $\omega/\omega_0$ in the limit $e_t\ll 1$, $e_0 \ll 1$ 
 that includes the next-to-leading order terms in $e_t$ and $e_0$.
 Thereafter, one need to employ
 the above $e_t$ expression at the sub-leading $e_t$ contributions and 
 invert the resulting $\omega/\omega_0$ expression for $e_t$. 
 This approach can in principle be extended to any higher order in $e_0$ and we list
below the Newtonian accurate 
expression for $e_t(e_0,\chi)$ that incorporates $\mathcal{O}(e_0^7)$  corrections 
as
\begin{widetext}
\begin{align} \label{eq:ete07}
e_t = &\, \frac{e_0}{\chi ^{19/18}}+\left(-\frac{3323}{1824 \, \chi ^{19/6}}+\frac{3323}{1824 \, \chi
   ^{19/18}}\right) e_0^3+\left(\frac{50259743}{6653952 \, \chi
   ^{95/18}}-\frac{11042329}{1108992 \, \chi ^{19/6}}+\frac{15994231}{6653952 \, \chi
   ^{19/18}}\right) e_0^5 \\ \nonumber & +\left(-\frac{1472105896313}{36410425344 \, \chi
   ^{133/18}}+\frac{835065629945}{12136808448 \, \chi
   ^{95/18}}-\frac{42178716049}{1348534272  \,\chi ^{19/6}}+\frac{105734339801}{36410425344 \,
   \chi ^{19/18}}\right) e_0^7+\mathcal{O}(e_0^9).    
\end{align}
\end{widetext}
We have verified that the above expression is identical to Eq.~(3.11) in  Ref.~\cite{YABW}.
Employing the above expression, 
it is straightforward to compute the extension of Eq.~(\ref{eq:Psi_NLO}) that incorporates  all $\mathcal{O}(e_0^8)$ corrections. These steps can be 
therefore extended to include still higher order $e_0$ contributions to the crucial  orbital eccentricity and Fourier phases expressions.

\begin{figure}[htp]
\begin{center}
     \includegraphics[width=0.5\textwidth,angle=0,trim=0 5 0 0, clip]{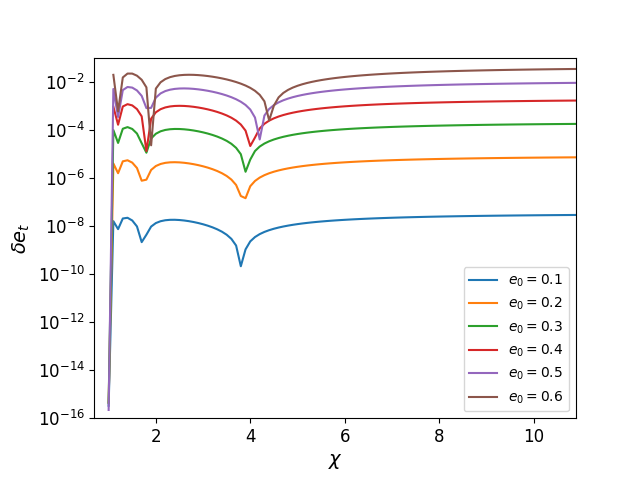}
\end{center}
 \caption{\label{fig:NumTay_e0_ordr7}
 Fractional $e_t$ errors, namely 
 $\delta e_t = |1- (e_t^{PC}/e_t^{Num}) |$, 
  as a function of $\chi = f/f_0$
  for three initial $e_t$ values at 
  $f_0 = 20\,$Hz. These plots are 
  for a BH binary with $m =20\,M_{\odot}$ as we terminate 
  the $f$ evolution at $f= c^3/(G\,m\,\pi\,6^{3/2})$Hz. 
 Clearly, the observed sharp rise in $\delta e_t$ 
 values depend critically on  $e_0$ values and 
 such observations essentially  prompted 
   Ref.~\cite{MRLY18} argue against the use of PC approach 
   to model eccentric inspirals having $e_0>0.2$.
 Specifically,  plots for $e_0>0.3$ have $\delta e_t > 10^{-4} $ and therefore 
 the PC approach should not be used to model such inspirals. Sharp dips in these plots are due to the chance cancellation of
  $e_0^3$ terms in the Eq.~(\ref{eq:ete07}) for $e_t$. }
\end{figure}

 The above $e_t$ expression and its extensions can be used to explore the validity of the PC scheme, as pursued in Ref.~\cite{MRLY18}.
The idea is to compare $e_t$ values that arise 
from the above $e_t$ expression (or its extensions)  with   
their counterparts that are obtained by 
 numerically inverting $\omega(e_t,e_0,\omega_0)$ in Eq.~(\ref{eq:omegaofet}) for various $e_0$ and a range of $\chi$ values.
In Fig.~\ref{fig:NumTay_e0_ordr7}, we plot 
$\delta e_t = |1- (e_t^{PC}/e_t^{Num}) |$,
where $e_t^{PC}$ values are associated with Eq.~(\ref{eq:ete07}) 
while  $e_t^{Num}$ values arise by inverting Eq.~(\ref{eq:omegaofet}) numerically.
Interestingly, both  $e_t$ estimates are independent 
of the intrinsic compact binaries parameters like their masses 
as we are dealing with the effect of quadrupolar order 
GW emission. 
However, the plots in  Fig.~\ref{fig:NumTay_e0_ordr7}
are for a $(10 M_\odot- 10 M_\odot ) $ BH binary as we terminate 
the GW emission induced $e_t$ evolution when 
the orbital frequency reaches 
$\omega = c^3/(G\,m\,6^{3/2})$.
This is of course the orbital frequency of the innermost stable circular orbit of a test particle moving along the geodesics in the Schwarzschild space-time. 
Further, we let $\omega_0$ to be $ 20\, \pi $ which 
corresponds to the lower frequency cut-off for the ground-based GW detectors like aLIGO.
    
We observe that the fractional relative errors between $e_t^{PC}$ and $e_t^{Num}$ grow rapidly from $10^{-8}$ to $10^{-4}$ as $e_0$ goes from $0.1$ to $0.3$.  It turned out that 
a fractional $e_t$ error $\sim 10^{-4} $ or higher can have 
undesirable data analysis implications at the quadrupolar order, 
as noted in Ref.~\cite{MRLY18}.
Our plots reveal that compact binaries with initial eccentricities above $0.3$ can develop relative errors  that are above 
$10^{-4}$.
This essentially prompted Ref.~\cite{MRLY18}
to question the usefulness of the PC scheme for constructing 
templates for eccentric inspirals.
In what follows, we summarize a rather  
semi-analytic approach of Ref.~\cite{MRLY18}
that allows one to construct
 quadrupolar order $\tilde{h}(f)$, valid for arbitrary initial eccentricities, influenced by Ref.~\cite{BBPM12}.

\subsection{\label{sec:level2B}
Moore-Robson-Loutrel-Yunes (MoRoLoYu) approach to improve the PC scheme}

 The new prescription of Ref.~\cite{MRLY18} crucially avoids the Taylor expansion of $\omega/\omega_0$ expression, given by Eq.~(\ref{eq:omegaofet}),  for obtaining an analytic expression for 
 $e_t$ in terms of $e_0, \chi$.
 This is essentially influenced by the fact that the PC scheme 
 does not provide an accurate prescription for the frequency 
 evolution of $e_t$ as evident from our 
  Fig.~\ref{fig:NumTay_e0_ordr7} and the associated discussions.
 Their approach employs the following orbital frequency version of 
 Eq.~(\ref{eq:omegaofet}) 
\begin{align}
\frac{F}{F_0} = &\, \frac{(1-e_t^2)^{3/2}e_0^{18/19}(304+121 e_0^2)^{1305/2299}}{(1-e_0^2)^{3/2} e_t^{18/19}(304+121 e_t^2)^{1305/2299}} \label{eq:Fofet},
\end{align} \\ \\ \\
where it is natural to define $F_0$ and $e_0$ using  the relation  $F(e_0) = F_0$.  In practice, $F_0$ provides the  orbital frequency of a compact binary whose dominant harmonic corresponds to the lower GW frequency cutoff of the detector and $e_0$ is the eccentricity of the system at $F_0$.
We employ numerical inversion of the above expression to obtain the GW frequency evolution of $e_t$ after imposing the SPA condition. This ensures that $e_t(f)$ prescription should be valid for all allowed $e_0$ values, namely  $0 < e_0 < 1$.
We note that certain analytic inversion approaches were provided in Ref.~\cite{MRLY18}
to avoid numerical inversion. However, we follow the straightforward 
numerical inversion to ensure no additional approximations are introduced while 
extracting $e_t(f)$ from Eq.~(\ref{eq:Fofet}).

   The MoRoLoYu approach provides a different prescription to compute the crucial Fourier phase of Eq.~(\ref{eq:Psi0}) to ensure that it is also valid for
 $0 < e_0 < 1$ cases.
This involves providing appropriate  expressions for the angular and temporal functions that appear in the definition of $\Psi_j$, namely  
$\Psi_j := j \phi(t^*_j) - 2 \pi f t^*_j$,  while not employing the PC scheme. It is straightforward to re-write these functions as 
\begin{align}
t-t_c = & \, \int_{0}^{e_t} \frac{de'_t}{\dot{e_t}(e'_t)}, \label{eq:tofe} \\ 
\phi-\phi_c = & \, 2 \pi \int_{0}^{e_t} \frac{F(e'_t)}{\dot{e_t}(e'_t)}de'_t. \label{eq:phiofe} 
\end{align}
To find closed form expressions for these integrals, we need 
a number of substitutions. First, we replace  
$F(e_t)$ in Eq.~(\ref{eq:phiofe}) with our quadrupolar order  Eq.~(\ref{eq:Fofet}). 
The $\dot{e}_t$ expression that appears in  Eqs.~(\ref{eq:tofe}) 
and (\ref{eq:phiofe})  is replaced by the quadrupolar order 
$de_t/dt$  equation while using Eq.~(\ref{eq:Fofet}) for $F$.
These substitutions ensure that the integrands of above two integrals
depend only on $e_t, e_0$ and $F_0$.
This leads to 

\begin{align}
t-t_c = & \, - \frac{15\, G\, m\, c^6}{ 304\, \eta\, (2\, \pi\, G\, m\, F_0)^{8/3}\, \sigma(e_0)^4}\,I_t(e_t), \label{eq:tofe1}\\
\phi-\phi_c = & \, - \frac{30\, \pi}{304\, \eta\, (2\, \pi\, G\, m\, F_0)^{5/3} \,\sigma(e_0)^{5/2}}\,I_l(e_t), \label{eq:phiofe1}
\end{align}
where the three new symbols are defined to be 
\begin{align}
\sigma(e_0) = & \, \frac{e_0^{12/19}}{1-e_0^2}\left(1+\frac{121}{304}e_0^2\right)^{870/2299}, \label{eq:sigmae0} \\
I_t(e_t) = & \, \frac{19}{48}\,e_t^{48/19}F_1\left(\frac{24}{19};-\frac{1181}{2299},\frac{3}{2};\frac{43}{19};-\frac{121}{304}e_t^2,e_t^2\right), \label{eq:Ite} \\
I_l(e_t) = & \, \frac{19}{30}\,e_t^{30/19} {}_2F_1\left(\frac{124}{2299},\frac{15}{19};\frac{34}{19};-\frac{121}{304}e_t^2\right), \label{eq:Ile} 
\end{align}
while $F_1$ and ${}_2F_1$ 
stand for the ApellF1 hypergeometric function  and 
the generalised hypergeometric function, respectively.
We now employ these integrals in the $\Psi_j$ equation, given by 
Eq.~(\ref{eq:Psi0}), and after a few straightforward 
simplifications  obtain 
\begin{widetext}
\begin{align}
\Psi_j = j \phi_c -2 \pi f t_c - j \frac{15}{304\, \eta}\left(\frac{c^3}{2\, \pi\, G\, m\, F_0}\right)^{5/3} \sigma(e_0)^{-5/2} e_t^{30/19} I(e_t), \label{eq:Psihgf}  \end{align}
\end{widetext}
where $I(e_t)$ is a combination of $I_t(e_t)$ and $I_l(e_t)$ and is given by
\begin{widetext}
\begin{align}
I(e_t) = & \, \frac{19}{48\left(1+\frac{121 e_t^2}{304}\right)^{124/2299}} \times F_1\left(1;-\frac{1181}{2299},\frac{3}{2};\frac{43}{19};\frac{121 e_t^2}{304+121 e_t^2},\frac{e_t^2}{e_t^2-1}\right)-\frac{19}{30}{}_2F_1\left(\frac{124}{2299},\frac{15}{19};\frac{34}{19};-\frac{121 e_t^2}{304}\right).  
\end{align}
\end{widetext}

 For our investigations, we followed few 
 additional steps to convert the above 
$\Psi_j(F_0,e_0,e_t)$ 
expression for obtaining 
the Fourier-domain phase that should depend on  $\Psi_j(f,f_0,e_0)$. 
These include first obtaining $e_t(F)$ by 
numerically inverting Eq.~(\ref{eq:Fofet}) at each desired value of frequency $F$ and employing it in 
Eq.~(\ref{eq:Psihgf}) to get $\Psi_j$ at that $F$ value. 
Thereafter, we 
invoked the stationary phase approximation which demands that the Fourier phase must be computed only at Fourier frequencies which are  integral multiples of the orbital frequency $F$.
 Note that the above approach to obtain $\Psi_j (f)$ treats orbital eccentricities 
 in an exact manner and therefore the MoRoLoYu approach is valid for 
 compact binaries of arbitrary bound  eccentricities : $0<(e_0, e_t)<1$. 
 In our implementation of the approach, we did not employ various fits and 
 approximations suggested in Sec.~IV~B of Ref.~\cite{MRLY18}.
 This is obviously to ensure that an accurate implementation of the NeF model is used
 for benchmarking our approaches. We would like to state that we Taylor expanded 
 the explicit 
expressions for $C_{+,\times}^{(j)}$ and $S_{+,\times}^{(j)}$,
expressed 
in terms of Bessel functions of first kind as given by Eqs.~(9) in Ref.~\cite{MRLY18},
while constructing the amplitudes of these templates.
However, we did perform 
 several numerical tests to ensure that such expansions in the amplitudes 
 do not affect any of our conclusions. In what follows, we describe a way to obtain analytically quadrupolar order Fourier domain GW response function that should be valid up to moderately high initial eccentricities like $e_0 \sim 0.6$.
 
\begin{figure}[htp]
\begin{center}
\includegraphics[width =0.5\textwidth,angle = 0,trim=0 0 0 0, clip]{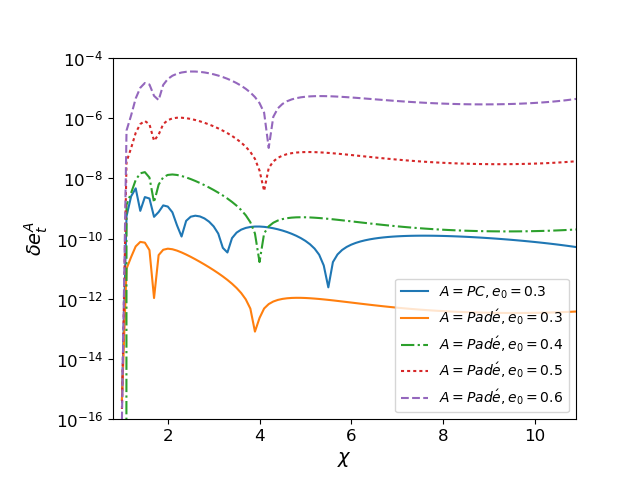} 
\caption{
Plots that mainly show fractional errors in $e_t$ values while employing our 
Newtonian Pad\'e approximant
for $e_t$ and Eq.~(\ref{eq:Fofet})
as a function of $\chi$ for various 
$e_0$ values.  We do not display $e_0=0.1$ and $0.2$ plots as their Pad\'e based fractional errors are below $10^{-16}$ and essentially represent the rounding errors generated by the computing algorithm. These plots are for  $(10 \, M_{\odot},10\, M_\odot )$ BH-BH binary
as in Fig.~\ref{fig:NumTay_e0_ordr7}.
For making easy comparisons, we
over plot  $\delta e_t =  |1- (e_t^{PC}/e_t^{Num})| $ that 
employs PC scheme based $e_t$ expression  that includes $\mathcal{O}(e_0^{19})$  corrections
for the $e_0=0.3$ case. 
It turns out that Pad\'e approximant usually provides two orders of magnitude improvements 
in these $\delta e_t$ estimates compared to their PC counterparts.
We find that our Pad\'e approximant is capable of smoothly following the exact quadrupolar order $e_t(f)$ evolution 
for compact binaries even with $e_0$ values around $0.6$.
Note that it is not computationally expensive to obtain higher order Pad\'e approximant
to improve $\delta e_t$ estimates for $e_0$ values around $0.6$.}
\label{fig:NumTayPde_e0_ordr19}
\end{center}
\end{figure}

\subsection{\label{sec:level2C}
Pad\'e approximation to model quadrupolar order eccentric inspirals
}

We now explore  the possibility of rescuing the PC scheme 
with the help of 
 an easy and elegant way of resumming a poorly converging 
power series. 
Clearly, our PC scheme based  
 analytical $e_t(f)$ expression of Sec.~\ref{sec:level2A} 
 that invoked Taylor expansion 
 does not converge to numerically computed 
 $e_t$ values, based on an exact $\omega( e_t,e_0, \omega_0)$ expression. This prompted us to employ the popular Pad\'e approximation, detailed 
in Ref.~\cite{BenOrz}, 
for computing the $e_t(f)$ and subsequently $\Psi_j(f)$ expressions analytically.

It turns out that Pad\'e approximation is helpful for obtaining time-domain inspiral templates for compact binaries 
in PN-accurate eccentric orbits \cite{THG}.
This approximation allowed us to obtain 
closed form expressions for the hereditary contributions to both GW energy and angular momentum 
fluxes, which are crucial for computing such templates.
Specifically, Pad\'e approximation can be employed 
to re-sum certain infinite series expressions 
for the PN-accurate hereditary contributions to GW fluxes from compact binaries in eccentric orbits.
Additionally, Pad\'e approximation was invoked to compute GW inspiral template families for quasi-circular inspirals from their Taylor expanded PN counterparts that incorporate 
higher order PN corrections in terms of the $x$ parameter
in Ref.~\cite{DIS98}. These template families,
referred to as the {\it Pad\'e approximants},
were shown to be more 
{\it effectual} and {\it faithful}  compared to their Taylor expanded PN counterparts\cite{DIS98}. 
We note in passing that 
Pad\'e approximation was employed to model neutron stars
and it converges faster to the
underlying general relativistic solution than the truncated post-Newtonian ones \cite{GGI_00}.

The simplest form of  Pad\'e approximation  
involves a {\it rational function of two polynomials} that provides the original 
truncated power series under Taylor expansion.
Formerly, the simplest Pad\'e approximant to a truncated power series 
$S_u(z)$ in the variable $z$ may be written as 
\begin{align}
P^m_s(z) = & \, \frac{N_m(z)}{D_s(z)},\label{eq:PadeDef}    
\end{align}
where $N_m(z) = \sum_{i=0}^m n_i z^i$ and $D_s(z) = \sum_{i=0}^s d_i z^i$ are polynomials in $z$ of order $m$ and $s$, respectively. 
To find the coefficients that define these polynomials, we Taylor expand the approximant $P^m_s(z)$ upto the same order in $z$ as the original 
truncated power series $S_u(z)$ and then solve the resulting set of linear 
equations. In other words, if
$T_u[...]$ denotes the operation of Taylor expanding any function upto an order $u$ of its variable and $S_u$ stands for the truncated Taylor series whose Pad\'e approximant we are seeking, we define $P^m_s$ such that,
\begin{align}
T_u[P^m_s(z)] = &\, S_u(z), \label{eq:PadeTay}     
\end{align}
where $m+s=u$ is a mandatory condition with $z^u$ being the highest order term in the Taylor series required for constructing Pad\'e approximant $P^m_s(z)$. 

It is now straightforward to employ the above detailed 
Pad\'e approximation on the PC scheme based analytical $e_t$ and $\Psi_j$ expressions, obtained in Sec.~\ref{sec:level2A}. For the present Pad\'e computations, we have obtained  Newtonian accurate $e_t(e_0,\chi)$ and $ \Psi_j (e_0,\chi)$ expressions that incorporate
$\mathcal{O}(e_0^{19})$ and $\mathcal{O}(e_0^{20})$ corrections, 
respectively in the initial orbital eccentricity.
This allows us to obtain the following fully analytic 
Pad\'e approximant for $e_t(e_0,\chi)$ as 
\begin{widetext}
\begin{align}
e_t = & \, e_0 \frac{\bar{n}_0+ \bar{n}_1 \, z+ \bar{n}_2 \, z^2+ \bar{n}_3 \, z^3+ \bar{n}_4 \, z^4+ \bar{n}_5\, z^5}{1+ \bar{d}_1 \, z+ \bar{d}_2 \, z^2+ \bar{d}_3 \,  z^3+ \bar{d}_4 \, z^4}, \label{eq:etPade}
\end{align}
\end{widetext}
where $z = e_0^2$. The coefficients $\bar{n}_j$ and $\bar{d}_k$, where $j$ runs 
from $0$ to $5$ while $k$ runs from $1$ to $4$, can easily be computed 
from the PC scheme based $e_t(e_0,\chi)$ expression that is 
$\mathcal{O}(e_0^{19})$ accurate, as noted earlier.
The explicit expressions for all these $10$ coefficients are available in 
the accompanying \texttt{Mathematica} notebook and we display few of them 
for the sake of introducing the inherent structure to the readers:
\begin{widetext}
\begin{subequations} \label{eq:nbar_dbar}
\begin{align}
\bar{n}_0 =&\, u^ {-1/2},\\ \nonumber \\
\bar{n}_1 =&\, u^{-3/2}\,\left\{\left(-0.382209 + 1.86199\, u - 3.89605\, u^2 + 4.72722\, u^3 - 3.85449\, u^4 +  2.27674\, u^5 - 0.939021\, u^6 + 0.22659\, u^7 \right. \right. \nonumber \\ &\, - 0.0177036\, u^8 - 0.00373226\, u^9 + 0.000682323\, u^{10} - 0.0000175432\, u^{11} + 5.63003 \times 10^{-7}\, u^{12} + 3.69953 \times 10^{-9}\, u^{13} \nonumber \\ &\,\left.  - 1.1001 \times 10^{-11}\, u^{14} -  1.0129 \times 10^{-15}\, u^{15}\right)/ \left(-0.0220143 + 0.100073\, u - 0.192567\, u^2 + 0.213443\, u^3 - 
 0.161031\, u^4  \right. \nonumber \\ &\, + 0.0889986\, u^5 - 0.0323116\, u^6 + 0.00527697\, u^7 + 0.000319898\, u^8 - 0.000193037\, u^9 + 5.35235 \times 10^{-6}\, u^{10} \nonumber \\ &\, \left.\left.-2.75227 \times 10^{-7}\, u^{11} - 2.38813 \times 10^{-9}\, u^{12} + 1.19436 \times 10^{-11}\, u^{13} + 2.38851 \times 10^{-15}\, u^{14}\right)\right\} \,, \\ \nonumber \\
\bar{d}_1 =&\, u^{-1}\,\left\{\left(-0.105579 + 0.521104\, u - 1.1073\, u^2 + 1.36672\, u^3 - 1.13418\, u^4 +  0.683062\, u^5 - 0.290007\, u^6 + 0.0737674\, u^7 \right. \right. \nonumber \\ &\,  - 0.00668361\, u^8 - 0.00116669\, u^9 + 0.000260938\, u^{10} - 6.94891 \times 10^{-6}\, u^{11} + 2.65017 \times 10^{-7}\, u^{12} + 2.01801 \times 10^{-9}\, u^{13} \nonumber \\ &\,  \left. - 8.18895 \times 10^{-12}\, u^{14} - 1.34108 \times 10^{-15}\, u^{15}\right)/\left(-0.00550358 + 0.0250184\, u - 0.0481419\, u^2 + 0.0533608\, u^3 \right. \nonumber \\ &\, - 0.0402577\, u^4 + 0.0222497\, u^5 - 0.00807791\, u^6 + 0.00131924\, u^7 +  0.0000799744\, u^8 - 0.0000482594\, u^9  \\ &\,  \left.\left. + 1.33809 \times 10^{-6}\, u^{10} -  6.88068 \times 10^{-8}\, u^{11} - 5.97032 \times 10^{-10}\, u^{12} + 2.9859 \times 10^{-12}\, u^{13} + 5.97128 \times 10^{-16}\, u^{14}\right)\right\} \,, \nonumber \\ \nonumber \\
\bar{d}_2 =&\, u^{-2}\,\left\{\left(-6685.94 + 35543.3\, u - 82569.5\, u^2 + 112439.\, u^3 - 102405.\, u^4 +  66990.1\, u^5 - 31807.1\, u^6 + 10322.2\, u^7 \right. \right. \nonumber \\ &\,  - 1965.07\, u^8 + 116.15\, u^9 + 26.8824\, u^{10} - 4.82738\, u^{11} + 0.108707\, u^{12} - 0.00375904\, u^{13} - 
 0.0000213561\, u^{14} \nonumber \\ &\, \left. + 6.78027 \times 10^{-8}\, u^{15} + 8.25127 \times 10^{-12}\, u^{16}\right)/\left(-53.5388 + 243.379\, u - 468.324\, u^2 + 519.094\, u^3 - 391.627\, u^4 \right. \nonumber \\ &\,  + 216.445\, u^5 - 78.5819\, u^6 + 12.8336\, u^7 + 0.777991\, u^8 - 0.469467\, u^9 + 0.0130169\, u^{10} - 0.000669353\, u^{11} \nonumber \\ &\, \left. \left. - 5.80793 \times 10^{-6}\, u^{12} + 2.90468 \times 10^{-8}\, u^{13} + 5.80886 \times 10^{-12}\, u^{14}\right)\right\}\,,  \\ \nonumber 
\end{align}
\end{subequations}
\end{widetext}

where $u = \chi^{19/9}$. It should be obvious that we restricted our attention to a very specific rational polynomial form. This was essentially the result of many numerical experiments that compared 
$e_t$ values from various Pad\'e approximations against the accurate numerical 
evaluations of Eq.~(\ref{eq:Fofet}) for $e_t$. The above form turned out to be the 
minimalistic
$e_t(e_0,\chi)$ expression that provided 
fractional relative errors, namely 
$ \delta e_t^{Pad\acute{e}} = |1-(e_t^{Pad\acute{e}}/e_t^{Num})| $,
  that are  $ \sim 10^{-5} $ even for 
$e_0 \sim 0.6$ cases as evident from Fig.~\ref{fig:NumTayPde_e0_ordr19}.
Further, the construction of 
higher order  Pad\'e approximants didn't 
necessarily produce fractional errors substantially below the 
threshold of $10^{-4}$ of Ref.~\cite{MRLY18}  at 
moderately high initial eccentricities like $e_0 \sim 0.6$.

 We now present a symbolic Pad\'e approximation based expression for 
 a crucial ingredient to compute frequency domain inspiral templates, namely
 the  Fourier phase $\Psi$ of Eq.~(\ref{eq:Psi0}).
 The PC ingredient for our computation involves quadrupolar order 
 $\Psi_j$ expression that includes $\mathcal{O}(e_0^{20})$ corrections in 
 initial eccentricity, as noted earlier.
The resulting Pad\'e approximant reads 
\begin{widetext}
\begin{align}
\Psi_j = &\,j \phi_c - 2 \pi f t_c -\frac{3\,j}{256\,\eta\,x^{5/2}} \frac{\hat{n}_0+ \hat{n}_1 \, z+ \hat{n}_2 \, z^2+ \hat{n}_3 \, z^3+ \hat{n}_4 \, z^4+ \hat{n}_5\, z^5+ \hat{n}_6 \, z^6}{1+ \hat{d}_1 \, z+ \hat{d}_2 \, z^2+ \hat{d}_3 \,  z^3+ \hat{d}_4 \, z^4}\,, \label{eq:PsiPade}
\end{align}
\end{widetext}
where explicit form of these new coefficients $\hat{n}_j$ and $\hat{d}_k$ 
are listed in the attached \texttt{Mathematica} notebook. Explicit form for 
few of these coefficients read 
\begin{widetext}
\begin{subequations} \label{eq:nhat_dhat}
\begin{align}
\hat{n}_0 =&\, 1  \,, \\ \nonumber \\
\hat{n}_1 =&\,u^{-1}\, \left\{\left(0.00372532 - 0.0413365\, u + 0.208387\, u^2 - 0.634955\, u^3 + 1.31478\, u^4 - 1.97388\, u^5 + 2.22821\, u^6 - 1.89854\, u^7 \right. \right. \nonumber  \\ &\, + 1.14607\, u^8 - 0.353449\, u^9 - 0.136235\, u^{10} + 0.249108\, u^{11} - 0.159735\, u^{12} + 0.0596684\, u^{13} - 0.0133939\, u^{14} \nonumber \\ &\,  + 0.0016679\, u^{15} - 0.000101623\, u^{16} + 8.60407 \times 10^{-6}\, u^{17} - 1.7774 \times 10^{-6}\, u^{18} + 4.74674 \times 10^{-8}\, u^{19}  \nonumber \\ &\, \left. + 1.08886 \times 10^{-13}\, u^{20} - 1.8703 \times 10^{-13}\, u^{21}\right)/\left(0.000213461 - 0.00225419\, u + 0.0107362\, u^2 - 0.0306904\, u^3 \right. \nonumber \\&\,  + 
 0.0593036\, u^4 - 0.0828513\, u^5 + 0.086702\, u^6 - 0.067027\, u^7 + 
 0.0331534\, u^8 - 0.00218905\, u^9 - 0.0125614\, u^{10} \nonumber \\ &\, + 0.0117481\, u^{11} - 0.00561934\, u^{12} + 0.00155413\, u^{13} - 0.000233167\, u^{14} + 
 0.000015674\, u^{15} - 1.06586 \times 10^{-6}\, u^{16} \nonumber \\&\, \left. \left.+ 3.42586 \times 10^{-7}\, u^{17} -  1.06549 \times 10^{-8}\, u^{18} - 6.35488 \times 10^{-15}\, u^{19} + 5.09185 \times 10^{-14}\, u^{20}\right)\right\}\,,  \\ \nonumber \\
\hat{d}_1 =&\,u^{-1}\,\left\{\left(0.00556658 - 0.0615151\, u + 0.308729\, u^2 - 0.936239\, u^3 +  1.92928\, u^4 - 2.88281\, u^5 + 3.23922\, u^6 - 2.74489\, u^7 \right. \right. \nonumber \\ &\, +  1.64087\, u^8 - 0.488338\, u^9 - 0.214048\, u^{10} + 0.366665\, u^{11} - 0.230898\, u^{12} + 0.0850503\, u^{13} - 0.0188365\, u^{14}  \nonumber \\ &\,  + 0.0023162\, u^{15} - 0.000141368\, u^{16} + 0.0000125252\, u^{17} - 2.45494 \times 10^{-6}\, u^{18} + 6.49348 \times 10^{-8}\, u^{19} \nonumber \\ &\, \left.+ 2.61157 \times 10^{-13}\, u^{20} - 2.55855 \times 10^{-13}\, u^{21}\right)/\left(0.000292012 - 0.00308371\, u + 0.0146869\, u^2 - 0.0419841\, u^3 \right. \nonumber \\ &\,  + 0.0811266\, u^4 - 0.11334\, u^5 + 0.118607\, u^6 - 0.0916922\, u^7 + 0.0453535\, u^8 - 0.0029946\, u^9  - 0.0171838\, u^{10} \nonumber \\ &\,+ 0.0160713\, u^{11} - 
 0.0076872\, u^{12} + 0.00212603\, u^{13} - 0.000318971\, u^{14} + 
 0.0000214419\, u^{15} - 1.45808 \times 10^{-6}\, u^{16} \nonumber \\ &\,  \left. \left.+ 4.68655 \times 10^{-7}\, u^{17} - 1.45759 \times 10^{-8}\, u^{18} - 8.69341 \times 10^{-15}\, u^{19} + 6.96559 \times 10^{-14}\, u^{20}\right)\right\} \,, \\ \nonumber \\
 \hat{d}_2 =&\,u^{-2} \left\{\left(0.00334745 - 0.0387328\, u + 0.204895\, u^2 - 0.659079\, u^3 +  1.44779\, u^4 - 2.31304\, u^5 + 2.78762\, u^6 - 2.56801\, u^7 \right. \right. \nonumber \\&\,  + 
 1.76572\, u^8 - 0.804281\, u^9 + 0.104142\, u^{10} + 0.179923\, u^{11} - 
 0.174374\, u^{12} + 0.0859044\, u^{13} - 0.0262981\, u^{14} \nonumber \\ &\,  + 0.00499171\, u^{15} - 0.0005403\, u^{16} + 0.0000300179\, u^{17} - 2.25185 \times 10^{-6}\, u^{18} + 4.10531 \times 10^{-7}\, u^{19} \nonumber \\ &\, \left. - 9.86595 \times 10^{-9}\, u^{20} - 5.54463 \times 10^{-15}\, u^{21} + 3.18323 \times 10^{-14}\,u^{22}\right)/\left(0.0000266315 - 0.000281234\, u \right. \nonumber \\ &\, + 0.00133945\, u^2   - 0.00382895\, u^3 + 0.00739875\, u^4 - 0.0103366\, u^5 + 0.010817\, u^6 - 0.00836233\, u^7 + 0.00413624\, u^8 \nonumber \\ &\, - 0.000273108\, u^9 - 0.00156716\, u^{10} +  0.0014657\, u^{11} - 0.000701072\, u^{12} + 0.000193894\, u^{13} - 
 0.0000290901\, u^{14} \nonumber \\ &\, + 1.9555 \times 10^{-6}\, u^{15} - 1.32977 \times 10^{-7}\, u^{16} + 4.27413 \times 10^{-8}\, u^{17} - 1.32932 \times 10^{-9}\, u^{18} - 7.92839 \times 10^{-16}\, u^{19}\nonumber \\ &\, \left. \left. + 6.35262 \times 10^{-15}\, u^{20}\right)\right\}\,. \\ \nonumber \end{align}
\end{subequations}
\end{widetext}

\begin{figure}[htp]
    \begin{center}
    \includegraphics[width=0.5\textwidth, angle=0,trim=0 0 0 0, clip]{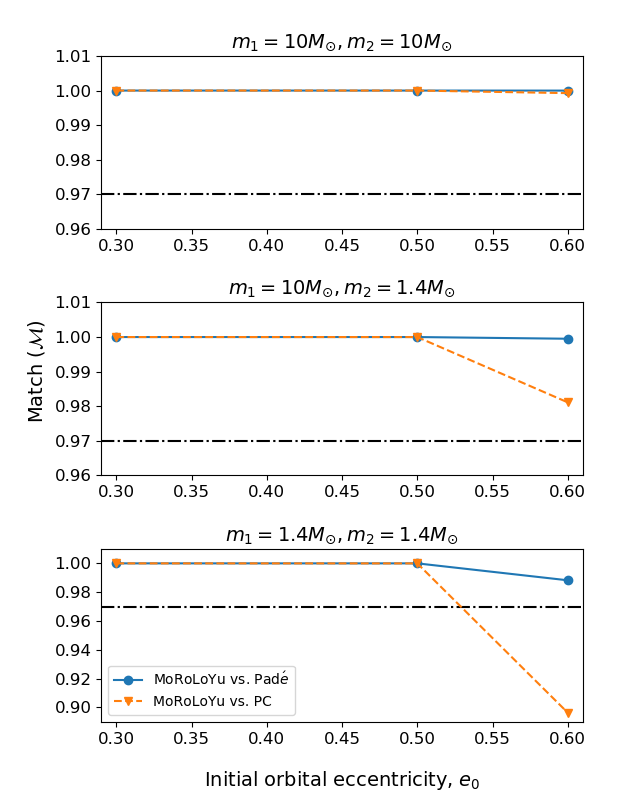}
     \end{center}
   { \caption{ \label{fig:Match_newt} 
    Match ($\mathcal{M}$) plots for the three traditional LIGO relevant compact binaries 
   having eccentricities up to $0.6$ at a GW frequency of $20$Hz.
   We let the expected eccentric GW signal to be described by the MoRoLoYu approach, 
   described in Sec.~\ref{sec:level2B}, while our 
   Pad\'e approximant provided the fully analytic eccentric inspiral templates.
   The dashed line marks the $0.97$ $\mathcal{M}$ value and it is evident that 
   our quadrupolar order Pad\'e approximant templates are both effectual and faithful 
   to our expected GW signal from data analysis considerations.
   The associated PC based templates show drop in match estimates around $e_0 \sim 0.5$. We have computed such computationally expensive $\mathcal{M}$ estimates at intermediate $e_0$ values randomly to ensure that these few point plots are representatives of a finely sampled $e_0$ match plots.
   }}
\end{figure}

  We are now in a position to compare our Pad\'e approximant $\tilde{h}(f)$ with the ones arising from our implementation of the MoRoLoYu and PC approaches. This is pursued with the help of 
$\mathcal{M}$ estimates. 
Recall that the $\mathcal{M}(h_s,h_t)$
 estimates provide certain {\it effectualness} and {\it faithfulness} criteria between the members of  two GW waveform families denoted here as $h_s$ and $h_t$ \cite{DIS98}.
 A template family $h_t$ is said to be {\it effectual} in detection and {\it faithful} for parameter estimation if it produces a match $\mathcal{M} \geq 0.97$ with a signal waveform $h_s$. An effectual template is desirable to ensure detection of more than $90\%$ of expected GW signals while a faithful template is mandatory to infer the signal parameters with smaller biases.
Following Ref.~\cite{DIS98}, we define 
\begin{align} \label{eq:Matchdef}
\mathcal{M} = & \, \max_{t_c,\phi_c} \frac{(h_s|h_t)}{\sqrt{(h_s|h_s)(h_t|h_t)}} ,  \end{align}
where the inner product $(a|b)$ is given as,
\begin{align} \label{eq:innerprod}
(a|b) = & \, 4\, Re \int_{f_l}^{f_u} \frac{\tilde{a}^*(f)\tilde{b}(f)}{S_n(f)} df.  
\end{align}
In practice, $h_s$ and $h_t$ may be treated as
the members of the expected GW signal and its approximate template families.
Further,  $S_n(f)$ stands for the one-sided noise spectral density of a GW detector and we use the zero-detuned high power (ZDHP) noise configuration of the Advanced LIGO at design sensitivity \cite{NoiseCurve}.
The limits of the above integral provide certain 
 lower and upper cut-off frequencies and we let $f_l=20$Hz.
For the present studies, we choose $f_u$ to be the popular GW frequency 
associated with the last stable circular orbit of a test particle in the Schwarzschild metric, namely $f_u = c^3/(G m \pi 6^{3/2}) $. Additionally, we have explored the effect of orbital eccentricity 
on the above $f_u$ estimate with the help of 
 Eqs.~(D1) and (D2) of Ref.~\cite{YABW}.
 The fact that orbital eccentricities were $ \sim 10^{-2}$ 
 at GW frequencies around $200$Hz 
 even for our $e_0 \sim 0.6$ systems justified the use 
of above expression for $f_u$ in our numerical experiments.
Additionally, we have explicitly verified that use of the above 
mentioned eccentric $f_u$ didn't affect our match estimates 
in high $e_0$ systems in any noticeably manner.
{In Fig.~\ref{fig:Match_newt}, we plot the $\mathcal{M}$-estimates 
for the three traditional compact binary systems having various values of initial orbital eccentricities.
The traditional binaries include 
 $1.4\,M_{\odot}-1.4\, M_{\odot}$ NS binaries,  $10\,M_{\odot}-10\, M_{\odot}$ BH binaries and their mixtures. 
We let $\tilde{h}(f)$ that arise from the MoRoLoYu approach to be the expected eccentric inspiral signal as detailed in 
Sec.~\ref{sec:level2B}.
The templates are provided by our Pad\'e approximant that employs Eq.~(\ref{eq:PsiPade}) for $\Psi_j$.
Further, we keep various amplitudes at the quadrupolar order and employ  Pad\'e 
approximation based expression for $e_t(f)$ (Eq.~(\ref{eq:etPade})) in both waveform families
for the ease of implementation.
The use of quadrupolar order amplitudes 
 is justifiable as match estimates crucially depend on the Fourier phase evolution differences and not on the amplitudes of underlying waveform families. Further, we usually included the first $22$ harmonics while pursuing 
our match computations. We have verified in many instances that 
the results were not sensitive to the number of harmonics used
by substantially increasing their numbers.

Plots in Fig.~\ref{fig:Match_newt} reveal that Pad\'e approximant is capable of providing 
$\mathcal{M}$ estimates that are $\geq 0.97$ even for $e_0$ values in the neighborhood of $0.6$.
This allows us to state that our quadrupolar order Pad\'e approximant should be 
both {\textit{effectual}} and {\textit{faithful}} to model inspiral GWs from compact binaries with moderately high initial eccentricities \cite{DIS98}.
In contrast, our numerical experiments show that the PC scheme based templates provide substantially lower $\mathcal{M}$ estimates especially for compact binaries that contain neutron stars. 
We now detail how to obtain 1PN extensions of these three approaches and their 
implications.
\newline
}

\section{\label{sec:level3}
Extending Eccentric Fourier-domain families to PN orders }

 We begin by summarizing how one extends the PC scheme to 1PN order, as detailed in Ref.~\cite{THG}. This is followed by a brief summary 
 of our detailed computations that essentially extend the MoRoLoYu approach 
 to 1PN order. Such a computation allows us to explore if the deficiencies
 of the PC scheme, evident at the quadrupolar order, persists even at the PN orders.
This is followed by a straightforward extension of our quadrupolar order 
Pad\'e approximant to 1PN order while keeping the amplitudes to the Newtonian order and exploration of its $\mathcal{M}$ estimate implications. Finally, we list subtleties of 
extending our Pad\'e approximant  to higher PN orders, influenced by Ref.~\cite{MY19}.

\subsection{1PN extension of the post-circular approximation} \label{sec:level3A}

We begin by describing how to compute 
an ingredient that is critical to extend the quadrupolar order PC scheme to 1PN order, namely 1PN accurate analytic $e_t$ expression with the leading order $e_0$ corrections.
This requires us to compute 1PN-accurate expression for $d \omega/ de_t$ 
by dividing  1PN-accurate expressions for   
$d\omega/dt$ and  $de_t/dt$, given by  Eqs.~(3.12) of Ref.~\cite{THG}.
This leads to
\begin{widetext}
\begin{align}
 d \omega/ d e_t
 =  \biggl \{ -\frac{18}{19 e_t} -\frac{3}{10108 e_t} \left(-2833+5516 \eta\right) \left(\frac{G m \omega}{c^3}\right)^{2/3} 
\biggr \}\, \omega 
  \,, \label{eq:domgbyomg}
\end{align}
\end{widetext}
where we have restricted our attention to the leading order $e_t$ contributions.
We now replace  $\omega$ that appears in the PN expansion parameter 
 by its Newtonian version, namely $\omega = \omega_0\, \left ( e_0/e_t \right )^{18/19}$.
 The resulting equation may be written as 
\begin{widetext}
\begin{align}
 d \omega/ \omega \sim \left\lbrace-\frac{18}{19 e_t} -  \frac{3}{10108}   \left(\frac{e_0^{12/19}}{e_t^{31/19}}\right)
    \left(-2833+5516\eta\right)\, x_0 
  \right\rbrace de_t\,,      \label{eq:domgbyomg1}
\end{align}
\end{widetext}
where  $ x_0= \left ( G\, m\, \omega_0/c^3 \right )^{2/3}$.
The above equation can be integrated to get $\ln(\omega/\omega_0)$ as a function of $\omega_0,e_t$ and $e_0$. The exponential of such an expression, followed by a bivariate expansion in $e_t$ and $x_0$ results in 
\begin{widetext}
\begin{align}
\omega & \sim \left\lbrace  \left(\frac{e_0}{e_t}\right)^{18/19} + x_0\left(\frac{2833-5516 }{2128}\eta   \right)     \left[  \left( \frac{e_0}{e_t}\right)^{18/19}     -       \left( \frac{e_0}{e_t}\right)^{30/19}      \right]             \right\rbrace   \omega_0  \,.
  \label{eq:domgbydomg2}
\end{align}
\end{widetext}
It should be obvious that we need to assume $x_0\ll1$ and $e_t\ll1$ during 
such a bivariate expansion and therefore we are implementing a PN-accurate version of the quadrupolar order PC scheme.
Explicit 1PN-accurate 
$e_t$ expression is obtained by first replacing $e_t$ terms that appear in the coefficients of the $x_0$ terms by its 
 Newtonian accurate expression, namely 
 $e_t = e_0\,\chi^{-19/18}$.
The resulting intermediate expression is inverted assuming 
 $x_0 \ll 1$ and $e_0 \ll 1$ which leads to
\begin{widetext}
\begin{align}
&e_t \sim e_0 
\left\lbrace  \chi ^{-19/18}+x_0      \left(\frac{2833}{2016}-\frac{197 }{72}\eta\right)    \left(- \chi ^{-7/18}   +  \chi
   ^{-19/18}   \right)    
\right\rbrace \,.                   \label{eq:et1PNx0}
\end{align}
\end{widetext}
We now re-write the above expression for $e_t$ in terms of usual PN parameter $x$ by noting that  $x_0 = x\, \chi^{-2/3}$. 
The resulting 1PN-accurate $e_t(e_0,x,\chi)$ 
that includes $\mathcal{O}(e_0)$ corrections reads
\begin{widetext}
\begin{align}
 &e_t \sim e_0 \left\lbrace   \chi^{-19/18}   +  x\left(   \frac{2833}{2016}-\frac{197 }{72}\eta    \right)\left( -\chi^{-19/18} +\chi^{-31/18} \right)              \right\rbrace         .   \label{eq:et1PNx}
\end{align}
\end{widetext}
It is fairly straightforward to repeat the computations of Sec.~\ref{sec:level2A} to obtain 
 $\Psi_j$ that incorporates $\mathcal{O}(e_0^2)$ eccentricity corrections with the help of Eq.~(\ref{eq:Psidef})  even at 1PN order, as detailed in Ref.~\cite{THG}.
 The final result is  
\begin{widetext}
\begin{align}
 &\Psi_j \sim     j \phi_c- 2\pi f t_c  -   \left(\frac{3 j  }{256 \eta \, x^{5/2}} \right)      \left\lbrace 1- \frac{2355 e_0^2 }{1462} \chi^{-19/9} +  x \left[ \frac{3715}{756}                 +\frac{55  
   }{9}\eta +   \left(   \left[   -\frac{2045665  }{348096}- \frac{128365 }{12432}\eta\right] \chi^{-19/9}   \right.\right.\right.\nonumber\\       
  &\qquad \left.\left.\left. {}      +  \left[ -\frac{2223905}{491232 }+\frac{154645  }{17544 }\eta\right]\chi^{-25/9}        \right)e_0^2 \right]\right\rbrace     \,,            \label{eq:Psi1PN}
 \end{align}
 \end{widetext}
 where $x= (\frac{G\,m\,2\,\pi\,F}{c^3})^{2/3}$ and $\chi=F/F_0$ have to be evaluated at the {\it stationary points} i.e. at $F = f/j$ and $F_0 = f_0/j$ with $j$ being the harmonic index.
 
 It is  straightforward but demanding to extend these calculations to include 
 higher order $e_0$ corrections. In fact, we have computed  1PN-accurate expressions for $e_t$ and $\Psi$ upto $\mathcal{O}(e_0^{19})$ and $\mathcal{O}(e_0^{20})$, respectively.
 The resulting 1PN-accurate PC scheme based $\tilde{h}(f)$ with quadrupolar order amplitudes will be used to explore the suitability of employing the PC scheme at 1PN order for eccentric inspirals. The other ingredient, namely 1PN extension of the MoRoLoYu approach will be discussed in the next subsection.

\begin{figure*}[htp] 
 \begin{center}
    \includegraphics[width=0.9\textwidth, angle=0,trim=0 0 0 0, clip]{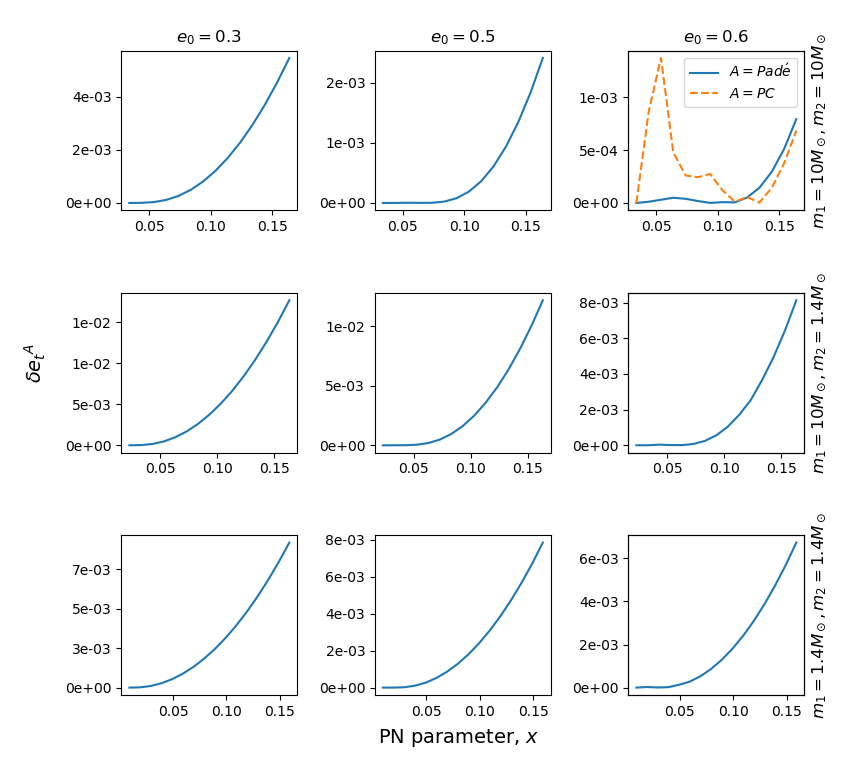}
     \end{center}
 \caption{
 Fractional errors in 1PN accurate $e_t$ values as a function of the 
 dimensionless $x$ parameter. The associated $e_t$ values are 
 obtained by i) numerically
 inverting Eq.~(\ref{eq:xofetexact}) for $x(e_t,e_0,x_0)$ 
 and ii) by using  our Eq.~(\ref{eq:etPade1PN})
 for the 1PN-accurate $e_t^{Pad\acute{e}}$.
 We restrict our attention to the traditional LIGO binaries with three 
 initial eccentricities of $0.3,0.5$ and $0.6$.
 In contrast to Fig.~\ref{fig:NumTayPde_e0_ordr19} plots, these fractional error plots 
 depend on intrinsic compact binary parameters like $m$ and $\eta$ and they begin at 
 $(G\,m\,20\,\pi/c^3)^{2/3}$ and end at $x = 1/6$.
 Sharp rise in $\delta e_t$ at higher $x$ values, visible in all plots,
 may be attributed to the different ways PN corrections are included 
 in our analytic and numerical approaches to obtain $e_t$ values.
 Additionally, our numerical experiments show similar behaviour 
 for $\delta e_t$ plots that employ 1PN-accurate PC scheme based expression 
 for $e_t$ as showed in the dashed plot of top right panel.
 Further, PC based $e_t$ values suffer from system dependent 
 sharp variations in $\delta e_t$ values.
 }
 \label{fig:1PN_et_relerr}
\end{figure*}

\subsection{1PN accurate  $e_t$ and $\Psi$
in our MoRoLoYu approach} \label{sec:level3B}

This subsection sketches a way to obtain 1PN accurate expressions for 
$e_t$ and $\Psi$ that are {\it exact} in $e_0$, influenced by 
ideas gathered from Refs.~\cite{BBPM12,MY19}.
This extension allows us to benchmark both our fully analytic 1PN accurate PC scheme and its Pad\'e approximation to model  frequency domain GW templates for eccentric inspirals.
We begin by extending our quadrupolar order $\omega( \omega_0, e_0, e_t)$ expression, given by Eq.~(\ref{eq:domgbydomg2}), to 1PN order. 

For practical reasons, we plan to compute  1PN-accurate $x(e_t, e_0, x_0)$ expression 
and the starting point of these computations involves 1PN-accurate equations for 
$\dot{x}$ and  $\dot{e_t}$, extracted from Eqs.~(3.12a),(3.12b),(B9a - B9d)  in Ref.~\cite{THG}.
It is straightforward to obtain 1PN-accurate expression for 
$dx/de_t = \dot{x}/\dot{e_t}$ and it reads 
\begin{widetext}
\begin{align} \label{eq:dxdet1PNexact}
\frac{dx}{de_t} = &\,x\left[-\frac{2\,(96 + 292 e_t^2 + 37 e_t^4)}{e_t\,(1 - e_t^2)\,(304 + 121 e_t^2)}-\frac{x}{42\,e_t\,(1-e_t^2)(304 + 121 e_t^2)^2}\left(-2175744+4236288\,\eta\right. \right. \\ \nonumber &\, \left.\left.+e_t^2 (11073288 - 6573728 \eta)  +e_t^4 (-4607952 + 3626672 \eta)+e_t^6 (192543 - 219632 \eta)\right) \right] .     
\end{align}
\end{widetext}
Thereafter, we write the above equation symbolically as
\begin{align} \label{eq:dxdetsymb}
\frac{dx}{de_t} = &\, x \left[a_0(e_t) + a_1(e_t)\,x\right]\,.    
\end{align}
We seek its 1PN-accurate solution in the form
\begin{align} \label{eq:xetsymb}
x(e_t) = &\, x_0 \left[b_0(e_t) + b_1(e_t)\,x_0 \right],
\end{align}
where $x_0 = (\frac{G m 2 \pi F_0}{c^3})^{2/3}$ and $b_0$, $b_1$ are certain functions of $e_0$ and $e_t$. 
The explicit expressions for these  
 functions are obtained by inserting Eq.~(\ref{eq:xetsymb}) into Eq.(\ref{eq:dxdetsymb}) and expanding the resulting equation while 
  incorporating all contributions accurate to $x_0^2$. 
 This leads to a set of coupled ordinary differential equations for the unknown 
 coefficient functions $b_0(e_t)$ and $b_1(e_t)$ and these equations may be 
 written as 
\begin{subequations} \label{eq:b0b1ofet}
\begin{align}
b_0'(e_t) = &\, a_0b_0 , \\
b_1'(e_t) = &\, a_0b_1 + a_1b_0^2\,,
\end{align}
\end{subequations}
where primes ($'$) denote differentiation w.r.t $e_t$. It is natural to impose 
the following constraints like $b_0(e_0)=1$ and $b_1(e_0)=0$, mainly 
to ensure that $x(e_0) = x_0$.
This allows us to obtain 
a 1PN-accurate expression for $x(e_t)$:
\begin{widetext}
\begin{align} \label{eq:xofetexact}
x(e_t,e_0,x_0) = &\, x_0\left\{\left(\frac{1-e_t^2}{1-e_0^2}\right)\left(\frac{e_0}{e_t}\right)^{12/19}\left(\frac{304+121 e_0^2}{304+121 e_t^2}\right)^{870/2299}+x_0 \frac{(1-e_t^2)\,e_0^{12/19} (304 + 121 e_0^2)^{1740/2299}}{16056942720\,(1-e_0^2)^2\,e_t^{12/19} (304 + 121 e_t^2)^{870/2299}} \right. \\ \nonumber &\, \left.\left[\left(\frac{e_0}{e_t}\right)^{12/19} \mathcal{G}(e_t)-\mathcal{G}(e_0)\right]\right\} \,, \nonumber
\end{align}
where
\begin{align*}
\mathcal{G}(e) = &\, \frac{52}{(304 + 121 e^2)^{3169/2299}}\left[29408320\, (-2833 + 5516 \eta) + 168\,e^2 (-1555687953 + 1605256000 \eta) \right.\\ \nonumber &\, \left.+ e^4 (-4472255861 + 16145243380 \eta) \right]+2^{1118/2299} 19^{1429/2299} e^2 (-37041343 + 14343420 \eta) \\ \nonumber &\,  {}_2F_1\left(\frac{870}{2299},\frac{13}{19};\frac{32}{19};-\frac{121}{304}e^2\right). \nonumber
\end{align*}
\end{widetext}
The above expression provides certain 1PN-accurate solution to Eq.~(\ref{eq:dxdet1PNexact}) while 
 the $_2F_1(...)$ in the $\mathcal{G}(e)$ expression stands for the computationally demanding - generalised Hypergeometric function.

We note that the present approach can, in principle, be extended to higher PN orders,
provided closed form expressions exist for higher PN order contributions to $\dot{x}$ and  $\dot{e_t}$. In other words, it will be difficult to extend the approach when we
 deal with hereditary contributions to GW fluxes that do not support closed 
form expressions \cite{RS96}.

\begin{figure}[htp]
    \begin{center}
    \includegraphics[width=0.5\textwidth, angle=0,trim=0 0 0 0, clip]{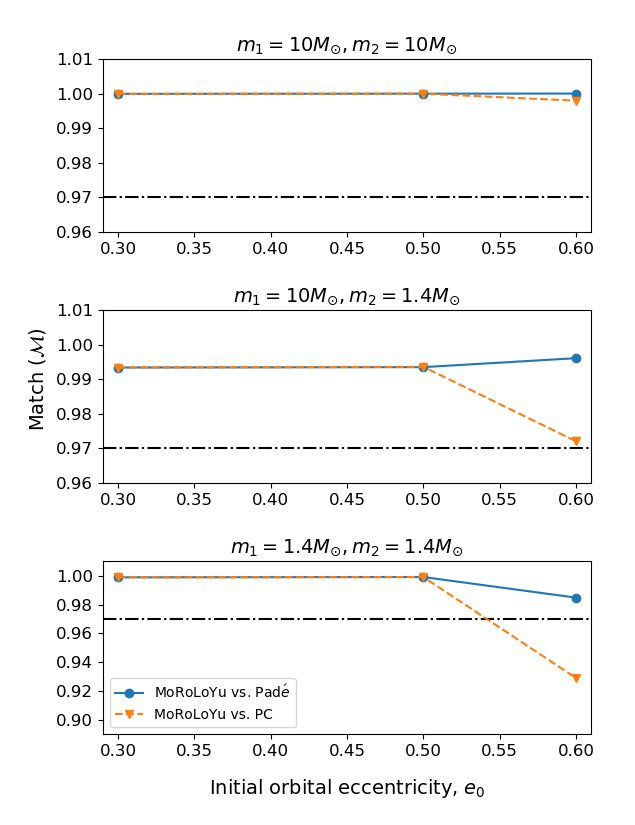}
    \end{center}
    \caption{ 
    Plots of 
    $\mathcal{M}$-values as a function of $e_0$ for the traditional binaries 
    entering the aLIGO frequency window and other specifications are
    similar to those in Fig.~\ref{fig:Match_newt}.
    We let the expected inspiral GW signal to be modeled by our 1PN-accurate 
    version of the  MoRoLoYu approach  while the templates belong to
    our 1PN-accurate $\tilde{h}^{Pad\acute{e}}$ waveform family.
    We find that our Pad\'e templates are fairly faithful to the GW signals
    that treat eccentricity in an exact manner up to $e_0 \sim 0.6$ while 
    the associated PC templates suffer drop in match numbers around 
    $e_0 \sim 0.5$.
    }
\label{fig:Match_1PN}
\end{figure} 

\begin{figure*}[htp]
 \begin{center}
    \includegraphics[width=0.9\textwidth, angle=0,trim=0 0 0 0, clip]{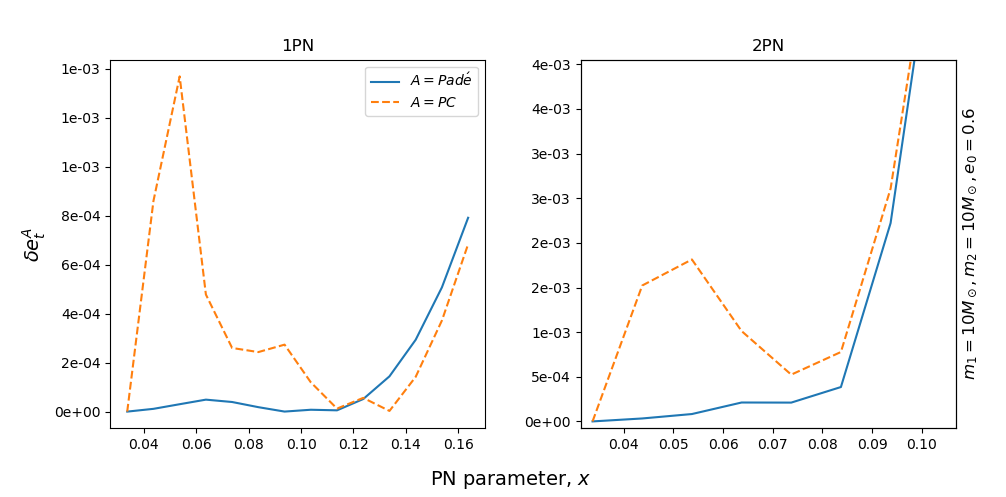}
     \end{center}
 \caption{ 
 Plots of fractional relative errors in $e_t$ at first and second post-Newtonian orders as a function of $x$ for a BBH system with $e_0 = 0.6$.
The orange plot in the left panel  employs  
 1PN accurate PC scheme based $e_t$ values, obtained by 
 an extension of Eq.~(\ref{eq:et1PNx}) that included 
 $\mathcal{O}(e_0^{19})$  corrections at each PN order 
 and its Pad\'e counterpart, the blue curve, employed Eq.~(\ref{eq:etPade1PN}).
 The right panel plots employ the 2PN-accurate extensions of both PC and 
 Pad\'e approaches for computing analytic $e_t$ expressions.
 Additionally, the numerical $e_t$ values that are required to compute these 
 $\delta e_t^A = |1-(e_t^A/e_t^{Num})|$ plots employ our 1PN-accurate 
 Eq.~(\ref{eq:xofetexact}) that provides 1PN extension of the MoRoLoYu approach.
 For the right panel plots, we numerically solve Eq.~(\ref{Eqs_T4omg2})
 to estimate $e_t^{Num}$ values.
  These plots reveal that 
  Pad\'e based $e_t$ values appear to smooth out the peculiarities in the fractional errors coming from the PC scheme based $e_t$ values at both 1PN and 2PN orders. However, the rapidly growing fractional errors with the PN expansion parameter suggests that a Pad\'e-ing on $x$ might also be required to precisely model $e_t(f)$.
  This is mainly because our numerical experiments indicate that the sharp variations 
  in $\delta e_t^A$ during the later part of the inspiral are rather independent of $e_0$ values and nature of the compact binaries, as evident from various subplots in our Fig.~\ref{fig:1PN_et_relerr}.
  }
\label{fig:2PN_et_relerr}
\end{figure*} 

We compute  1PN-accurate Fourier phase $\Psi$ of the MoRoLoYu approach 
by obtaining 1PN-accurate versions of 
the time and orbital phase evolution functions, namely Eq.~(\ref{eq:tofe}) and (\ref{eq:phiofe}). This requires us to employ 1PN-accurate version of 
 $\dot{e_t}(e_t')$ in both these integrals and we use 
\begin{widetext}
\begin{align} \label{eq:dedt1PN}
\frac{de_t}{dt} = &\, -\frac{c^3 x^4 \eta e_t}{G m \left(1-e_t^2\right)^{5/2}}\left\{\frac{304+121e_t^2}{15}+\frac{x \left(-67608-228704\eta+e_t^2(718008-651252\eta)+e_t^4(125361-93184\eta)\right)}{2520\left(1-e_t^2\right)}\right\}.
\end{align}
\end{widetext}

The above expression is identical to Eq.~(3.12b) in Ref.~\cite{THG} and 
we need to use $F= c^3\,x^{3/2}/(G\,m\,2\,\pi)$ in Eq.~(\ref{eq:phiofe}) for 
$\phi$ to be consistent.
Thereafter, we employ our 1PN-accurate expression for $x(e_t,e_0,x_0)$, given by
Eq.~(\ref{eq:xofetexact}), in these two integrals and this allows us 
to express their integrands  in terms of $x_0,e_t,e_0$. 
Next step involves expansion of these integrands in terms of $x_0$ 
up to 1PN order but the
resulting $t$ and $\phi$ integrals still remain non-trivial to evaluate analytically due to complex dependence on the variable $e_t'$. 
We perform these integrations 
by first expanding coefficient of each $x_0$ term in terms of $e_t$ without expanding the $e_0$ terms and this is influenced by Ref.~\cite{MY19}.
In our computations, we keep $e_t$ contributions accurate to 
  $\mathcal{O}(e_t^{40})$ and this is again influenced by
  the detailed analysis provided in Sec.~(5.1),(5.2) of \cite{MY19}. 
  The integration of resulting expressions with respect to $e_t$ provided us with 1PN-accurate time and phase functions. We list below expressions for these time and phase functions that incorporate only  the leading order corrections in $e_t$ as 
 Eqs.~(\ref{eq:t1PNMRLY}) and (\ref{eq:Phi1PNMRLY}), respectively.
 It is important to note that $e_0$ contributions are treated 
 in an exact manner in these two expressions. A few comments are in order. The 1PN-accurate expressions for time and phase functions which treat both $e_0$ and $e_t$ in an exact manner using Hypergeometric functions, were first provided by 
Eq.~(59) of Ref.~\cite{BPM15}. 
We have verified that our Eqs.~(51) and (52) are  in agreement with 
the $t$ and $\phi$ expressions of Ref.~[61], given by their Eq.~(59), 
when  Taylor expanding relevant expressions around $e_t = 0$ while not expanding in $e_0$. We now can obtain with the help of 
 these 1PN-accurate expressions  
  a 1PN-accurate version of $\Psi$ as $\Psi_j := j \phi(t^*_j) - 2 \pi f t^*_j$.
The fact that we have treated the initial eccentricity in an exact manner 
in our 1PN accurate $\Psi_j$ expression makes it suitable to model
eccentric inspirals with moderately high initial eccentricities.   

\begin{widetext}
\begin{align} \label{eq:t1PNMRLY}
t-t_c = &\, -\frac{95 \times 19^{1181/2299} \left(1-e_0^2\right)^4 e_t^{48/19}  G m}{2 \times 2^{2173/2299} e_0^{48/19} \left(304 + 121 e_0^2\right)^{3480/2299}  c^3 \, x_0^4\, \eta}-\frac{25 \times 19^{311/2299} \left(1-e_0^2\right)^3 (889-444 \eta )\, e_t^{36/19} G m}{288 \times 2^{1055/2299}  e_0^{36/19} \left(304+121 e_0^2\right)^{2610/2299} c^3\,x_0^3\,\eta } \nonumber \\ &\, \bigg\{1-\frac{10334784 \times 2^{1181/2299} 19^{870/2299} e_t^{12/19}}{35\, e_0^{12/19} \left(304+121 e_0^2\right)^{3169/2299} (889-444 \eta )}\bigg[1-\frac{5516 \eta }{2833}+e_0^2\left(\frac{32669447013}{10414221320}-\frac{842759400 \eta}{260355533}\right) \nonumber \\  &\, +e_0^4 \left(\frac{4472255861}{83313770560}-\frac{807262169 \eta }{4165688528}\right)+\frac{19^{1429/2299} \left(304+121 e_0^2\right)^{870/2299}}{2^{1181/2299}} \bigg(e_0^2 \left(\frac{703785517}{135384877160}-\frac{1048173 \eta
   }{520711066}\right) \nonumber \\ &\, +e_0^4 \left(\frac{4482002503}{2166158034560}-\frac{6675207 \eta }{8331377056}\right)\bigg){}_2F_1\left(\frac{870}{2299},\frac{13}{19};\frac{32}{19};-\frac{121}{304}e_0^2\right) \bigg] \bigg\}\,,
\end{align}
\end{widetext}

\begin{widetext}
\begin{align} \label{eq:Phi1PNMRLY}
\phi-\phi_c = &\, -\frac{19^{2175/2299} \left(1-e_0^2\right)^{5/2} e_t^{30/19}}{2^{2795/2299} e_0^{30/19} \left(304+121 e_0^2\right)^{2175/2299} x_0^{5/2}\eta }-\frac{e_t^{18/19} \left(1-e_0^2\right)^{3/2} (14135+630 \eta)}{504 \times 2^{1677/2299} 19^{994/2299} e_0^{18/19} \left(304+121 e_0^2\right)^{1305/2299} x_0^{3/2}\eta} \nonumber \\ &\, \bigg\{1-\frac{645924 \times 2^{1181/2299} 19^{870/2299} e_t^{12/19}}{e_0^{12/19} \left(304+121 e_0^2\right)^{3169/2299} (2827 + 126 \eta)} \bigg[1-\frac{5516 \eta }{2833}+e_0^2\left(\frac{32669447013}{10414221320}-\frac{842759400 \eta }{260355533}\right) \nonumber \\ &\, + e_0^4 \left(\frac{4472255861}{83313770560}-\frac{807262169 \eta }{4165688528}\right)+ \frac{19^{1429/2299} (304 + 121 e_0^2)^{870/2299}}{2^{1181/2299}} \bigg(e_0^2 \left(\frac{703785517}{135384877160}-\frac{1048173 \eta }{520711066}\right) \nonumber \\ &\, +e_0^4 \left(\frac{4482002503}{2166158034560}-\frac{6675207 \eta }{8331377056}\right)\bigg) {}_2F_1\left(\frac{870}{2299},\frac{13}{19};\frac{32}{19};-\frac{121}{304}e_0^2\right)\bigg] \bigg\}\,.
\end{align}
\end{widetext}

However, few more steps are required to fully operationalize the above computed 
PN-accurate $\Psi(x_0,e_t,e_0)$ expression. First, one needs to numerically invert our 1PN-accurate 
expression
for $x(x_0,e_t,e_0)$, namely Eq.~(\ref{eq:xofetexact}), for extracting
$e_t(x,x_0,e_0)$
that leads to a chart between $e_t$ and $F$ values.
Thereafter,  the {\it stationary phase condition} should be invoked to
replace $F$ and $F_0$ by their  Fourier frequency  counterparts  $f/j$ and $f_0/j$, respectively.
We refrain from showing the lengthy expression for the resulting 1PN-accurate $\Psi_j$ that extends its Newtonian counterpart. 
It is obvious that the resulting ready-to-use template family will be computationally expensive due to presence of these special functions and numerical treatments.
In the next subsection, we 
 outline steps to obtain 
  1PN-accurate 
 Pad\'e approximants that provide fully analytic $e_t$ and $\Psi$
 expressions.
 \\ 

\subsection{Our 1PN-accurate $e_t$ and $\Psi$ using Pad\'e approximants} \label{sec:level3C}

 We provide here a brief description for computing 1PN-accurate 
fully analytic Pad\'e approximant associated with our 1PN-accurate PC scheme based $\tilde{h}(f)$, detailed in Sec.~\ref{sec:level3A}.
Clearly, this is pursued to probe the ability of such an approximant 
to model eccentric inspirals in comparison with our 1PN-accurate 
extension of the MoRoLoYu approach that treats $e_0$ effects in an exact manner.
Our Pad\'e approximant, as expected, requires PC scheme based expressions for $e_t$ and $\Psi_j$ and 
we specifically employ such 1PN accurate $e_t$ and $\Psi_j$ expressions that incorporate $\mathcal{O}(e_0^{19})$ and $\mathcal{O}(e_0^{20})$ corrections in initial eccentricity.
We obtain Pad\'e approximations of these quantities by applying the 
resummation technique individually to Newtonian and 1PN contributions. 
This allows us to propose the following expression to 
obtain a simplistic 1PN-accurate Pad\'e approximation for $e_t$
\begin{widetext}
\begin{align} \label{eq:etPade1PN}
e_t = & \, e_0 \bigg\{ \frac{\bar{n}_0+ \bar{n}_1 \, z+ \bar{n}_2 \, z^2+ \bar{n}_3 \, z^3+ \bar{n}_4 \, z^4+ \bar{n}_5\, z^5}{1+ \bar{d}_1 \, z+ \bar{d}_2 \, z^2+ \bar{d}_3 \,  z^3+ \bar{d}_4 \, z^4} + x\, \frac{\bar{n}'_0+ \bar{n}'_1 \, z+ \bar{n}'_2 \, z^2+ \bar{n}'_3 \, z^3+ \bar{n}'_4 \, z^4+ \bar{n}'_5\, z^5}{1+ \bar{d}'_1 \, z+ \bar{d}'_2 \, z^2+ \bar{d}'_3 \,  z^3+ \bar{d}'_4 \, z^4}\bigg\} \,.\nonumber \\
\end{align}
\end{widetext}
For the sake of simplicity, we denote 1PN order coefficients 
with the help of $'$ symbols. 
These coefficients can be obtained from their 1PN order
counterparts, present in our 1PN accurate PC scheme based $e_t$ expression.
Further, the Newtonian order coefficients like $\bar{n}_0...\bar{n}_5$, $\bar{d}_1...\bar{d}_4$ are identical to those present in 
Eq.~(\ref{eq:etPade}). 
The resulting expression allows us to compute 
the fractional differences between  $e_t$ values 
that are based on 
our 1PN-accurate extension of the MoRoLoYu and 
Pad\'e approximations for $e_t$.
These differences are expected to depend on  both total mass and 
mass ratio as Eq.~(\ref{eq:domgbydomg2}) for 1PN-accurate $\omega$ depends on these quantities.
In Fig.~\ref{fig:1PN_et_relerr}, we plot fractional errors in $e_t$ as a function of the PN expansion parameter $x$. We find that $\delta e_t$ values are essentially independent of $e_0$ values and sharp rises 
 in $\delta e_t$ values are observed  when  $x$ values cross $0.1$.
 This may be attributable to the differences in the way PN corrections are incorporated in Eqs.~(\ref{eq:xofetexact}) and (\ref{eq:etPade1PN}). We found similar behaviour for 1PN-accurate fractional errors up to $e_0 = 0.6$, for systems having  $ m < 50\, M_{\odot}$. The curves follow similar pattern up to mild eccentricities $e_0 \sim 0.3$ for systems with  $m > 50\, M_{\odot}$. However, more heavier systems with $ m > 50\,M_{\odot}$ and with $e_0 > 0.3$ do not display similar increases in fractional errors  with the PN expansion parameter, $x$. This is expected as such systems will evolve rapidly from $f_0 = 20$ Hz to the ISCO frequency without causing any noticeable disagreement between our Pad\'e approximant for $e_t$ and its numerical counterpart.
 Further, our numerical experiments reveal that  $\delta e_t$  plots created with
  the PC scheme based 1PN-accurate $e_t$ expression show similar $x$ variations though these plots are spikey at higher $e_0$ values.
  These considerations suggest that multi-Pad\'e expression that perform Pad\'e-ing on both $x$ and $e_0$ values may be required while constructing PN extensions of 
Eq.~(\ref{eq:etPade1PN}). 
This issue requires further investigations.
 We proceed to list our 1PN accurate Pad\'e approximated Fourier phases expression, computed 
 from the 1PN-accurate  PC scheme based $\Psi_j$ expression that 
 includes $\mathcal{O}(e_0^{20})$ order corrections in $e_0$. The symbolic 
expression for $\Psi_j$ reads 

\begin{widetext}
\begin{align}\label{eq:PsiPade1PN}
\Psi_j = &\,j \phi_c - 2 \pi f t_c -\frac{3\,j}{256\,\eta\,x^{5/2}} \bigg\{\frac{\hat{n}_0+ \hat{n}_1 \, z+ \hat{n}_2 \, z^2+ \hat{n}_3 \, z^3+ \hat{n}_4 \, z^4+ \hat{n}_5\, z^5+ \hat{n}_6 \, z^6}{1+ \hat{d}_1 \, z+ \hat{d}_2 \, z^2+ \hat{d}_3 \,  z^3+ \hat{d}_4 \, z^4} \nonumber \\ &+ x\,\frac{\hat{n}'_0+ \hat{n}'_1 \, z+ \hat{n}'_2 \, z^2+ \hat{n}'_3 \, z^3+ \hat{n}'_4 \, z^4+ \hat{n}'_5\, z^5+ \hat{n}'_6 \, z^6}{1+ \hat{d}'_1 \, z+ \hat{d}'_2 \, z^2+ \hat{d}'_3 \,  z^3+ \hat{d}'_4 \, z^4}\bigg\}, 
\end{align}
\end{widetext}
where the explicit expressions for these 
$\bar{n}_0...\bar{n}_5$, $\bar{n}'_0...\bar{n}'_5$, $\bar{d}_1...\bar{d}_4$, $\bar{d}'_1...\bar{d}'_4$, $\hat{n}_0...\hat{n}_6$, $\hat{n}'_0...\hat{n}'_6$, $\hat{d}_1...\hat{d}_4$ and $\hat{d}'_1...\hat{d}'_4$ are provided 
in the accompanying  \texttt{Mathematica} notebook.

We are now in a position to obtain match $(\mathcal{M})$ estimates, outlined in Sec.~\ref{sec:level2C}, that probe the ability of 
our 1PN-accurate Pad\'e approximant to capture
inspiral $\tilde{h}(f)$  arising from our improved 1PN order MoRoLoYu approach.
In Fig.~\ref{fig:Match_1PN}, we plot $\mathcal{M}$ estimates
as a function of $e_0$ for the classical aLIGO compact binaries.
For these $\mathcal{M}$ plots, we employ quadrupolar order amplitudes 
in $\tilde h(f)$ while the Fourier phases are 1PN-accurate.
 Additionally, we employ  1PN-accurate Pad\'e approximant for $e_t(f)$,
given by Eq.~(\ref{eq:etPade1PN}), in
these GW amplitude expressions for computational ease and
our results are not sensitive to such a choice.
Plots in  Fig.~\ref{fig:Match_1PN} reveal that our eccentric Pad\'e approximant 
is quite capable of faithfully capturing expected GW inspiral waveforms
where eccentricity effects are modeled in an exact manner up to 
initial orbital eccentricities $\sim 0.6$.
The sharp drop in $\mathcal{M}$ values for the NS-NS systems may be attributed 
to their comparatively longer inspiral durations in the aLIGO frequency window.
These plots suggest that 
fully analytic Pad\'e approximant may be useful to model eccentric 
inspirals with $e_0 \sim 0.6$ 
when general relativistic effects are included. Further, 
it is capable of extending the 
validity of the PN-accurate PC approach to higher $e_0$ values.
Therefore, it is natural to explore possible subtleties one may face 
while modeling eccentric inspirals using higher PN order Pad\'e approximants.
This is what we pursue in the next subsection.

\subsection{\label{sec:level3D}
On constructing eccentric Pad\'e approximants at higher PN orders
}
It is important to  extend our Pad\'e approximant
to higher PN orders.
This is because 
 the widely employed \texttt{TaylorF2} approximant for quasi-circular inspiral incorporates Fourier phase to 3.5PN order \cite{Buonanno09}. In contrast, various eccentric inspiral template families 
employ 3PN accurate GW phase evolution \cite{Moore16,TGMH}.
This  subsection  explores the difficulties that we may face while extending our Pad\'e approach to higher PN orders.
We will focus our attention on the secular orbital evolution for eccentric binaries while restricting our attention to 2PN accurate 
radiation reaction effects.
This is because $\Delta \phi$,
the accumulated orbital phase provides a data analysis relevant tool to compare various eccentric approximants \cite{THG}. 
There exists several ways to obtain $\Delta \phi$ estimates in PN approach and we will focus on few relevant ones.
The first approach is influenced by the GW phasing approach, detailed in Refs.~\cite{DGI,KG06,THG}.
In this approach, we obtain the secular orbital phase evolution by solving numerically the following three 
coupled differential equations \cite{THG}:
\begin{widetext}
\begin{subequations} \label{Eqs_T4t}
\begin{align}
\frac{d \phi}{dt} =&\, \omega   \label{Eqs_T4t1}\,,\\
\frac{d \omega }{dt} =&\, \frac{ c^6\, \eta\, x^{11/2}}{G^2\, m^2}\left\{\frac{96}{5 \left(1-e_t^2\right)^{7/2}}\left[1+\frac{73\, e_t^2}{24}+\frac{37\, e_t^4}{96}\right]-\frac{1486\, x}{35\, \left(1-e_t^2\right)^{9/2}} \left[1+\frac{924\,
   \eta }{743}+e_t^2 \left(-\frac{10965}{1486}+\frac{9975\, \eta }{743}\right) \right. \right. \label{Eqs_T4t2} \\ \nonumber  &\,  \left. +e_t^4 \left(-\frac{85519}{5944}+\frac{35427\, \eta }{2972}\right)+e_t^6 \left(-\frac{11717}{11888}+\frac{518\, \eta }{743}\right)\right]-\frac{11257\,x^2}{945 \left(1-e_t^2\right)^{11/2}}\left[1-\frac{141093\, \eta }{11257}-\frac{59472\, \eta ^2}{11257} \right.\\ \nonumber &\, \left. +e_t^2 \left(\frac{2901455}{11257}-\frac{483273\, \eta }{11257}-\frac{3830127\, \eta ^2}{22514}\right)+ e_t^4 \left(-\frac{97971}{45028}+\frac{25900533\, \eta }{45028}-\frac{41626515\, \eta ^2}{90056}\right) \right. \\ \nonumber &\,   +e_t^6 \left(-\frac{41712201}{180112}+\frac{61554213\, \eta }{180112}-\frac{4051803\, \eta ^2}{22514}\right)+e_t^8 \left(-\frac{3523113}{360224}+\frac{814995\, \eta }{90056}-\frac{61383\, \eta ^2}{11257}\right) \\ \nonumber &\,+ \sqrt{1-e_t^2} \left(-\frac{45360}{11257}+\frac{18144\, \eta }{11257}+e_t^2 \left(-\frac{2016630}{11257}+\frac{806652\, \eta }{11257}\right)+e_t^4 \left(-\frac{2072385}{11257}+\frac{828954\, \eta }{11257}\right) \right. \\ \nonumber &\, \left. \left. \left.+e_t^6 \left(-\frac{165375}{22514}+\frac{33075\, \eta }{11257}\right) \right)\right]+ x^{3/2} \left[ \frac{384}{5}\,\pi\,\phi(e_t)\right] \right\}    \,, \\
\frac{d e_t}{dt}  =&\,-\frac{c^3\,e_t\,\eta\,x^4}{G\, m}\left\{ \frac{304}{15 \left(1-e_t)^2\right)^{5/2}}\left[1+\frac{121\,e_t^2}{304}\right]-\frac{939\, x}{35 \left(1-e_t^2\right)^{7/2}}\left[1+\frac{28588\, \eta }{8451}+e_t^2 \left(-\frac{29917}{2817}+\frac{54271\, \eta }{5634}\right) \right. \right.  \label{Eqs_T4t3} \\ \nonumber &\, \left. \left.+ e_t^4 \left(-\frac{4643}{2504}+\frac{11648\, \eta }{8451}\right)\right]-\frac{949877\, x^2}{1890 \left(1-e_t^2\right)^{9/2}}\left[1-\frac{844335\, \eta }{949877}-\frac{284256\, \eta ^2}{949877}+e_t^2 \left(\frac{9248349}{3799508}+\frac{8895807\, \eta }{3799508} \right. \right. \right.\\ \nonumber &\,  \left.-\frac{12177837\, \eta ^2}{3799508}\right)+e_t^4 \left(-\frac{23289859}{7599016}+\frac{39056133\, \eta }{7599016}-\frac{2675631\, \eta ^2}{949877}\right)+e_t^6 \left(-\frac{3786543}{15198032}+\frac{1086213\, \eta }{3799508} \right. \\ \nonumber &\, \left.-\frac{172410\, \eta ^2}{949877}\right)+\sqrt{1-e_t^2}\left(-\frac{841680}{949877}+\frac{336672\, \eta }{949877}+e_t^2 \left(-\frac{2193345}{949877}+\frac{877338\, \eta }{949877}\right)+e_t^4 \left(-\frac{177975}{949877} \right. \right.\\ \nonumber &\, \left. \left. \left. \left.+\frac{71190\, \eta }{949877}\right)\right)\right]+x^{3/2}\left[\frac{394}{3}\pi\,\phi_e(e_t)\right]\right\}  \,, 
\end{align}
\end{subequations}
\end{widetext}
where the explicit expressions for various PN contributions are also listed as Eqs.~(3.12a), (3.12b) and (B9) in Ref.~\cite{THG}. 
The enhancement functions that appear at the relative 1.5PN order are
also adapted from Ref.~\cite{THG} and are accurate enough to 
model binaries with very high eccentricities like $e_0 \sim 0.9$.
The plan is to evolve the above equation set during a time interval when the $\omega$ varies from $\omega_0$ to $\omega_{LSO}$ for 
compact binaries, specified by certain $m,\eta$ and $e_0$ values.
Note that in the original GW phasing approach, we have $\phi = \lambda + W$, where $W$ provides certain PN accurate 
quasi-periodic contributions to the orbital phase.
We have ignored these sub-dominant contributions to the orbital phase 
evolution and write 
$ d \phi/ dt = d \lambda /dt \equiv \omega$. Further,
this approach provides  secular GW phase evolution in the 
time-domain \texttt{Taylor} approximant, 
available in the \texttt{LSC Algorithm Library} and leads to the popular \texttt{TaylorT4} approximant
in the circular limit \cite{Buonanno09}.
This approximant was called \texttt{TaylorT4t} approximant in Ref.~\cite{MY19}.

 The second approach is influenced by the \texttt{TaylorT4y} approximant of Ref.~\cite{MY19}. In our case, 
this involves obtaining differential equations for $d \phi/ d \omega$ and $ d e_t/d \omega$ to 2PN order 
while keeping $e_t$ contributions in an exact manner. These 2PN-accurate differential equations are obtainable 
from Eqs.~(\ref{Eqs_T4t}) such that $ d \phi /d \omega = \dot{\phi}/ \dot \omega$ and $ d e_t/ d \omega = \dot{e_t}/ \dot \omega$,
where an overdot stands for the time derivative. The resulting 2PN accurate equations read
\begin{widetext}
\begin{subequations} \label{Eqs_T4omg}
\begin{align}
\frac{d \phi}{d \omega} =&\, \frac{G\,m}{c^3\,x^4\,\eta}\left\{\frac{5 (1-e_t^2)^{7/2}}{(96 + 292 e_t^2 + 37 e_t^4)}+  \phi_{\omega}^{1PN}\,x + \phi_{\omega}^{1.5PN}\,x^{3/2}\,  + \phi_{\omega}^{2PN}\,x^2\right\} \,, \label{Eqs_T4omg1}\\
 \frac{d e_t}{d \omega} =&\, \frac{G\,m}{c^3\,x^{3/2}}\left\{-\frac{\left(1-e_t^2\right) \left(304 e_t+121 e_t^3\right)}{3 \left(96+292 e_t^2+37 e_t^4\right)}+ e_{\omega}^{1PN}\,x + e_{\omega}^{1.5PN}\,x^{3/2}+e_{\omega}^{2PN}\,x^2\right\}  \,, \label{Eqs_T4omg2}
\end{align}
\end{subequations}
\end{widetext}
where $\phi^{1PN}_{\omega},\phi^{1.5PN}_{\omega},\phi^{2PN}_{\omega}$ and $e^{1PN}_{\omega},e^{1.5PN}_{\omega},e^{2PN}_{\omega}$ are explicitly given in Appendix \ref{appendixA}.
We obtain the accumulated orbital phase in a given $\omega$ interval by numerically solving the above set of two coupled 
differential equations and the resulting GW cycles are denoted by ${\mathcal N}_{\rm GW}^{\rm TaylorT4\omega}$ in Table~\ref{Table:NOC}.
This approximant is influenced by the \texttt{TaylorT4y} approximant of Ref.~\cite{MY19} as that approximant
solves numerically  PN-accurate $d \phi/ d y, d e_t/d y$ and $d t/dy$ ,
where $y = (G\,m\,\omega/c^3)^{1/3}/\sqrt{1-e_t^2}$, to obtain temporally evolving GW polarization states. A close inspection reveals that our two equations, namely
 $d \phi/ d \omega$ and $ d e_t/d \omega$, are structurally identical to $d \phi/ d y$ and $ d e_t/d y$ equations under 
 PN considerations. We also list in our Table.~\ref{Table:NOC}, ${\mathcal N}_{\rm GW}^{\rm TaylorT4t}$ - the number of gravitational wave cycles obtained from the TaylorT4t approximant described above.
 
The remaining two approaches are purely analytic in nature. The third approximant computes $\Delta \phi$ using analytic 
expressions for $\phi$ as detailed in Ref.~\cite{THG}. 
This approach employs the PN-accurate PC scheme to obtain PN-accurate expression for $e_t$ in terms of $e_0,\omega, \omega_0$ \cite{THG}.
Thereafter, it is fairly straightforward to obtain analytic expression for $\phi$ with the help of the following 
equations, namely $ \phi = \int \omega\, dt = \int ( \omega/ \dot \omega) d\omega $.

This ensures that PN-accurate $\omega/ \dot \omega$ becomes a function of $\omega$ which can be integrated.
The resulting 2PN-accurate expression for $\phi$ is  given by Eqs.~(2.25) in Ref.~\cite{THG}. We have extended this computation to 
incorporate $\mathcal{O}(e_0^{20})$ order $e_0$ corrections. The associated GW cycle estimates are obtained by  evaluating  $ [ \phi (\omega_f) - \phi ( \omega_i)]$ and diving it by $\pi$ for compact binaries specified by $e_0, \omega_0, m$ and $\eta$.  We compute the accumulated number of GW cycles within aLIGO's frequency window, starting from an orbital frequency of $\omega_i = 20\, \pi$ Hz to a final orbital frequency of $\omega_f = c^3/(G\,m\,6^{3/2})$ Hz, corresponding to the last stable orbit of a compact binary.The resulting entries are denoted by ${\mathcal N}_{\rm GW}^{\rm PC}$ in Table.~\ref{Table:NOC}.
The fourth and final estimate is based on our Pad\'e approximation, influenced by the fact that we have Taylor expansion, accurate to
${\mathcal O}(e_0^{20})$ for 2PN-accurate $\phi$. We construct Pad\'e approximant using the rational polynomial approach with polynomials 
of order $6$ and $4$ in the numerator and the denominator. The associated ${\mathcal N}_{\rm GW}$ are listed in Table.~\ref{Table:NOC} as ${\mathcal N}_{\rm GW}^{\rm Pad\acute{e}}$.
Further, we plot relative fractional errors 
at second post-Newtonian orders as a function of $x$ parameter for a BBH system with $e_0 = 0.6$ in Fig.~\ref{fig:2PN_et_relerr}. We employ both the PC and Pad\'e based 
$e_t (\omega)$ expressions that incorporate ${\mathcal O}(e_0^{19})$ eccentricity corrections. The numerical $e_t$ values are obtained by solving Eq.~(\ref{Eqs_T4omg2}) and therefore treats orbital eccentricity in an exact manner. We infer that the sharp variations in $\delta e_t$ values during the late inspiral are essentially independent of $e_0$ values similar to 1PN $\delta e_t$ in Fig.~\ref{fig:1PN_et_relerr}.

A close look at various entries of the Table.~\ref{Table:NOC} and the $\delta e_t$ plot in Fig.~\ref{fig:2PN_et_relerr} presents a possible way to obtain 
fully analytic ready-to-use $\tilde{h}(f)$ for compact binaries inspiraling along PN-accurate eccentric orbits.
The idea involves Pad\'e approximant version of $e_t(\omega)$ that incorporates ${\mathcal O}(e_0^{19})$ eccentricity corrections
or its extensions with inputs from Refs.~\cite{THG,KBGJ_18}.
This ensures smooth and accurate 
$e_t(\omega, \omega_0,e_0)$ expression, required to obtain amplitudes of $\tilde{h}(f)$ as evident from Eq.~(\ref{eq:etPade}) or its PN extension, given by Eq.~(\ref{eq:etPade1PN}).
However, it may be desirable to employ
Pad\'e approximation additionally  on the $x$ parameter.
This is to essentially  probe if the resulting 
multivariate Pad\'e approximation for $e_t(\omega, \omega_0,e_0)$  
follows closely the numerically obtained $e_t$ values even during the 
late stages of compact binary inspiral.
Clearly, it will be desirable to do such an exploration at a 3PN-accurate 
 $e_t(\omega, \omega_0,e_0)$ that provides $x$ corrections at five distinct orders.
For the Fourier phase, we suggest the use of PN-accurate PC scheme that incorporates eccentricity corrections accurate to $\mathcal{O}(e_0^{20})$ or its higher order extensions.
Additionally, we may probe the possibility of introducing 
multivariate Pad\'e approximation for 
$\Psi_j$ in both 
$x$ and $e_0$. 
Obviously, this is  motivated by the possibility that such a 
multivariate Pad\'e approximant can be more closer to \texttt{TaylorT4$\omega$} approximant 
from the perspective of the accumulated orbital phase in a given $x$ window.
It will be interesting to probe if these modifications can lead to 
orbital phase evolution  similar to the one based on the \texttt{TaylorT4t} approximant.
This is of course influenced the observation that this Taylor 
approximant showed remarkable closeness to fully NR simulations during the quasi-circular inspiral \cite{Boyle07,Buonanno09}.
Of course, detailed comparisons of Numerical Relativity  based GW phase evolution to its counterparts  under various PN-accurate 
eccentric approximants will be crucial to choose the best strategy for computing fully analytic inspiral templates for 
compact binaries spiraling along PN-accurate eccentric orbits. 
Such comparisons will also help us to estimate  the minimum order
of $e_0$ corrections that are required to construct 
efficient eccentric inspiral $\tilde{h}(f)$.
These efforts are being pursued and their results will be reported elsewhere.

 \begin{widetext}
 \begin{center}
 \begin{table}[!htbp]
\begin{tabular}{|c|l|l|l|l|}
\hline
\boldsymbol{$(m_1, m_2)$}       & \multicolumn{1}{c|}{\boldsymbol{$(1.4 M_\odot, 1.4 M_\odot)$}} & \multicolumn{1}{c|}{\boldsymbol{$(10 M_\odot, 1.4 M_\odot)$}} & \multicolumn{1}{c|}{\boldsymbol{$(10 M_\odot, 10 M_\odot)$}} & \multicolumn{1}{c|}{\boldsymbol{$(30 M_\odot, 30 M_\odot)$}} \\ \hline
\multicolumn{5}{|c|}{\boldsymbol{$e_0 = 0.1$}}                                                                                                                                                                                                \\ \hline
\textbf{$\mathcal{N}^{\rm TaylorT4t}_{\rm GW}$}             &          \multicolumn{1}{c|}{4980.31}                      & \multicolumn{1}{c|}{1078.29}                     & \multicolumn{1}{c|}{178.03}                       &        \multicolumn{1}{c|}{22.88}                                         \\ \hline
\textbf{$\mathcal{N}^{\rm TaylorT4\omega}_{\rm GW}$}             &         \multicolumn{1}{c|}{4991.35}                      & \multicolumn{1}{c|}{1087.71}                     & \multicolumn{1}{c|}{182.20}                          &        \multicolumn{1}{c|}{24.33}                                         \\ \hline
\textbf{$\mathcal{N}^{\rm PC}_{\rm GW}$}         &               \multicolumn{1}{c|}{4991.31}                      & \multicolumn{1}{c|}{1087.63}                     & \multicolumn{1}{c|}{182.17}                   &       \multicolumn{1}{c|}{24.30}                                          \\ \hline
\textbf{$\mathcal{N}^{\rm Pad\acute{e}}_{\rm GW}$} &                                   \multicolumn{1}{c|}{4991.31}                      & \multicolumn{1}{c|}{1087.63}                     & \multicolumn{1}{c|}{182.17}         &          \multicolumn{1}{c|}{24.30}                                       \\ \hline
\multicolumn{5}{|c|}{\boldsymbol{$e_0 = 0.3$}}                                                                                                                                                                                                \\ \hline
\textbf{$\mathcal{N}^{\rm TaylorT4t}_{\rm GW}$}            &           \multicolumn{1}{c|}{3884.20}                      & \multicolumn{1}{c|}{823.22}                     & \multicolumn{1}{c|}{134.14}                          &    \multicolumn{1}{c|}{16.08}                                             \\ \hline
\textbf{$\mathcal{N}^{\rm TaylorT4\omega}_{\rm GW}$}                &      \multicolumn{1}{c|}{3893.70}                      & \multicolumn{1}{c|}{828.93}                     & \multicolumn{1}{c|}{136.84}                       &     \multicolumn{1}{c|}{16.38}                                            \\ \hline
\textbf{$\mathcal{N}^{\rm PC}_{\rm GW}$}   &   \multicolumn{1}{c|}{3893.40}                      & \multicolumn{1}{c|}{828.38}                     & \multicolumn{1}{c|}{136.60}                                             &            \multicolumn{1}{c|}{16.19}                                     \\ \hline
\textbf{$\mathcal{N}^{\rm Pad\acute{e}}_{\rm GW}$} &     \multicolumn{1}{c|}{3893.40}                       & \multicolumn{1}{c|}{828.38}                     & \multicolumn{1}{c|}{136.60}                                              &              \multicolumn{1}{c|}{16.19}                     \\ \hline
\multicolumn{5}{|c|}{\boldsymbol{$e_0 = 0.5$}}                                                                                                                                                                                                \\ \hline
\textbf{$\mathcal{N}^{\rm TaylorT4t}_{\rm GW}$}            &         \multicolumn{1}{c|}{2215.53}                      & \multicolumn{1}{c|}{444.59}                     & \multicolumn{1}{c|}{70.00}                            &      \multicolumn{1}{c|}{4.20}                                           \\ \hline
\textbf{$\mathcal{N}^{\rm TaylorT4\omega}_{\rm GW}$}                &         \multicolumn{1}{c|}{2221.56}                      & \multicolumn{1}{c|}{442.81}                     & \multicolumn{1}{c|}{69.80}                            &   \multicolumn{1}{c|}{5.43}                                              \\ \hline
\textbf{$\mathcal{N}^{\rm PC}_{\rm GW}$}    &   \multicolumn{1}{c|}{2220.95}                      & \multicolumn{1}{c|}{441.84}                     & \multicolumn{1}{c|}{69.37}                                         &       \multicolumn{1}{c|}{5.23}                                          \\ \hline
\textbf{$\mathcal{N}^{\rm Pad\acute{e}}_{\rm GW}$} &   \multicolumn{1}{c|}{2220.95}                      & \multicolumn{1}{c|}{441.84}                     & \multicolumn{1}{c|}{69.37}                              &      \multicolumn{1}{c|}{5.23}                                        \\ \hline
\multicolumn{5}{|c|}{\boldsymbol{$e_0 = 0.6$}}                                                                                                                                                                                                \\ \hline
\textbf{$\mathcal{N}^{\rm TaylorT4t}_{\rm GW}$}            &        \multicolumn{1}{c|}{1406.60}                      & \multicolumn{1}{c|}{229.63}                     & \multicolumn{1}{c|}{40.78}                            &    \multicolumn{1}{c|}{0.57}                                             \\ \hline
\textbf{$\mathcal{N}^{\rm TaylorT4\omega}_{\rm GW}$}                   &            \multicolumn{1}{c|}{1410.06}                      & \multicolumn{1}{c|}{261.50}                     & \multicolumn{1}{c|}{38.85}                       &    \multicolumn{1}{c|}{1.14}                                             \\ \hline
\textbf{$\mathcal{N}^{\rm PC}_{\rm GW}$}          &                                     \multicolumn{1}{c|}{1409.35}                      & \multicolumn{1}{c|}{260.49}                     & \multicolumn{1}{c|}{38.41}                                 & \multicolumn{1}{c|}{1.10}                                                \\ \hline
\textbf{$\mathcal{N}^{\rm Pad\acute{e}}_{\rm GW}$} &    \multicolumn{1}{c|}{1409.35}                      & \multicolumn{1}{c|}{260.49}                     & \multicolumn{1}{c|}{38.41}                         &      \multicolumn{1}{c|}{1.11}                                           \\ \hline
\end{tabular}
\caption{{\label{Table:NOC}}
Values of $\mathcal{N}_{\rm GW}$, the accumulated number of GW cycles 
in the aLIGO frequency window for four distinct compact binaries with 
four different $e_0$ values at the 2PN order.
These $\mathcal{N}_{\rm GW}$ estimates arise from
four approaches, namely 
 TaylorT4t, TaylorT4$\omega$, Post-circular and Pad\'e approximants 
 as denoted by the superscripts (how to obtain these four types of 
 $\mathcal{N}_{\rm GW}$ are 
 detailed in Sec.~\ref{sec:level3D}).
 It is clear that TaylorT4t approximant leads to very different 
 $\mathcal{N}_{\rm GW}$  estimates while the other three approaches 
 provide fairly similar estimates for the accumulated GW cycles.
 Note that our 2PN-accurate Pad\'e approximant arises 
 from  the 2PN-accurate Post-circular approach that incorporated ${\cal O}(e_0^{20})$
 order corrections at every PN order.
 }
\end{table}
\end{center}
\end{widetext}
 
\section{\label{sec:level4} Summary and Discussion}
 We explored the possibility of resumming the PC scheme that provided 
 analytic expressions for the frequency evolution of orbital eccentricity 
 and Fourier phases of  GW response function, 
 associated with eccentric inspirals.
The simplest form of Pad\'e approximation, namely the ratio of rational polynomials, for the quadrupolar order PC scheme based 
$e_t$ expression 
provided relative fractional $e_t$ errors $\sim 10^{-5}$ in the aLIGO frequency 
window even for initial $e_t$ values $\sim 0.6$.
These error estimates employed numerical inversion of an analytic expression for the orbital frequency while treating both $e_t$ and $e_0$ contributions in an 
exact manner. 
Preliminary aLIGO relevant match estimates reveal that the associated
quadrupolar order Pad\'e approximant $\tilde{h}(f)$ is faithful to MoRoLoYu approach based $\tilde{h}(f)$ for $e_0$ values $\sim 0.6$ (recall that the quadrupolar 
order MoRoLoYu  approach of Ref.~\cite{MRLY18}
essentially treats orbital eccentricity parameters in an exact manner).

 Encouraged by our quadrupolar order results, we obtained a similar 
 Pad\'e approximation to the 1PN-accurate PC scheme based $e_t(f)$ expression.
Additionally, we computed 1PN-accurate expression for the dimensionless PN expansion parameter $x$ that incorporated $e_t$ and $e_0$ contributions in an
exact manner and this is, of course, for making comparisons between analytically and numerically computed frequency evolution for $e_t$.
 It turns out that our Pad\'e approximation for $e_t(f)$ does include $e_0$ contributions more accurately and smoothly compared to the PC scheme. However, 
differences in the way of incorporating PN corrections  ensure 
that fractional differences in 1PN-accurate $e_t$ estimates do depend on the 
$x$ parameter. Specifically, we observe $e_0$ independent sharp rises in our $\delta e_t$ values for $x$ values that characterize later part of the compact binary inspiral.
Thereafter, we developed a 1PN-accurate extension of the MoRoLoYu approach 
to compute  eccentric $\tilde{h}(f)$ that includes 
$e_0$ contributions in an exact manner though in a semi-analytic fashion. We showed that our analytic 
$\tilde{h}(f)$, improved by employing Pad\'e approximation for $e_t(f)$ and $\Psi_j(f)$, is faithful 
to our 1PN extension of the MoRoLoYu $\tilde{h}(f)$ for $e_0$ values $\sim 0.6$ for 
the traditional aLIGO compact binaries.
Interestingly, our Pad\'e approximation for $e_t(f)$ provides smooth 
evolution of orbital eccentricity even at higher PN orders.
We additionally probed the ability of our Pad\'e approximation and its underlying 
PC scheme to track accurately the orbital phase evolution 
 at 2PN order for eccentric inspirals in the aLIGO frequency window.
It turns out that both 2PN accurate PC based $\phi(x,f,f_0, e_0)$  expression
which includes ${\cal O}(e_0^{20})$ contributions 
 and its Pad\'e variant are capable of obtaining ${\cal N}_{\rm GW}$, based 
 on numerical  \texttt{TaylorT4$\omega$} prescription that 
 incorporates eccentricity effects exactly. 
 
   These considerations and observations suggest that it may be 
   possible to devise an improved PC scheme to compute fully analytic 
Fourier domain inspiral template family for eccentric inspirals 
with initial eccentricities up to $0.6$.
However, additional investigations will be required to implement several improvements 
to the present results.
These include extending the computations of Ref.~\cite{TGMH} to include 3PN accurate eccentricity contributions, accurate up to  
 ${\cal O}(e_0^{40})$ order.
Additionally, it may be required to pursue 
multivariate Pad\'e approximation of 3PN accurate PC based $e_t(f)$ expression 
 while employing both $e_0$ and $x$ parameters.
This is to obtain smoothly varying $e_t(f)$ expression that will have 
small relative fractional errors compared to numerically obtained 
frequency evolution for $e_t$, based on 3PN-accurate 
$\dot x$ and $\dot e_t$ expressions of Ref.~\cite{ABIS}.
Further, we will require to probe how eccentric 
\texttt{TaylorT4$\omega$} based GW phase evolution compares with its Numerical Relativity counterpart during the inspiral phase that extends what were pursued 
in Ref.~\cite{GHHB}.
It will also be interesting to apply Pad\'e approximation to the amplitudes of the two GW polarization states while incorporating 
PN-accurate corrections, as pursued in Ref.~\cite{BSGKJ_17}.

\section{Acknowledgements}

We thank Gihyuk Cho and Sourav Chatterjee for their helpful comments. We acknowledge support of the Department of Atomic Energy, Government of India, under project no. 12-R\&D-TFR-5.02-0200.
The use of open software packages from {\tt PyCBC}  \cite{alex_nitz_2019_2556644} and  {\tt Matplotlib} \cite{matplotlib}
is warmly acknowledged.

\twocolumngrid

\appendix
\section{PN correction terms in $d\phi/d\omega$ and $de_t/d\omega$} \label{appendixA}
In Sec.~\ref{sec:level3D}, we presented the differential equations for the evolution of $\phi$ and $e_t$ with respect to orbital frequency $\omega$
while displaying explicitly only Newtonian-accurate contributions.
Here we explicitly list the 1PN, 1.5PN and 2PN order contributions appearing in our  Eq.~(\ref{Eqs_T4omg1}) and (\ref{Eqs_T4omg2}) for $\frac{d\phi}{d\omega}$ and $\frac{de_t}{d\omega}$, respectively. Following are the various PN terms appearing in Eq.~(\ref{Eqs_T4omg1}) which are exact in eccentricity:
\begin{widetext}
\begin{subequations} \label{eq:phiomega}
\begin{align}
\phi_{\omega}^{1PN} =&\, \frac{5\, (1-e_t^2)^{5/2}}{56 (96 + 292 e_t^2 + 37 e_t^4)^2}\left[11888+14784\,\eta+e_t^2(-87720+159600\,\eta) +e_t^4(-171038+141708\,\eta)\right.  \nonumber \\ &\, \left.+e_t^6(-11717+8288\,\eta)\right] \,,  \\ \nonumber \\
\phi_{\omega}^{1.5PN} =&\, -\frac{1920\,(1-e_t^2)^7\,\pi}{(96 + 292 e_t^2 + 37 e_t^4)^2}\, \phi(e_t)\,, \\ \nonumber \\ 
\phi_{\omega}^{2PN} =&\, \frac{5\,(1-e_t^2)^{3/2}}{84672 (96 + 292 e_t^2 + 37 e_t^4)^3}\left[4299903744+3422490624\,\eta  +3343527936\,\eta^2 +e_t^2(69946342912\right. \nonumber \\  &\,-6816321792\,\eta +37271485440\,\eta^2)+e_t^4(476651319744-588658174272\,\eta +325588018560\,\eta^2) \nonumber \\ &\,+e_t^6(735432808064 -1144863272448\,\eta +428361998400\,\eta^2) +e_t^8(499179942876  -834083043696\,\eta \nonumber \\ &\,+259710115560\,\eta^2)+e_t^{10}(50602495104-89112638412\,\eta+21810477024\,\eta^2) +e_t^{12}(1881805869  \nonumber \\ &\,-3555297144\,\eta+837170880\,\eta^2 )+\sqrt{1-e_t^2}\left(-1950842880 + 780337152\,\eta +e_t^2(-92665036800  \right. \nonumber \\ &\,+ 37066014720\,\eta)+e_t^4(-353688491520 + 141475396608\,\eta)+e_t^6(-308084999040 + 123233999616\,\eta)  \nonumber \\ &\,\left.\left.+e_t^8(-45168701760 + 18067480704 \,\eta)+e_t^{10}(-1370628000 + 548251200\,\eta)\right)\right].  \\ \nonumber
\end{align}
\end{subequations}
\end{widetext}

We now list various PN terms appearing in Eq.~(\ref{Eqs_T4omg2}) which are also exact in eccentricity.

\begin{widetext}
\begin{subequations} \label{eq:eomega}
\begin{align}
e_{\omega}^{1PN} = &\, \frac{1}{252\,(96 + 292 e_t^2 + 37 e_t^4)^2}\left[e_t\left(-2175744+4236288\,\eta\right)+e_t^3\left(13249032-10810016\,\eta\right) +e_t^5\left(-15681240 \right. \right. \nonumber \\  &\, \left. \left. +10200400\,\eta\right)+e_t^7\left(4800495-3846304\,\eta\right)+e_t^9\left(-192543+219632 \,\eta\right)\right] \,,  \\ \nonumber \\
e_{\omega}^{1.5PN} = &\, -\frac{2\,(1-e_t^2)^{7/2}\,\pi}{3\,(96+292 e_t^2+37 e_t^4)^2}\left[(-58368 e_t + 35136 e_t^3 + 23232 e_t^5)\phi(e_t)+(94560 e_t + 287620 e_t^3 + 36445 e_t^5)\phi_e(e_t)\right] \,, \\ \nonumber \\
e_{\omega}^{2PN} = &\,-\frac{1}{\left(127008  (1 - e_t^2) (96 + 292 e_t^2 + 37 e_t^4)^3\right)}\left[e_t(-2634989678592+1528438947840\,\eta-72224538624\,\eta^2) \right.  \nonumber \\  &\, +e_t^3(-8967549348736+8550074244096\,\eta  -3721065707520\,\eta^2)+e_t^5(9968753953856-18090544550400\,\eta \nonumber \\  &\,+5526952080384\,\eta^2)+e_t^7(-3117120147776 +7289636256000\,\eta-3469472530944\,\eta^2)+e_t^9(1968660609712 \nonumber \\  &\,+5502284032896\,\eta-352446993024\,\eta^2)  +e_t^{11}(3300114491838-5318475912288\,\eta+2269744327200\,\eta^2) \nonumber \\  &\,+e_t^{13}(-496014129723+523809598032\,\eta  -182851869984\,\eta^2)+e_t^{15}(-21855750579+14777383824\,\eta \nonumber \\  &\,+1365232512\,\eta^2)+\sqrt{1-e_t^2}\left(e_t(2309797969920  -923919187968\,\eta)+e_t^3(8443898265600-3377559306240\,\eta) \right. \nonumber \\  &\,+e_t^5(8623457095680-3449382838272\,\eta)+ e_t^7(9308253749760-3723301499904\,\eta)+e_t^9(-3669739940160 \nonumber \\  &\, \left.\left.+1467895976064\,\eta)+e_t^{11}(-639988549200  +255995419680\,\eta)+e_t^{13}(-1057341600+422936640\,\eta) \right)\right]. \\ \nonumber 
\end{align}
\end{subequations}
\end{widetext}
 
The symbols $\phi(e_t)$ and $\phi_e(e_t)$ that appear in above equations for $\phi_\omega^{1.5PN}$ and $e_\omega^{1.5PN}$ are the tail enhancement functions at 1.5PN order. We note that these functions first appeared in the temporal evolution of $\omega$ and $e_t$ in our Eqs.~(\ref{Eqs_T4t2}) and (\ref{Eqs_T4t3}). The expressions for these enhancement functions, which could model binaries with very high eccentricities like $e_0 \sim 0.9$, were extracted from Eqs.~(3.14a), (3.14b) and (3.16) of Ref.~\cite{THG}. 
For the present effort, we
 Taylor expanded the original expressions for  $\phi(e_t)$ and $\phi_e(e_t)$ expressions, given in Ref.~\cite{RS96}, to desired order to construct our 2PN-accurate post-circular $\tilde{h}(f)$ and it's Pad\'e approximants.

\bibliography{MS_STAG}

\begin{thebibliography}{78}%
\makeatletter
\providecommand \@ifxundefined [1]{%
 \@ifx{#1\undefined}
}%
\providecommand \@ifnum [1]{%
 \ifnum #1\expandafter \@firstoftwo
 \else \expandafter \@secondoftwo
 \fi
}%
\providecommand \@ifx [1]{%
 \ifx #1\expandafter \@firstoftwo
 \else \expandafter \@secondoftwo
 \fi
}%
\providecommand \natexlab [1]{#1}%
\providecommand \enquote  [1]{``#1''}%
\providecommand \bibnamefont  [1]{#1}%
\providecommand \bibfnamefont [1]{#1}%
\providecommand \citenamefont [1]{#1}%
\providecommand \href@noop [0]{\@secondoftwo}%
\providecommand \href [0]{\begingroup \@sanitize@url \@href}%
\providecommand \@href[1]{\@@startlink{#1}\@@href}%
\providecommand \@@href[1]{\endgroup#1\@@endlink}%
\providecommand \@sanitize@url [0]{\catcode `\\12\catcode `\$12\catcode
  `\&12\catcode `\#12\catcode `\^12\catcode `\_12\catcode `\%12\relax}%
\providecommand \@@startlink[1]{}%
\providecommand \@@endlink[0]{}%
\providecommand \url  [0]{\begingroup\@sanitize@url \@url }%
\providecommand \@url [1]{\endgroup\@href {#1}{\urlprefix }}%
\providecommand \urlprefix  [0]{URL }%
\providecommand \Eprint [0]{\href }%
\providecommand \doibase [0]{http://dx.doi.org/}%
\providecommand \selectlanguage [0]{\@gobble}%
\providecommand \bibinfo  [0]{\@secondoftwo}%
\providecommand \bibfield  [0]{\@secondoftwo}%
\providecommand \translation [1]{[#1]}%
\providecommand \BibitemOpen [0]{}%
\providecommand \bibitemStop [0]{}%
\providecommand \bibitemNoStop [0]{.\EOS\space}%
\providecommand \EOS [0]{\spacefactor3000\relax}%
\providecommand \BibitemShut  [1]{\csname bibitem#1\endcsname}%
\let\auto@bib@innerbib\@empty
\bibitem [{\citenamefont {Aasi}\ \emph {et~al.}(2015)\citenamefont {Aasi},
  \citenamefont {Abbott}, \citenamefont {Abbott}, \citenamefont {Abbott},
  \citenamefont {Abernathy}, \citenamefont {Ackley}, \citenamefont {Adams},
  \citenamefont {Adams}, \citenamefont {Addesso},\ and\ \citenamefont
  {et~al.}}]{ALIGO}%
  \BibitemOpen
  \bibfield  {author} {\bibinfo {author} {\bibfnamefont {J.}~\bibnamefont
  {Aasi}}, \bibinfo {author} {\bibfnamefont {B.~P.}\ \bibnamefont {Abbott}},
  \bibinfo {author} {\bibfnamefont {R.}~\bibnamefont {Abbott}}, \bibinfo
  {author} {\bibfnamefont {T.}~\bibnamefont {Abbott}}, \bibinfo {author}
  {\bibfnamefont {M.~R.}\ \bibnamefont {Abernathy}}, \bibinfo {author}
  {\bibfnamefont {K.}~\bibnamefont {Ackley}}, \bibinfo {author} {\bibfnamefont
  {C.}~\bibnamefont {Adams}}, \bibinfo {author} {\bibfnamefont
  {T.}~\bibnamefont {Adams}}, \bibinfo {author} {\bibfnamefont
  {P.}~\bibnamefont {Addesso}}, \ and\ \bibinfo {author} {\bibnamefont
  {et~al.}},\ }\href {\doibase 10.1088/0264-9381/32/7/074001} {\bibfield
  {journal} {\bibinfo  {journal} {Classical and Quantum Gravity}\ }\textbf
  {\bibinfo {volume} {32}},\ \bibinfo {pages} {074001} (\bibinfo {year}
  {2015})}\BibitemShut {NoStop}%
\bibitem [{\citenamefont {Acernese}\ \emph {et~al.}(2014)\citenamefont
  {Acernese}, \citenamefont {Agathos}, \citenamefont {Agatsuma}, \citenamefont
  {Aisa}, \citenamefont {Allemandou}, \citenamefont {Allocca}, \citenamefont
  {Amarni}, \citenamefont {Astone}, \citenamefont {Balestri}, \citenamefont
  {Ballardin},\ and\ \citenamefont {et~al.}}]{AVirgo}%
  \BibitemOpen
  \bibfield  {author} {\bibinfo {author} {\bibfnamefont {F.}~\bibnamefont
  {Acernese}}, \bibinfo {author} {\bibfnamefont {M.}~\bibnamefont {Agathos}},
  \bibinfo {author} {\bibfnamefont {K.}~\bibnamefont {Agatsuma}}, \bibinfo
  {author} {\bibfnamefont {D.}~\bibnamefont {Aisa}}, \bibinfo {author}
  {\bibfnamefont {N.}~\bibnamefont {Allemandou}}, \bibinfo {author}
  {\bibfnamefont {A.}~\bibnamefont {Allocca}}, \bibinfo {author} {\bibfnamefont
  {J.}~\bibnamefont {Amarni}}, \bibinfo {author} {\bibfnamefont
  {P.}~\bibnamefont {Astone}}, \bibinfo {author} {\bibfnamefont
  {G.}~\bibnamefont {Balestri}}, \bibinfo {author} {\bibfnamefont
  {G.}~\bibnamefont {Ballardin}}, \ and\ \bibinfo {author} {\bibnamefont
  {et~al.}},\ }\href {\doibase 10.1088/0264-9381/32/2/024001} {\bibfield
  {journal} {\bibinfo  {journal} {Classical and Quantum Gravity}\ }\textbf
  {\bibinfo {volume} {32}},\ \bibinfo {pages} {024001} (\bibinfo {year}
  {2014})}\BibitemShut {NoStop}%
\bibitem [{\citenamefont {{KAGRA collaboration}}(2019)}]{KAGRA}%
  \BibitemOpen
  \bibfield  {author} {\bibinfo {author} {\bibnamefont {{KAGRA
  collaboration}}},\ }\href {\doibase 10.1038/s41550-018-0658-y} {\bibfield
  {journal} {\bibinfo  {journal} {Nature Astronomy}\ }\textbf {\bibinfo
  {volume} {3}},\ \bibinfo {pages} {35–40} (\bibinfo {year}
  {2019})}\BibitemShut {NoStop}%
\bibitem [{\citenamefont {Abbott}\ \emph
  {et~al.}(2019{\natexlab{a}})\citenamefont {Abbott}, \citenamefont {Abbott},
  \citenamefont {Abbott}, \citenamefont {Abraham}, \citenamefont {Acernese},
  \citenamefont {Ackley}, \citenamefont {Adams}, \citenamefont {Adhikari},
  \citenamefont {Adya}, \citenamefont {Affeldt},\ and\ \citenamefont
  {et~al.}}]{GWTC-1}%
  \BibitemOpen
  \bibfield  {author} {\bibinfo {author} {\bibfnamefont {B.}~\bibnamefont
  {Abbott}}, \bibinfo {author} {\bibfnamefont {R.}~\bibnamefont {Abbott}},
  \bibinfo {author} {\bibfnamefont {T.}~\bibnamefont {Abbott}}, \bibinfo
  {author} {\bibfnamefont {S.}~\bibnamefont {Abraham}}, \bibinfo {author}
  {\bibfnamefont {F.}~\bibnamefont {Acernese}}, \bibinfo {author}
  {\bibfnamefont {K.}~\bibnamefont {Ackley}}, \bibinfo {author} {\bibfnamefont
  {C.}~\bibnamefont {Adams}}, \bibinfo {author} {\bibfnamefont
  {R.}~\bibnamefont {Adhikari}}, \bibinfo {author} {\bibfnamefont
  {V.}~\bibnamefont {Adya}}, \bibinfo {author} {\bibfnamefont {C.}~\bibnamefont
  {Affeldt}}, \ and\ \bibinfo {author} {\bibnamefont {et~al.}},\ }\href
  {\doibase 10.1103/physrevx.9.031040} {\bibfield  {journal} {\bibinfo
  {journal} {Physical Review X}\ }\textbf {\bibinfo {volume} {9}} (\bibinfo
  {year} {2019}{\natexlab{a}}),\ 10.1103/physrevx.9.031040}\BibitemShut
  {NoStop}%
\bibitem [{\citenamefont {GraceDB}(2020)}]{O3_public}%
  \BibitemOpen
  \bibfield  {author} {\bibinfo {author} {\bibnamefont {GraceDB}},\ }\href
  {https://gracedb.ligo.org/superevents/public/O3/} {\enquote {\bibinfo {title}
  {Ligo/virgo o3 public alerts},}\ } (\bibinfo {year} {2020})\BibitemShut
  {NoStop}%
\bibitem [{\citenamefont {Abbott}\ \emph
  {et~al.}(2019{\natexlab{b}})\citenamefont {Abbott}, \citenamefont {Abbott},
  \citenamefont {Abbott}, \citenamefont {Abraham}, \citenamefont {Acernese},
  \citenamefont {Ackley}, \citenamefont {Adams}, \citenamefont {Adhikari},
  \citenamefont {Adya}, \citenamefont {Affeldt},\ and\ \citenamefont
  {et~al.}}]{eBBH_19}%
  \BibitemOpen
  \bibfield  {author} {\bibinfo {author} {\bibfnamefont {B.~P.}\ \bibnamefont
  {Abbott}}, \bibinfo {author} {\bibfnamefont {R.}~\bibnamefont {Abbott}},
  \bibinfo {author} {\bibfnamefont {T.~D.}\ \bibnamefont {Abbott}}, \bibinfo
  {author} {\bibfnamefont {S.}~\bibnamefont {Abraham}}, \bibinfo {author}
  {\bibfnamefont {F.}~\bibnamefont {Acernese}}, \bibinfo {author}
  {\bibfnamefont {K.}~\bibnamefont {Ackley}}, \bibinfo {author} {\bibfnamefont
  {C.}~\bibnamefont {Adams}}, \bibinfo {author} {\bibfnamefont {R.~X.}\
  \bibnamefont {Adhikari}}, \bibinfo {author} {\bibfnamefont {V.~B.}\
  \bibnamefont {Adya}}, \bibinfo {author} {\bibfnamefont {C.}~\bibnamefont
  {Affeldt}}, \ and\ \bibinfo {author} {\bibnamefont {et~al.}},\ }\href
  {\doibase 10.3847/1538-4357/ab3c2d} {\bibfield  {journal} {\bibinfo
  {journal} {The Astrophysical Journal}\ }\textbf {\bibinfo {volume} {883}},\
  \bibinfo {pages} {149} (\bibinfo {year} {2019}{\natexlab{b}})}\BibitemShut
  {NoStop}%
\bibitem [{\citenamefont {Laine}\ \emph {et~al.}(2020)\citenamefont {Laine},
  \citenamefont {Dey}, \citenamefont {Valtonen}, \citenamefont {Gopakumar},
  \citenamefont {Zola}, \citenamefont {Komossa}, \citenamefont {Kidger},
  \citenamefont {Pihajoki}, \citenamefont {Gómez}, \citenamefont {Caton},\
  and\ \citenamefont {et~al.}}]{Laine_2020}%
  \BibitemOpen
  \bibfield  {author} {\bibinfo {author} {\bibfnamefont {S.}~\bibnamefont
  {Laine}}, \bibinfo {author} {\bibfnamefont {L.}~\bibnamefont {Dey}}, \bibinfo
  {author} {\bibfnamefont {M.}~\bibnamefont {Valtonen}}, \bibinfo {author}
  {\bibfnamefont {A.}~\bibnamefont {Gopakumar}}, \bibinfo {author}
  {\bibfnamefont {S.}~\bibnamefont {Zola}}, \bibinfo {author} {\bibfnamefont
  {S.}~\bibnamefont {Komossa}}, \bibinfo {author} {\bibfnamefont
  {M.}~\bibnamefont {Kidger}}, \bibinfo {author} {\bibfnamefont
  {P.}~\bibnamefont {Pihajoki}}, \bibinfo {author} {\bibfnamefont {J.~L.}\
  \bibnamefont {Gómez}}, \bibinfo {author} {\bibfnamefont {D.}~\bibnamefont
  {Caton}}, \ and\ \bibinfo {author} {\bibnamefont {et~al.}},\ }\href {\doibase
  10.3847/2041-8213/ab79a4} {\bibfield  {journal} {\bibinfo  {journal} {The
  Astrophysical Journal}\ }\textbf {\bibinfo {volume} {894}},\ \bibinfo {pages}
  {L1} (\bibinfo {year} {2020})}\BibitemShut {NoStop}%
\bibitem [{\citenamefont {Perera}\ \emph {et~al.}(2019)\citenamefont {Perera},
  \citenamefont {DeCesar}, \citenamefont {Demorest}, \citenamefont {Kerr},
  \citenamefont {Lentati}, \citenamefont {Nice}, \citenamefont {Osłowski},
  \citenamefont {Ransom}, \citenamefont {Keith}, \citenamefont {Arzoumanian},\
  and\ \citenamefont {et~al.}}]{ipta}%
  \BibitemOpen
  \bibfield  {author} {\bibinfo {author} {\bibfnamefont {B.~B.~P.}\
  \bibnamefont {Perera}}, \bibinfo {author} {\bibfnamefont {M.~E.}\
  \bibnamefont {DeCesar}}, \bibinfo {author} {\bibfnamefont {P.~B.}\
  \bibnamefont {Demorest}}, \bibinfo {author} {\bibfnamefont {M.}~\bibnamefont
  {Kerr}}, \bibinfo {author} {\bibfnamefont {L.}~\bibnamefont {Lentati}},
  \bibinfo {author} {\bibfnamefont {D.~J.}\ \bibnamefont {Nice}}, \bibinfo
  {author} {\bibfnamefont {S.}~\bibnamefont {Osłowski}}, \bibinfo {author}
  {\bibfnamefont {S.~M.}\ \bibnamefont {Ransom}}, \bibinfo {author}
  {\bibfnamefont {M.~J.}\ \bibnamefont {Keith}}, \bibinfo {author}
  {\bibfnamefont {Z.}~\bibnamefont {Arzoumanian}}, \ and\ \bibinfo {author}
  {\bibnamefont {et~al.}},\ }\href {\doibase 10.1093/mnras/stz2857} {\bibfield
  {journal} {\bibinfo  {journal} {Monthly Notices of the Royal Astronomical
  Society}\ }\textbf {\bibinfo {volume} {490}},\ \bibinfo {pages} {4666–4687}
  (\bibinfo {year} {2019})}\BibitemShut {NoStop}%
\bibitem [{\citenamefont {Susobhanan}\ \emph {et~al.}(2020)\citenamefont
  {Susobhanan}, \citenamefont {Gopakumar}, \citenamefont {Hobbs},\ and\
  \citenamefont {Taylor}}]{AGGT}%
  \BibitemOpen
  \bibfield  {author} {\bibinfo {author} {\bibfnamefont {A.}~\bibnamefont
  {Susobhanan}}, \bibinfo {author} {\bibfnamefont {A.}~\bibnamefont
  {Gopakumar}}, \bibinfo {author} {\bibfnamefont {G.}~\bibnamefont {Hobbs}}, \
  and\ \bibinfo {author} {\bibfnamefont {S.~R.}\ \bibnamefont {Taylor}},\
  }\href {\doibase 10.1103/physrevd.101.043022} {\bibfield  {journal} {\bibinfo
   {journal} {Physical Review D}\ }\textbf {\bibinfo {volume} {101}} (\bibinfo
  {year} {2020}),\ 10.1103/physrevd.101.043022}\BibitemShut {NoStop}%
\bibitem [{\citenamefont {Baibhav}\ \emph {et~al.}(2019)\citenamefont
  {Baibhav}, \citenamefont {Barack}, \citenamefont {Berti}, \citenamefont
  {Bonga}, \citenamefont {Brito}, \citenamefont {Cardoso}, \citenamefont
  {Compère}, \citenamefont {Das}, \citenamefont {Doneva}, \citenamefont
  {Garcia-Bellido}, \citenamefont {Heisenberg}, \citenamefont {Hughes},
  \citenamefont {Isi}, \citenamefont {Jani}, \citenamefont {Kavanagh},
  \citenamefont {Lukes-Gerakopoulos}, \citenamefont {Mueller}, \citenamefont
  {Pani}, \citenamefont {Petiteau}, \citenamefont {Rajendran}, \citenamefont
  {Sotiriou}, \citenamefont {Stergioulas}, \citenamefont {Taylor},
  \citenamefont {Vagenas}, \citenamefont {van~de Meent}, \citenamefont
  {Warburton}, \citenamefont {Wardell}, \citenamefont {Witzany},\ and\
  \citenamefont {Zimmerman}}]{mHZ-GW}%
  \BibitemOpen
  \bibfield  {author} {\bibinfo {author} {\bibfnamefont {V.}~\bibnamefont
  {Baibhav}}, \bibinfo {author} {\bibfnamefont {L.}~\bibnamefont {Barack}},
  \bibinfo {author} {\bibfnamefont {E.}~\bibnamefont {Berti}}, \bibinfo
  {author} {\bibfnamefont {B.}~\bibnamefont {Bonga}}, \bibinfo {author}
  {\bibfnamefont {R.}~\bibnamefont {Brito}}, \bibinfo {author} {\bibfnamefont
  {V.}~\bibnamefont {Cardoso}}, \bibinfo {author} {\bibfnamefont
  {G.}~\bibnamefont {Compère}}, \bibinfo {author} {\bibfnamefont
  {S.}~\bibnamefont {Das}}, \bibinfo {author} {\bibfnamefont {D.}~\bibnamefont
  {Doneva}}, \bibinfo {author} {\bibfnamefont {J.}~\bibnamefont
  {Garcia-Bellido}}, \bibinfo {author} {\bibfnamefont {L.}~\bibnamefont
  {Heisenberg}}, \bibinfo {author} {\bibfnamefont {S.~A.}\ \bibnamefont
  {Hughes}}, \bibinfo {author} {\bibfnamefont {M.}~\bibnamefont {Isi}},
  \bibinfo {author} {\bibfnamefont {K.}~\bibnamefont {Jani}}, \bibinfo {author}
  {\bibfnamefont {C.}~\bibnamefont {Kavanagh}}, \bibinfo {author}
  {\bibfnamefont {G.}~\bibnamefont {Lukes-Gerakopoulos}}, \bibinfo {author}
  {\bibfnamefont {G.}~\bibnamefont {Mueller}}, \bibinfo {author} {\bibfnamefont
  {P.}~\bibnamefont {Pani}}, \bibinfo {author} {\bibfnamefont {A.}~\bibnamefont
  {Petiteau}}, \bibinfo {author} {\bibfnamefont {S.}~\bibnamefont {Rajendran}},
  \bibinfo {author} {\bibfnamefont {T.~P.}\ \bibnamefont {Sotiriou}}, \bibinfo
  {author} {\bibfnamefont {N.}~\bibnamefont {Stergioulas}}, \bibinfo {author}
  {\bibfnamefont {A.}~\bibnamefont {Taylor}}, \bibinfo {author} {\bibfnamefont
  {E.}~\bibnamefont {Vagenas}}, \bibinfo {author} {\bibfnamefont
  {M.}~\bibnamefont {van~de Meent}}, \bibinfo {author} {\bibfnamefont
  {N.}~\bibnamefont {Warburton}}, \bibinfo {author} {\bibfnamefont
  {B.}~\bibnamefont {Wardell}}, \bibinfo {author} {\bibfnamefont
  {V.}~\bibnamefont {Witzany}}, \ and\ \bibinfo {author} {\bibfnamefont
  {A.}~\bibnamefont {Zimmerman}},\ }\href@noop {} {\enquote {\bibinfo {title}
  {Probing the nature of black holes: Deep in the mhz gravitational-wave
  sky},}\ } (\bibinfo {year} {2019}),\ \Eprint
  {http://arxiv.org/abs/1908.11390} {arXiv:1908.11390 [astro-ph.HE]}
  \BibitemShut {NoStop}%
\bibitem [{\citenamefont {{Zwick}}\ \emph {et~al.}(2020)\citenamefont
  {{Zwick}}, \citenamefont {{Capelo}}, \citenamefont {{Bortolas}},
  \citenamefont {{Mayer}},\ and\ \citenamefont {{Amaro-Seoane}}}]{Zwick_2020}%
  \BibitemOpen
  \bibfield  {author} {\bibinfo {author} {\bibfnamefont {L.}~\bibnamefont
  {{Zwick}}}, \bibinfo {author} {\bibfnamefont {P.~R.}\ \bibnamefont
  {{Capelo}}}, \bibinfo {author} {\bibfnamefont {E.}~\bibnamefont
  {{Bortolas}}}, \bibinfo {author} {\bibfnamefont {L.}~\bibnamefont {{Mayer}}},
  \ and\ \bibinfo {author} {\bibfnamefont {P.}~\bibnamefont {{Amaro-Seoane}}},\
  }\href {\doibase 10.1093/mnras/staa1314} {\bibfield  {journal} {\bibinfo
  {journal} {\mnras}\ }\textbf {\bibinfo {volume} {495}},\ \bibinfo {pages}
  {2321} (\bibinfo {year} {2020})},\ \Eprint {http://arxiv.org/abs/1911.06024}
  {arXiv:1911.06024 [astro-ph.GA]} \BibitemShut {NoStop}%
\bibitem [{\citenamefont {Sato}\ \emph {et~al.}(2017)\citenamefont {Sato},
  \citenamefont {Kawamura}, \citenamefont {Ando}, \citenamefont {Nakamura},
  \citenamefont {Tsubono}, \citenamefont {Araya}, \citenamefont {Funaki},
  \citenamefont {Ioka}, \citenamefont {Kanda}, \citenamefont {Moriwaki},\ and\
  \citenamefont {et~al.}}]{DECIGO}%
  \BibitemOpen
  \bibfield  {author} {\bibinfo {author} {\bibfnamefont {S.}~\bibnamefont
  {Sato}}, \bibinfo {author} {\bibfnamefont {S.}~\bibnamefont {Kawamura}},
  \bibinfo {author} {\bibfnamefont {M.}~\bibnamefont {Ando}}, \bibinfo {author}
  {\bibfnamefont {T.}~\bibnamefont {Nakamura}}, \bibinfo {author}
  {\bibfnamefont {K.}~\bibnamefont {Tsubono}}, \bibinfo {author} {\bibfnamefont
  {A.}~\bibnamefont {Araya}}, \bibinfo {author} {\bibfnamefont
  {I.}~\bibnamefont {Funaki}}, \bibinfo {author} {\bibfnamefont
  {K.}~\bibnamefont {Ioka}}, \bibinfo {author} {\bibfnamefont {N.}~\bibnamefont
  {Kanda}}, \bibinfo {author} {\bibfnamefont {S.}~\bibnamefont {Moriwaki}}, \
  and\ \bibinfo {author} {\bibnamefont {et~al.}},\ }\href {\doibase
  10.1088/1742-6596/840/1/012010} {\bibfield  {journal} {\bibinfo  {journal}
  {Journal of Physics: Conference Series}\ }\textbf {\bibinfo {volume} {840}},\
  \bibinfo {pages} {012010} (\bibinfo {year} {2017})}\BibitemShut {NoStop}%
\bibitem [{\citenamefont {{Romero-Shaw}}\ \emph {et~al.}(2019)\citenamefont
  {{Romero-Shaw}}, \citenamefont {{Lasky}},\ and\ \citenamefont
  {{Thrane}}}]{RLT_2019}%
  \BibitemOpen
  \bibfield  {author} {\bibinfo {author} {\bibfnamefont {I.~M.}\ \bibnamefont
  {{Romero-Shaw}}}, \bibinfo {author} {\bibfnamefont {P.~D.}\ \bibnamefont
  {{Lasky}}}, \ and\ \bibinfo {author} {\bibfnamefont {E.}~\bibnamefont
  {{Thrane}}},\ }\href {\doibase 10.1093/mnras/stz2996} {\bibfield  {journal}
  {\bibinfo  {journal} {\mnras}\ }\textbf {\bibinfo {volume} {490}},\ \bibinfo
  {pages} {5210} (\bibinfo {year} {2019})},\ \Eprint
  {http://arxiv.org/abs/1909.05466} {arXiv:1909.05466 [astro-ph.HE]}
  \BibitemShut {NoStop}%
\bibitem [{\citenamefont {{Moore}}\ and\ \citenamefont
  {{Yunes}}(2020)}]{MY_2020}%
  \BibitemOpen
  \bibfield  {author} {\bibinfo {author} {\bibfnamefont {B.}~\bibnamefont
  {{Moore}}}\ and\ \bibinfo {author} {\bibfnamefont {N.}~\bibnamefont
  {{Yunes}}},\ }\href@noop {} {\bibfield  {journal} {\bibinfo  {journal} {arXiv
  e-prints}\ ,\ \bibinfo {eid} {arXiv:2002.05775}} (\bibinfo {year} {2020})},\
  \Eprint {http://arxiv.org/abs/2002.05775} {arXiv:2002.05775 [gr-qc]}
  \BibitemShut {NoStop}%
\bibitem [{\citenamefont {{Belczynski}}\ \emph {et~al.}(2002)\citenamefont
  {{Belczynski}}, \citenamefont {{Kalogera}},\ and\ \citenamefont
  {{Bulik}}}]{BKT_2002}%
  \BibitemOpen
  \bibfield  {author} {\bibinfo {author} {\bibfnamefont {K.}~\bibnamefont
  {{Belczynski}}}, \bibinfo {author} {\bibfnamefont {V.}~\bibnamefont
  {{Kalogera}}}, \ and\ \bibinfo {author} {\bibfnamefont {T.}~\bibnamefont
  {{Bulik}}},\ }\href {\doibase 10.1086/340304} {\bibfield  {journal} {\bibinfo
   {journal} {\apj}\ }\textbf {\bibinfo {volume} {572}},\ \bibinfo {pages}
  {407} (\bibinfo {year} {2002})},\ \Eprint
  {http://arxiv.org/abs/astro-ph/0111452} {arXiv:astro-ph/0111452 [astro-ph]}
  \BibitemShut {NoStop}%
\bibitem [{\citenamefont {Kruckow}\ \emph {et~al.}(2018)\citenamefont
  {Kruckow}, \citenamefont {Tauris}, \citenamefont {Langer}, \citenamefont
  {Kramer},\ and\ \citenamefont {Izzard}}]{Kruckow_2018}%
  \BibitemOpen
  \bibfield  {author} {\bibinfo {author} {\bibfnamefont {M.~U.}\ \bibnamefont
  {Kruckow}}, \bibinfo {author} {\bibfnamefont {T.~M.}\ \bibnamefont {Tauris}},
  \bibinfo {author} {\bibfnamefont {N.}~\bibnamefont {Langer}}, \bibinfo
  {author} {\bibfnamefont {M.}~\bibnamefont {Kramer}}, \ and\ \bibinfo {author}
  {\bibfnamefont {R.~G.}\ \bibnamefont {Izzard}},\ }\href {\doibase
  10.1093/mnras/sty2190} {\bibfield  {journal} {\bibinfo  {journal} {Monthly
  Notices of the Royal Astronomical Society}\ }\textbf {\bibinfo {volume}
  {481}},\ \bibinfo {pages} {1908–1949} (\bibinfo {year} {2018})}\BibitemShut
  {NoStop}%
\bibitem [{\citenamefont {Kowalska}\ \emph {et~al.}(2011)\citenamefont
  {Kowalska}, \citenamefont {Bulik}, \citenamefont {Belczynski}, \citenamefont
  {Dominik},\ and\ \citenamefont {Gondek-Rosinska}}]{Kowalska_2011}%
  \BibitemOpen
  \bibfield  {author} {\bibinfo {author} {\bibfnamefont {I.}~\bibnamefont
  {Kowalska}}, \bibinfo {author} {\bibfnamefont {T.}~\bibnamefont {Bulik}},
  \bibinfo {author} {\bibfnamefont {K.}~\bibnamefont {Belczynski}}, \bibinfo
  {author} {\bibfnamefont {M.}~\bibnamefont {Dominik}}, \ and\ \bibinfo
  {author} {\bibfnamefont {D.}~\bibnamefont {Gondek-Rosinska}},\ }\href
  {\doibase 10.1051/0004-6361/201015777} {\bibfield  {journal} {\bibinfo
  {journal} {Astronomy \& Astrophysics}\ }\textbf {\bibinfo {volume} {527}},\
  \bibinfo {pages} {A70} (\bibinfo {year} {2011})}\BibitemShut {NoStop}%
\bibitem [{\citenamefont {Abbott}\ \emph {et~al.}(2016)\citenamefont {Abbott},
  \citenamefont {Abbott}, \citenamefont {Abbott}, \citenamefont {Abernathy},
  \citenamefont {Acernese}, \citenamefont {Ackley}, \citenamefont {Adams},
  \citenamefont {Adams}, \citenamefont {Addesso}, \citenamefont {Adhikari},\
  and\ \citenamefont {et~al.}}]{GW150914}%
  \BibitemOpen
  \bibfield  {author} {\bibinfo {author} {\bibfnamefont {B.}~\bibnamefont
  {Abbott}}, \bibinfo {author} {\bibfnamefont {R.}~\bibnamefont {Abbott}},
  \bibinfo {author} {\bibfnamefont {T.}~\bibnamefont {Abbott}}, \bibinfo
  {author} {\bibfnamefont {M.}~\bibnamefont {Abernathy}}, \bibinfo {author}
  {\bibfnamefont {F.}~\bibnamefont {Acernese}}, \bibinfo {author}
  {\bibfnamefont {K.}~\bibnamefont {Ackley}}, \bibinfo {author} {\bibfnamefont
  {C.}~\bibnamefont {Adams}}, \bibinfo {author} {\bibfnamefont
  {T.}~\bibnamefont {Adams}}, \bibinfo {author} {\bibfnamefont
  {P.}~\bibnamefont {Addesso}}, \bibinfo {author} {\bibfnamefont
  {R.}~\bibnamefont {Adhikari}}, \ and\ \bibinfo {author} {\bibnamefont
  {et~al.}},\ }\href {\doibase 10.1103/physrevlett.116.241102} {\bibfield
  {journal} {\bibinfo  {journal} {Physical Review Letters}\ }\textbf {\bibinfo
  {volume} {116}} (\bibinfo {year} {2016}),\
  10.1103/physrevlett.116.241102}\BibitemShut {NoStop}%
\bibitem [{\citenamefont {{Huerta}}\ \emph {et~al.}(2018)\citenamefont
  {{Huerta}}, \citenamefont {{Moore}}, \citenamefont {{Kumar}}, \citenamefont
  {{George}}, \citenamefont {{Chua}}, \citenamefont {{Haas}}, \citenamefont
  {{Wessel}}, \citenamefont {{Johnson}}, \citenamefont {{Glennon}},
  \citenamefont {{Rebei}}, \citenamefont {{Holgado}}, \citenamefont {{Gair}},\
  and\ \citenamefont {{Pfeiffer}}}]{ENIGMA}%
  \BibitemOpen
  \bibfield  {author} {\bibinfo {author} {\bibfnamefont {E.~A.}\ \bibnamefont
  {{Huerta}}}, \bibinfo {author} {\bibfnamefont {C.~J.}\ \bibnamefont
  {{Moore}}}, \bibinfo {author} {\bibfnamefont {P.}~\bibnamefont {{Kumar}}},
  \bibinfo {author} {\bibfnamefont {D.}~\bibnamefont {{George}}}, \bibinfo
  {author} {\bibfnamefont {A.~J.~K.}\ \bibnamefont {{Chua}}}, \bibinfo {author}
  {\bibfnamefont {R.}~\bibnamefont {{Haas}}}, \bibinfo {author} {\bibfnamefont
  {E.}~\bibnamefont {{Wessel}}}, \bibinfo {author} {\bibfnamefont
  {D.}~\bibnamefont {{Johnson}}}, \bibinfo {author} {\bibfnamefont
  {D.}~\bibnamefont {{Glennon}}}, \bibinfo {author} {\bibfnamefont
  {A.}~\bibnamefont {{Rebei}}}, \bibinfo {author} {\bibfnamefont {A.~M.}\
  \bibnamefont {{Holgado}}}, \bibinfo {author} {\bibfnamefont {J.~R.}\
  \bibnamefont {{Gair}}}, \ and\ \bibinfo {author} {\bibfnamefont {H.~P.}\
  \bibnamefont {{Pfeiffer}}},\ }\href {\doibase 10.1103/PhysRevD.97.024031}
  {\bibfield  {journal} {\bibinfo  {journal} {\prd}\ }\textbf {\bibinfo
  {volume} {97}},\ \bibinfo {eid} {024031} (\bibinfo {year} {2018})},\ \Eprint
  {http://arxiv.org/abs/1711.06276} {arXiv:1711.06276 [gr-qc]} \BibitemShut
  {NoStop}%
\bibitem [{\citenamefont {Fragione}\ and\ \citenamefont
  {Bromberg}(2019)}]{Fragione_2019}%
  \BibitemOpen
  \bibfield  {author} {\bibinfo {author} {\bibfnamefont {G.}~\bibnamefont
  {Fragione}}\ and\ \bibinfo {author} {\bibfnamefont {O.}~\bibnamefont
  {Bromberg}},\ }\href {\doibase 10.1093/mnras/stz2024} {\bibfield  {journal}
  {\bibinfo  {journal} {Monthly Notices of the Royal Astronomical Society}\
  }\textbf {\bibinfo {volume} {488}},\ \bibinfo {pages} {4370–4377} (\bibinfo
  {year} {2019})}\BibitemShut {NoStop}%
\bibitem [{\citenamefont {{Samsing}}(2018)}]{Samsing_2018}%
  \BibitemOpen
  \bibfield  {author} {\bibinfo {author} {\bibfnamefont {J.}~\bibnamefont
  {{Samsing}}},\ }\href {\doibase 10.1103/PhysRevD.97.103014} {\bibfield
  {journal} {\bibinfo  {journal} {\prd}\ }\textbf {\bibinfo {volume} {97}},\
  \bibinfo {eid} {103014} (\bibinfo {year} {2018})},\ \Eprint
  {http://arxiv.org/abs/1711.07452} {arXiv:1711.07452 [astro-ph.HE]}
  \BibitemShut {NoStop}%
\bibitem [{\citenamefont {{Kumamoto}}\ \emph {et~al.}(2020)\citenamefont
  {{Kumamoto}}, \citenamefont {{Fujii}},\ and\ \citenamefont
  {{Tanikawa}}}]{KFT_2020}%
  \BibitemOpen
  \bibfield  {author} {\bibinfo {author} {\bibfnamefont {J.}~\bibnamefont
  {{Kumamoto}}}, \bibinfo {author} {\bibfnamefont {M.~S.}\ \bibnamefont
  {{Fujii}}}, \ and\ \bibinfo {author} {\bibfnamefont {A.}~\bibnamefont
  {{Tanikawa}}},\ }\href {\doibase 10.1093/mnras/staa1440} {\bibfield
  {journal} {\bibinfo  {journal} {\mnras}\ }\textbf {\bibinfo {volume} {495}},\
  \bibinfo {pages} {4268} (\bibinfo {year} {2020})},\ \Eprint
  {http://arxiv.org/abs/2001.10690} {arXiv:2001.10690 [astro-ph.HE]}
  \BibitemShut {NoStop}%
\bibitem [{\citenamefont {{O'Leary}}\ \emph {et~al.}(2009)\citenamefont
  {{O'Leary}}, \citenamefont {{Kocsis}},\ and\ \citenamefont
  {{Loeb}}}]{OKL_2009}%
  \BibitemOpen
  \bibfield  {author} {\bibinfo {author} {\bibfnamefont {R.~M.}\ \bibnamefont
  {{O'Leary}}}, \bibinfo {author} {\bibfnamefont {B.}~\bibnamefont {{Kocsis}}},
  \ and\ \bibinfo {author} {\bibfnamefont {A.}~\bibnamefont {{Loeb}}},\ }\href
  {\doibase 10.1111/j.1365-2966.2009.14653.x} {\bibfield  {journal} {\bibinfo
  {journal} {\mnras}\ }\textbf {\bibinfo {volume} {395}},\ \bibinfo {pages}
  {2127} (\bibinfo {year} {2009})},\ \Eprint {http://arxiv.org/abs/0807.2638}
  {arXiv:0807.2638 [astro-ph]} \BibitemShut {NoStop}%
\bibitem [{\citenamefont {Kremer}\ \emph {et~al.}(2019)\citenamefont {Kremer},
  \citenamefont {Rodriguez}, \citenamefont {Amaro-Seoane}, \citenamefont
  {Breivik}, \citenamefont {Chatterjee}, \citenamefont {Katz}, \citenamefont
  {Larson}, \citenamefont {Rasio}, \citenamefont {Samsing}, \citenamefont
  {Ye},\ and\ \citenamefont {Zevin}}]{Kremer_2019}%
  \BibitemOpen
  \bibfield  {author} {\bibinfo {author} {\bibfnamefont {K.}~\bibnamefont
  {Kremer}}, \bibinfo {author} {\bibfnamefont {C.~L.}\ \bibnamefont
  {Rodriguez}}, \bibinfo {author} {\bibfnamefont {P.}~\bibnamefont
  {Amaro-Seoane}}, \bibinfo {author} {\bibfnamefont {K.}~\bibnamefont
  {Breivik}}, \bibinfo {author} {\bibfnamefont {S.}~\bibnamefont {Chatterjee}},
  \bibinfo {author} {\bibfnamefont {M.~L.}\ \bibnamefont {Katz}}, \bibinfo
  {author} {\bibfnamefont {S.~L.}\ \bibnamefont {Larson}}, \bibinfo {author}
  {\bibfnamefont {F.~A.}\ \bibnamefont {Rasio}}, \bibinfo {author}
  {\bibfnamefont {J.}~\bibnamefont {Samsing}}, \bibinfo {author} {\bibfnamefont
  {C.~S.}\ \bibnamefont {Ye}}, \ and\ \bibinfo {author} {\bibfnamefont
  {M.}~\bibnamefont {Zevin}},\ }\href {\doibase 10.1103/PhysRevD.99.063003}
  {\bibfield  {journal} {\bibinfo  {journal} {Phys. Rev. D}\ }\textbf {\bibinfo
  {volume} {99}},\ \bibinfo {pages} {063003} (\bibinfo {year}
  {2019})}\BibitemShut {NoStop}%
\bibitem [{\citenamefont {Peters}(1964)}]{Peters_64}%
  \BibitemOpen
  \bibfield  {author} {\bibinfo {author} {\bibfnamefont {P.~C.}\ \bibnamefont
  {Peters}},\ }\href {\doibase 10.1103/PhysRev.136.B1224} {\bibfield  {journal}
  {\bibinfo  {journal} {Phys. Rev.}\ }\textbf {\bibinfo {volume} {136}},\
  \bibinfo {pages} {B1224} (\bibinfo {year} {1964})}\BibitemShut {NoStop}%
\bibitem [{\citenamefont {{Kozai}}(1962)}]{Kozai}%
  \BibitemOpen
  \bibfield  {author} {\bibinfo {author} {\bibfnamefont {Y.}~\bibnamefont
  {{Kozai}}},\ }\href {\doibase 10.1086/108790} {\bibfield  {journal} {\bibinfo
   {journal} {\aj}\ }\textbf {\bibinfo {volume} {67}},\ \bibinfo {pages} {591}
  (\bibinfo {year} {1962})}\BibitemShut {NoStop}%
\bibitem [{\citenamefont {Antonini}\ \emph {et~al.}(2014)\citenamefont
  {Antonini}, \citenamefont {Murray},\ and\ \citenamefont
  {Mikkola}}]{Antonini_2014}%
  \BibitemOpen
  \bibfield  {author} {\bibinfo {author} {\bibfnamefont {F.}~\bibnamefont
  {Antonini}}, \bibinfo {author} {\bibfnamefont {N.}~\bibnamefont {Murray}}, \
  and\ \bibinfo {author} {\bibfnamefont {S.}~\bibnamefont {Mikkola}},\ }\href
  {\doibase 10.1088/0004-637x/781/1/45} {\bibfield  {journal} {\bibinfo
  {journal} {The Astrophysical Journal}\ }\textbf {\bibinfo {volume} {781}},\
  \bibinfo {pages} {45} (\bibinfo {year} {2014})}\BibitemShut {NoStop}%
\bibitem [{\citenamefont {{Randall}}\ and\ \citenamefont
  {{Xianyu}}(2018)}]{Randall_2018}%
  \BibitemOpen
  \bibfield  {author} {\bibinfo {author} {\bibfnamefont {L.}~\bibnamefont
  {{Randall}}}\ and\ \bibinfo {author} {\bibfnamefont {Z.-Z.}\ \bibnamefont
  {{Xianyu}}},\ }\href {\doibase 10.3847/1538-4357/aaa1a2} {\bibfield
  {journal} {\bibinfo  {journal} {\apj}\ }\textbf {\bibinfo {volume} {853}},\
  \bibinfo {eid} {93} (\bibinfo {year} {2018})},\ \Eprint
  {http://arxiv.org/abs/1708.08569} {arXiv:1708.08569 [gr-qc]} \BibitemShut
  {NoStop}%
\bibitem [{\citenamefont {{Farr}}\ \emph {et~al.}(2017)\citenamefont {{Farr}},
  \citenamefont {{Stevenson}}, \citenamefont {{Miller}}, \citenamefont {{Mand
  el}}, \citenamefont {{Farr}},\ and\ \citenamefont {{Vecchio}}}]{Farr_2017}%
  \BibitemOpen
  \bibfield  {author} {\bibinfo {author} {\bibfnamefont {W.~M.}\ \bibnamefont
  {{Farr}}}, \bibinfo {author} {\bibfnamefont {S.}~\bibnamefont {{Stevenson}}},
  \bibinfo {author} {\bibfnamefont {M.~C.}\ \bibnamefont {{Miller}}}, \bibinfo
  {author} {\bibfnamefont {I.}~\bibnamefont {{Mand el}}}, \bibinfo {author}
  {\bibfnamefont {B.}~\bibnamefont {{Farr}}}, \ and\ \bibinfo {author}
  {\bibfnamefont {A.}~\bibnamefont {{Vecchio}}},\ }\href {\doibase
  10.1038/nature23453} {\bibfield  {journal} {\bibinfo  {journal} {\nat}\
  }\textbf {\bibinfo {volume} {548}},\ \bibinfo {pages} {426} (\bibinfo {year}
  {2017})},\ \Eprint {http://arxiv.org/abs/1706.01385} {arXiv:1706.01385
  [astro-ph.HE]} \BibitemShut {NoStop}%
\bibitem [{\citenamefont {{Arca Sedda}}\ \emph {et~al.}(2020)\citenamefont
  {{Arca Sedda}}, \citenamefont {{Mapelli}}, \citenamefont {{Spera}},
  \citenamefont {{Benacquista}},\ and\ \citenamefont {{Giacobbo}}}]{Arca_2020}%
  \BibitemOpen
  \bibfield  {author} {\bibinfo {author} {\bibfnamefont {M.}~\bibnamefont
  {{Arca Sedda}}}, \bibinfo {author} {\bibfnamefont {M.}~\bibnamefont
  {{Mapelli}}}, \bibinfo {author} {\bibfnamefont {M.}~\bibnamefont {{Spera}}},
  \bibinfo {author} {\bibfnamefont {M.}~\bibnamefont {{Benacquista}}}, \ and\
  \bibinfo {author} {\bibfnamefont {N.}~\bibnamefont {{Giacobbo}}},\ }\href
  {\doibase 10.3847/1538-4357/ab88b2} {\bibfield  {journal} {\bibinfo
  {journal} {\apj}\ }\textbf {\bibinfo {volume} {894}},\ \bibinfo {eid} {133}
  (\bibinfo {year} {2020})},\ \Eprint {http://arxiv.org/abs/2003.07409}
  {arXiv:2003.07409 [astro-ph.GA]} \BibitemShut {NoStop}%
\bibitem [{\citenamefont {Tiwari}\ \emph {et~al.}(2016)\citenamefont {Tiwari},
  \citenamefont {Klimenko}, \citenamefont {Christensen}, \citenamefont
  {Huerta}, \citenamefont {Mohapatra}, \citenamefont {Gopakumar}, \citenamefont
  {Haney}, \citenamefont {Ajith}, \citenamefont {McWilliams}, \citenamefont
  {Vedovato},\ and\ \citenamefont {et~al.}}]{Tiwari_2016}%
  \BibitemOpen
  \bibfield  {author} {\bibinfo {author} {\bibfnamefont {V.}~\bibnamefont
  {Tiwari}}, \bibinfo {author} {\bibfnamefont {S.}~\bibnamefont {Klimenko}},
  \bibinfo {author} {\bibfnamefont {N.}~\bibnamefont {Christensen}}, \bibinfo
  {author} {\bibfnamefont {E.}~\bibnamefont {Huerta}}, \bibinfo {author}
  {\bibfnamefont {S.}~\bibnamefont {Mohapatra}}, \bibinfo {author}
  {\bibfnamefont {A.}~\bibnamefont {Gopakumar}}, \bibinfo {author}
  {\bibfnamefont {M.}~\bibnamefont {Haney}}, \bibinfo {author} {\bibfnamefont
  {P.}~\bibnamefont {Ajith}}, \bibinfo {author} {\bibfnamefont
  {S.}~\bibnamefont {McWilliams}}, \bibinfo {author} {\bibfnamefont
  {G.}~\bibnamefont {Vedovato}}, \ and\ \bibinfo {author} {\bibnamefont
  {et~al.}},\ }\href {\doibase 10.1103/physrevd.93.043007} {\bibfield
  {journal} {\bibinfo  {journal} {Physical Review D}\ }\textbf {\bibinfo
  {volume} {93}} (\bibinfo {year} {2016}),\
  10.1103/physrevd.93.043007}\BibitemShut {NoStop}%
\bibitem [{\citenamefont {Abbott}\ \emph
  {et~al.}(2019{\natexlab{c}})\citenamefont {Abbott}, \citenamefont {Abbott},
  \citenamefont {Abbott}, \citenamefont {Abraham}, \citenamefont {Acernese},
  \citenamefont {Ackley}, \citenamefont {Adams}, \citenamefont {Adhikari},
  \citenamefont {Adya}, \citenamefont {Affeldt},\ and\ \citenamefont
  {et~al.}}]{Abbott_2019}%
  \BibitemOpen
  \bibfield  {author} {\bibinfo {author} {\bibfnamefont {B.~P.}\ \bibnamefont
  {Abbott}}, \bibinfo {author} {\bibfnamefont {R.}~\bibnamefont {Abbott}},
  \bibinfo {author} {\bibfnamefont {T.~D.}\ \bibnamefont {Abbott}}, \bibinfo
  {author} {\bibfnamefont {S.}~\bibnamefont {Abraham}}, \bibinfo {author}
  {\bibfnamefont {F.}~\bibnamefont {Acernese}}, \bibinfo {author}
  {\bibfnamefont {K.}~\bibnamefont {Ackley}}, \bibinfo {author} {\bibfnamefont
  {C.}~\bibnamefont {Adams}}, \bibinfo {author} {\bibfnamefont {R.~X.}\
  \bibnamefont {Adhikari}}, \bibinfo {author} {\bibfnamefont {V.~B.}\
  \bibnamefont {Adya}}, \bibinfo {author} {\bibfnamefont {C.}~\bibnamefont
  {Affeldt}}, \ and\ \bibinfo {author} {\bibnamefont {et~al.}},\ }\href
  {\doibase 10.3847/1538-4357/ab3c2d} {\bibfield  {journal} {\bibinfo
  {journal} {The Astrophysical Journal}\ }\textbf {\bibinfo {volume} {883}},\
  \bibinfo {pages} {149} (\bibinfo {year} {2019}{\natexlab{c}})}\BibitemShut
  {NoStop}%
\bibitem [{\citenamefont {{Nitz}}\ \emph {et~al.}(2020)\citenamefont {{Nitz}},
  \citenamefont {{Lenon}},\ and\ \citenamefont {{Brown}}}]{Nitz_2020}%
  \BibitemOpen
  \bibfield  {author} {\bibinfo {author} {\bibfnamefont {A.~H.}\ \bibnamefont
  {{Nitz}}}, \bibinfo {author} {\bibfnamefont {A.}~\bibnamefont {{Lenon}}}, \
  and\ \bibinfo {author} {\bibfnamefont {D.~A.}\ \bibnamefont {{Brown}}},\
  }\href {\doibase 10.3847/1538-4357/ab6611} {\bibfield  {journal} {\bibinfo
  {journal} {\apj}\ }\textbf {\bibinfo {volume} {890}},\ \bibinfo {eid} {1}
  (\bibinfo {year} {2020})},\ \Eprint {http://arxiv.org/abs/1912.05464}
  {arXiv:1912.05464 [astro-ph.HE]} \BibitemShut {NoStop}%
\bibitem [{\citenamefont {{Lenon}}\ \emph {et~al.}(2020)\citenamefont
  {{Lenon}}, \citenamefont {{Nitz}},\ and\ \citenamefont {{Brown}}}]{LNB_2020}%
  \BibitemOpen
  \bibfield  {author} {\bibinfo {author} {\bibfnamefont {A.~K.}\ \bibnamefont
  {{Lenon}}}, \bibinfo {author} {\bibfnamefont {A.~H.}\ \bibnamefont {{Nitz}}},
  \ and\ \bibinfo {author} {\bibfnamefont {D.~A.}\ \bibnamefont {{Brown}}},\
  }\href@noop {} {\bibfield  {journal} {\bibinfo  {journal} {arXiv e-prints}\
  ,\ \bibinfo {eid} {arXiv:2005.14146}} (\bibinfo {year} {2020})},\ \Eprint
  {http://arxiv.org/abs/2005.14146} {arXiv:2005.14146 [astro-ph.HE]}
  \BibitemShut {NoStop}%
\bibitem [{\citenamefont {{Rodriguez}}\ \emph {et~al.}(2018)\citenamefont
  {{Rodriguez}}, \citenamefont {{Amaro-Seoane}}, \citenamefont {{Chatterjee}},
  \citenamefont {{Kremer}}, \citenamefont {{Rasio}}, \citenamefont {{Samsing}},
  \citenamefont {{Ye}},\ and\ \citenamefont {{Zevin}}}]{Rodriguez_2018}%
  \BibitemOpen
  \bibfield  {author} {\bibinfo {author} {\bibfnamefont {C.~L.}\ \bibnamefont
  {{Rodriguez}}}, \bibinfo {author} {\bibfnamefont {P.}~\bibnamefont
  {{Amaro-Seoane}}}, \bibinfo {author} {\bibfnamefont {S.}~\bibnamefont
  {{Chatterjee}}}, \bibinfo {author} {\bibfnamefont {K.}~\bibnamefont
  {{Kremer}}}, \bibinfo {author} {\bibfnamefont {F.~A.}\ \bibnamefont
  {{Rasio}}}, \bibinfo {author} {\bibfnamefont {J.}~\bibnamefont {{Samsing}}},
  \bibinfo {author} {\bibfnamefont {C.~S.}\ \bibnamefont {{Ye}}}, \ and\
  \bibinfo {author} {\bibfnamefont {M.}~\bibnamefont {{Zevin}}},\ }\href
  {\doibase 10.1103/PhysRevD.98.123005} {\bibfield  {journal} {\bibinfo
  {journal} {\prd}\ }\textbf {\bibinfo {volume} {98}},\ \bibinfo {eid} {123005}
  (\bibinfo {year} {2018})},\ \Eprint {http://arxiv.org/abs/1811.04926}
  {arXiv:1811.04926 [astro-ph.HE]} \BibitemShut {NoStop}%
\bibitem [{\citenamefont {Hannam}(2014)}]{Hannam_2013}%
  \BibitemOpen
  \bibfield  {author} {\bibinfo {author} {\bibfnamefont {M.}~\bibnamefont
  {Hannam}},\ }\href {\doibase 10.1007/s10714-014-1767-2} {\bibfield  {journal}
  {\bibinfo  {journal} {General Relativity and Gravitation}\ }\textbf {\bibinfo
  {volume} {46}} (\bibinfo {year} {2014}),\
  10.1007/s10714-014-1767-2}\BibitemShut {NoStop}%
\bibitem [{\citenamefont {{Damour}}\ and\ \citenamefont
  {{Nagar}}(2016)}]{DN_2014}%
  \BibitemOpen
  \bibfield  {author} {\bibinfo {author} {\bibfnamefont {T.}~\bibnamefont
  {{Damour}}}\ and\ \bibinfo {author} {\bibfnamefont {A.}~\bibnamefont
  {{Nagar}}},\ }\enquote {\bibinfo {title} {{The Effective-One-Body Approach to
  the General Relativistic Two Body Problem}},}\ in\ \href {\doibase
  10.1007/978-3-319-19416-5_7} {\emph {\bibinfo {booktitle} {Lecture Notes in
  Physics, Berlin Springer Verlag}}},\ Vol.\ \bibinfo {volume} {905},\ \bibinfo
  {editor} {edited by\ \bibinfo {editor} {\bibfnamefont {F.}~\bibnamefont
  {{Haardt}}}, \bibinfo {editor} {\bibfnamefont {V.}~\bibnamefont {{Gorini}}},
  \bibinfo {editor} {\bibfnamefont {U.}~\bibnamefont {{Moschella}}}, \bibinfo
  {editor} {\bibfnamefont {A.}~\bibnamefont {{Treves}}}, \ and\ \bibinfo
  {editor} {\bibfnamefont {M.}~\bibnamefont {{Colpi}}}}\ (\bibinfo {year}
  {2016})\ p.\ \bibinfo {pages} {273}\BibitemShut {NoStop}%
\bibitem [{\citenamefont {Hinder}\ \emph {et~al.}(2018)\citenamefont {Hinder},
  \citenamefont {Kidder},\ and\ \citenamefont {Pfeiffer}}]{Hinder_2018}%
  \BibitemOpen
  \bibfield  {author} {\bibinfo {author} {\bibfnamefont {I.}~\bibnamefont
  {Hinder}}, \bibinfo {author} {\bibfnamefont {L.~E.}\ \bibnamefont {Kidder}},
  \ and\ \bibinfo {author} {\bibfnamefont {H.~P.}\ \bibnamefont {Pfeiffer}},\
  }\href {\doibase 10.1103/physrevd.98.044015} {\bibfield  {journal} {\bibinfo
  {journal} {Physical Review D}\ }\textbf {\bibinfo {volume} {98}} (\bibinfo
  {year} {2018}),\ 10.1103/physrevd.98.044015}\BibitemShut {NoStop}%
\bibitem [{\citenamefont {{Chiaramello}}\ and\ \citenamefont
  {{Nagar}}(2020)}]{DCAN_2020}%
  \BibitemOpen
  \bibfield  {author} {\bibinfo {author} {\bibfnamefont {D.}~\bibnamefont
  {{Chiaramello}}}\ and\ \bibinfo {author} {\bibfnamefont {A.}~\bibnamefont
  {{Nagar}}},\ }\href@noop {} {\bibfield  {journal} {\bibinfo  {journal} {arXiv
  e-prints}\ ,\ \bibinfo {eid} {arXiv:2001.11736}} (\bibinfo {year} {2020})},\
  \Eprint {http://arxiv.org/abs/2001.11736} {arXiv:2001.11736 [gr-qc]}
  \BibitemShut {NoStop}%
\bibitem [{\citenamefont {{Ramos-Buades}}\ \emph {et~al.}(2020)\citenamefont
  {{Ramos-Buades}}, \citenamefont {{Husa}}, \citenamefont {{Pratten}},
  \citenamefont {{Estell{\'e}s}}, \citenamefont {{Garc{\'\i}a-Quir{\'o}s}},
  \citenamefont {{Mateu-Lucena}}, \citenamefont {{Colleoni}},\ and\
  \citenamefont {{Jaume}}}]{Ramos_2020}%
  \BibitemOpen
  \bibfield  {author} {\bibinfo {author} {\bibfnamefont {A.}~\bibnamefont
  {{Ramos-Buades}}}, \bibinfo {author} {\bibfnamefont {S.}~\bibnamefont
  {{Husa}}}, \bibinfo {author} {\bibfnamefont {G.}~\bibnamefont {{Pratten}}},
  \bibinfo {author} {\bibfnamefont {H.}~\bibnamefont {{Estell{\'e}s}}},
  \bibinfo {author} {\bibfnamefont {C.}~\bibnamefont
  {{Garc{\'\i}a-Quir{\'o}s}}}, \bibinfo {author} {\bibfnamefont
  {M.}~\bibnamefont {{Mateu-Lucena}}}, \bibinfo {author} {\bibfnamefont
  {M.}~\bibnamefont {{Colleoni}}}, \ and\ \bibinfo {author} {\bibfnamefont
  {R.}~\bibnamefont {{Jaume}}},\ }\href {\doibase 10.1103/PhysRevD.101.083015}
  {\bibfield  {journal} {\bibinfo  {journal} {\prd}\ }\textbf {\bibinfo
  {volume} {101}},\ \bibinfo {eid} {083015} (\bibinfo {year} {2020})},\ \Eprint
  {http://arxiv.org/abs/1909.11011} {arXiv:1909.11011 [gr-qc]} \BibitemShut
  {NoStop}%
\bibitem [{\citenamefont {Ramos-Buades}\ \emph {et~al.}(2020)\citenamefont
  {Ramos-Buades}, \citenamefont {Tiwari}, \citenamefont {Haney},\ and\
  \citenamefont {Husa}}]{Antoni_2020}%
  \BibitemOpen
  \bibfield  {author} {\bibinfo {author} {\bibfnamefont {A.}~\bibnamefont
  {Ramos-Buades}}, \bibinfo {author} {\bibfnamefont {S.}~\bibnamefont
  {Tiwari}}, \bibinfo {author} {\bibfnamefont {M.}~\bibnamefont {Haney}}, \
  and\ \bibinfo {author} {\bibfnamefont {S.}~\bibnamefont {Husa}},\ }\href@noop
  {} {\enquote {\bibinfo {title} {Impact of eccentricity on the gravitational
  wave searches for binary black holes: High mass case},}\ } (\bibinfo {year}
  {2020}),\ \Eprint {http://arxiv.org/abs/2005.14016} {arXiv:2005.14016
  [gr-qc]} \BibitemShut {NoStop}%
\bibitem [{\citenamefont {Husa}\ \emph {et~al.}(2016)\citenamefont {Husa},
  \citenamefont {Khan}, \citenamefont {Hannam}, \citenamefont {P\"urrer},
  \citenamefont {Ohme}, \citenamefont {Forteza},\ and\ \citenamefont
  {Boh\'e}}]{Husa_2016}%
  \BibitemOpen
  \bibfield  {author} {\bibinfo {author} {\bibfnamefont {S.}~\bibnamefont
  {Husa}}, \bibinfo {author} {\bibfnamefont {S.}~\bibnamefont {Khan}}, \bibinfo
  {author} {\bibfnamefont {M.}~\bibnamefont {Hannam}}, \bibinfo {author}
  {\bibfnamefont {M.}~\bibnamefont {P\"urrer}}, \bibinfo {author}
  {\bibfnamefont {F.}~\bibnamefont {Ohme}}, \bibinfo {author} {\bibfnamefont
  {X.~J.}\ \bibnamefont {Forteza}}, \ and\ \bibinfo {author} {\bibfnamefont
  {A.}~\bibnamefont {Boh\'e}},\ }\href {\doibase 10.1103/PhysRevD.93.044006}
  {\bibfield  {journal} {\bibinfo  {journal} {Phys. Rev. D}\ }\textbf {\bibinfo
  {volume} {93}},\ \bibinfo {pages} {044006} (\bibinfo {year}
  {2016})}\BibitemShut {NoStop}%
\bibitem [{\citenamefont {Khan}\ \emph {et~al.}(2016)\citenamefont {Khan},
  \citenamefont {Husa}, \citenamefont {Hannam}, \citenamefont {Ohme},
  \citenamefont {P\"urrer}, \citenamefont {Forteza},\ and\ \citenamefont
  {Boh\'e}}]{Khan_2016}%
  \BibitemOpen
  \bibfield  {author} {\bibinfo {author} {\bibfnamefont {S.}~\bibnamefont
  {Khan}}, \bibinfo {author} {\bibfnamefont {S.}~\bibnamefont {Husa}}, \bibinfo
  {author} {\bibfnamefont {M.}~\bibnamefont {Hannam}}, \bibinfo {author}
  {\bibfnamefont {F.}~\bibnamefont {Ohme}}, \bibinfo {author} {\bibfnamefont
  {M.}~\bibnamefont {P\"urrer}}, \bibinfo {author} {\bibfnamefont {X.~J.}\
  \bibnamefont {Forteza}}, \ and\ \bibinfo {author} {\bibfnamefont
  {A.}~\bibnamefont {Boh\'e}},\ }\href {\doibase 10.1103/PhysRevD.93.044007}
  {\bibfield  {journal} {\bibinfo  {journal} {Phys. Rev. D}\ }\textbf {\bibinfo
  {volume} {93}},\ \bibinfo {pages} {044007} (\bibinfo {year}
  {2016})}\BibitemShut {NoStop}%
\bibitem [{\citenamefont {{Yunes}}\ \emph {et~al.}(2009)\citenamefont
  {{Yunes}}, \citenamefont {{Arun}}, \citenamefont {{Berti}},\ and\
  \citenamefont {{Will}}}]{YABW}%
  \BibitemOpen
  \bibfield  {author} {\bibinfo {author} {\bibfnamefont {N.}~\bibnamefont
  {{Yunes}}}, \bibinfo {author} {\bibfnamefont {K.~G.}\ \bibnamefont {{Arun}}},
  \bibinfo {author} {\bibfnamefont {E.}~\bibnamefont {{Berti}}}, \ and\
  \bibinfo {author} {\bibfnamefont {C.~M.}\ \bibnamefont {{Will}}},\ }\href
  {\doibase 10.1103/PhysRevD.80.084001} {\bibfield  {journal} {\bibinfo
  {journal} {\prd}\ }\textbf {\bibinfo {volume} {80}},\ \bibinfo {eid} {084001}
  (\bibinfo {year} {2009})},\ \Eprint {http://arxiv.org/abs/0906.0313}
  {arXiv:0906.0313 [gr-qc]} \BibitemShut {NoStop}%
\bibitem [{\citenamefont {{Tanay}}\ \emph {et~al.}(2016)\citenamefont
  {{Tanay}}, \citenamefont {{Haney}},\ and\ \citenamefont {{Gopakumar}}}]{THG}%
  \BibitemOpen
  \bibfield  {author} {\bibinfo {author} {\bibfnamefont {S.}~\bibnamefont
  {{Tanay}}}, \bibinfo {author} {\bibfnamefont {M.}~\bibnamefont {{Haney}}}, \
  and\ \bibinfo {author} {\bibfnamefont {A.}~\bibnamefont {{Gopakumar}}},\
  }\href {\doibase 10.1103/PhysRevD.93.064031} {\bibfield  {journal} {\bibinfo
  {journal} {\prd}\ }\textbf {\bibinfo {volume} {93}},\ \bibinfo {eid} {064031}
  (\bibinfo {year} {2016})},\ \Eprint {http://arxiv.org/abs/1602.03081}
  {arXiv:1602.03081 [gr-qc]} \BibitemShut {NoStop}%
\bibitem [{\citenamefont {Moore}\ \emph {et~al.}(2016)\citenamefont {Moore},
  \citenamefont {Favata}, \citenamefont {Arun},\ and\ \citenamefont
  {Mishra}}]{Moore16}%
  \BibitemOpen
  \bibfield  {author} {\bibinfo {author} {\bibfnamefont {B.}~\bibnamefont
  {Moore}}, \bibinfo {author} {\bibfnamefont {M.}~\bibnamefont {Favata}},
  \bibinfo {author} {\bibfnamefont {K.}~\bibnamefont {Arun}}, \ and\ \bibinfo
  {author} {\bibfnamefont {C.~K.}\ \bibnamefont {Mishra}},\ }\href {\doibase
  10.1103/physrevd.93.124061} {\bibfield  {journal} {\bibinfo  {journal}
  {Physical Review D}\ }\textbf {\bibinfo {volume} {93}} (\bibinfo {year}
  {2016}),\ 10.1103/physrevd.93.124061}\BibitemShut {NoStop}%
\bibitem [{\citenamefont {Tiwari}\ \emph {et~al.}(2019)\citenamefont {Tiwari},
  \citenamefont {Gopakumar}, \citenamefont {Haney},\ and\ \citenamefont
  {Hemantakumar}}]{TGMH}%
  \BibitemOpen
  \bibfield  {author} {\bibinfo {author} {\bibfnamefont {S.}~\bibnamefont
  {Tiwari}}, \bibinfo {author} {\bibfnamefont {A.}~\bibnamefont {Gopakumar}},
  \bibinfo {author} {\bibfnamefont {M.}~\bibnamefont {Haney}}, \ and\ \bibinfo
  {author} {\bibfnamefont {P.}~\bibnamefont {Hemantakumar}},\ }\href {\doibase
  10.1103/PhysRevD.99.124008} {\bibfield  {journal} {\bibinfo  {journal} {Phys.
  Rev. D}\ }\textbf {\bibinfo {volume} {99}},\ \bibinfo {pages} {124008}
  (\bibinfo {year} {2019})}\BibitemShut {NoStop}%
\bibitem [{\citenamefont {Bender}\ and\ \citenamefont {Orszag}(1999)}]{BenOrz}%
  \BibitemOpen
  \bibfield  {author} {\bibinfo {author} {\bibfnamefont {C.~M.}\ \bibnamefont
  {Bender}}\ and\ \bibinfo {author} {\bibfnamefont {S.~A.}\ \bibnamefont
  {Orszag}},\ }\href@noop {} {\emph {\bibinfo {title} {Advanced mathematical
  methods for scientists and engineers}}}\ (\bibinfo  {publisher} {Springer},\
  \bibinfo {address} {New York},\ \bibinfo {year} {1999})\BibitemShut {NoStop}%
\bibitem [{\citenamefont {Moore}\ \emph {et~al.}(2018)\citenamefont {Moore},
  \citenamefont {Robson}, \citenamefont {Loutrel},\ and\ \citenamefont
  {Yunes}}]{MRLY18}%
  \BibitemOpen
  \bibfield  {author} {\bibinfo {author} {\bibfnamefont {B.}~\bibnamefont
  {Moore}}, \bibinfo {author} {\bibfnamefont {T.}~\bibnamefont {Robson}},
  \bibinfo {author} {\bibfnamefont {N.}~\bibnamefont {Loutrel}}, \ and\
  \bibinfo {author} {\bibfnamefont {N.}~\bibnamefont {Yunes}},\ }\href
  {\doibase 10.1088/1361-6382/aaea00} {\bibfield  {journal} {\bibinfo
  {journal} {Class. Quant. Grav.}\ }\textbf {\bibinfo {volume} {35}},\ \bibinfo
  {pages} {235006} (\bibinfo {year} {2018})},\ \Eprint
  {http://arxiv.org/abs/1807.07163} {arXiv:1807.07163 [gr-qc]} \BibitemShut
  {NoStop}%
\bibitem [{\citenamefont {{Moore}}\ and\ \citenamefont {{Yunes}}(2019)}]{MY19}%
  \BibitemOpen
  \bibfield  {author} {\bibinfo {author} {\bibfnamefont {B.}~\bibnamefont
  {{Moore}}}\ and\ \bibinfo {author} {\bibfnamefont {N.}~\bibnamefont
  {{Yunes}}},\ }\href@noop {} {\bibfield  {journal} {\bibinfo  {journal} {arXiv
  e-prints}\ ,\ \bibinfo {eid} {arXiv:1903.05203}} (\bibinfo {year} {2019})},\
  \Eprint {http://arxiv.org/abs/1903.05203} {arXiv:1903.05203 [gr-qc]}
  \BibitemShut {NoStop}%
\bibitem [{\citenamefont {{Colwell}}(1993)}]{KE_textbook}%
  \BibitemOpen
  \bibfield  {author} {\bibinfo {author} {\bibfnamefont {P.}~\bibnamefont
  {{Colwell}}},\ }\href@noop {} {\emph {\bibinfo {title} {{Solving Kepler's
  equation over three centuries}}}}\ (\bibinfo {year} {1993})\BibitemShut
  {NoStop}%
\bibitem [{\citenamefont {{Memmesheimer}}\ \emph {et~al.}(2004)\citenamefont
  {{Memmesheimer}}, \citenamefont {{Gopakumar}},\ and\ \citenamefont
  {{Sch{\"a}fer}}}]{GMS}%
  \BibitemOpen
  \bibfield  {author} {\bibinfo {author} {\bibfnamefont {R.-M.}\ \bibnamefont
  {{Memmesheimer}}}, \bibinfo {author} {\bibfnamefont {A.}~\bibnamefont
  {{Gopakumar}}}, \ and\ \bibinfo {author} {\bibfnamefont {G.}~\bibnamefont
  {{Sch{\"a}fer}}},\ }\href {\doibase 10.1103/PhysRevD.70.104011} {\bibfield
  {journal} {\bibinfo  {journal} {\prd}\ }\textbf {\bibinfo {volume} {70}},\
  \bibinfo {eid} {104011} (\bibinfo {year} {2004})},\ \Eprint
  {http://arxiv.org/abs/gr-qc/0407049} {arXiv:gr-qc/0407049 [gr-qc]}
  \BibitemShut {NoStop}%
\bibitem [{\citenamefont {Boetzel}\ \emph
  {et~al.}(2017{\natexlab{a}})\citenamefont {Boetzel}, \citenamefont
  {Susobhanan}, \citenamefont {Gopakumar}, \citenamefont {Klein},\ and\
  \citenamefont {Jetzer}}]{Boetzel_17}%
  \BibitemOpen
  \bibfield  {author} {\bibinfo {author} {\bibfnamefont {Y.}~\bibnamefont
  {Boetzel}}, \bibinfo {author} {\bibfnamefont {A.}~\bibnamefont {Susobhanan}},
  \bibinfo {author} {\bibfnamefont {A.}~\bibnamefont {Gopakumar}}, \bibinfo
  {author} {\bibfnamefont {A.}~\bibnamefont {Klein}}, \ and\ \bibinfo {author}
  {\bibfnamefont {P.}~\bibnamefont {Jetzer}},\ }\href {\doibase
  10.1103/physrevd.96.044011} {\bibfield  {journal} {\bibinfo  {journal}
  {Physical Review D}\ }\textbf {\bibinfo {volume} {96}} (\bibinfo {year}
  {2017}{\natexlab{a}}),\ 10.1103/physrevd.96.044011}\BibitemShut {NoStop}%
\bibitem [{\citenamefont {{Peters}}\ and\ \citenamefont
  {{Mathews}}(1963)}]{PM63}%
  \BibitemOpen
  \bibfield  {author} {\bibinfo {author} {\bibfnamefont {P.~C.}\ \bibnamefont
  {{Peters}}}\ and\ \bibinfo {author} {\bibfnamefont {J.}~\bibnamefont
  {{Mathews}}},\ }\href {\doibase 10.1103/PhysRev.131.435} {\bibfield
  {journal} {\bibinfo  {journal} {Phys. Rev.}\ }\textbf {\bibinfo {volume}
  {131}},\ \bibinfo {pages} {435} (\bibinfo {year} {1963})}\BibitemShut
  {NoStop}%
\bibitem [{\citenamefont {Kr\'olak}\ \emph {et~al.}(1995)\citenamefont
  {Kr\'olak}, \citenamefont {Kokkotas},\ and\ \citenamefont {Sch\"afer}}]{KKS}%
  \BibitemOpen
  \bibfield  {author} {\bibinfo {author} {\bibfnamefont {A.}~\bibnamefont
  {Kr\'olak}}, \bibinfo {author} {\bibfnamefont {K.~D.}\ \bibnamefont
  {Kokkotas}}, \ and\ \bibinfo {author} {\bibfnamefont {G.}~\bibnamefont
  {Sch\"afer}},\ }\href {\doibase 10.1103/PhysRevD.52.2089} {\bibfield
  {journal} {\bibinfo  {journal} {Phys. Rev. D}\ }\textbf {\bibinfo {volume}
  {52}},\ \bibinfo {pages} {2089} (\bibinfo {year} {1995})}\BibitemShut
  {NoStop}%
\bibitem [{\citenamefont {Mik\'oczi}\ \emph {et~al.}(2012)\citenamefont
  {Mik\'oczi}, \citenamefont {Kocsis}, \citenamefont {Forg\'acs},\ and\
  \citenamefont {Vas\'uth}}]{BBPM12}%
  \BibitemOpen
  \bibfield  {author} {\bibinfo {author} {\bibfnamefont {B.}~\bibnamefont
  {Mik\'oczi}}, \bibinfo {author} {\bibfnamefont {B.}~\bibnamefont {Kocsis}},
  \bibinfo {author} {\bibfnamefont {P.}~\bibnamefont {Forg\'acs}}, \ and\
  \bibinfo {author} {\bibfnamefont {M.}~\bibnamefont {Vas\'uth}},\ }\href
  {\doibase 10.1103/PhysRevD.86.104027} {\bibfield  {journal} {\bibinfo
  {journal} {Phys. Rev. D}\ }\textbf {\bibinfo {volume} {86}},\ \bibinfo
  {pages} {104027} (\bibinfo {year} {2012})}\BibitemShut {NoStop}%
\bibitem [{\citenamefont {Damour}\ \emph {et~al.}(1998)\citenamefont {Damour},
  \citenamefont {Iyer},\ and\ \citenamefont {Sathyaprakash}}]{DIS98}%
  \BibitemOpen
  \bibfield  {author} {\bibinfo {author} {\bibfnamefont {T.}~\bibnamefont
  {Damour}}, \bibinfo {author} {\bibfnamefont {B.~R.}\ \bibnamefont {Iyer}}, \
  and\ \bibinfo {author} {\bibfnamefont {B.~S.}\ \bibnamefont
  {Sathyaprakash}},\ }\href {\doibase 10.1103/PhysRevD.57.885} {\bibfield
  {journal} {\bibinfo  {journal} {Phys. Rev. D}\ }\textbf {\bibinfo {volume}
  {57}},\ \bibinfo {pages} {885} (\bibinfo {year} {1998})}\BibitemShut
  {NoStop}%
\bibitem [{\citenamefont {Gupta}\ \emph {et~al.}(2000)\citenamefont {Gupta},
  \citenamefont {Gopakumar}, \citenamefont {Iyer},\ and\ \citenamefont
  {Iyer}}]{GGI_00}%
  \BibitemOpen
  \bibfield  {author} {\bibinfo {author} {\bibfnamefont {A.}~\bibnamefont
  {Gupta}}, \bibinfo {author} {\bibfnamefont {A.}~\bibnamefont {Gopakumar}},
  \bibinfo {author} {\bibfnamefont {B.~R.}\ \bibnamefont {Iyer}}, \ and\
  \bibinfo {author} {\bibfnamefont {S.}~\bibnamefont {Iyer}},\ }\href {\doibase
  10.1103/physrevd.62.044038} {\bibfield  {journal} {\bibinfo  {journal}
  {Physical Review D}\ }\textbf {\bibinfo {volume} {62}} (\bibinfo {year}
  {2000}),\ 10.1103/physrevd.62.044038}\BibitemShut {NoStop}%
\bibitem [{\citenamefont {{Harry}}\ and\ \citenamefont {{LIGO Scientific
  Collaboration}}(2010)}]{NoiseCurve}%
  \BibitemOpen
  \bibfield  {author} {\bibinfo {author} {\bibfnamefont {G.~M.}\ \bibnamefont
  {{Harry}}}\ and\ \bibinfo {author} {\bibnamefont {{LIGO Scientific
  Collaboration}}},\ }\href {\doibase 10.1088/0264-9381/27/8/084006} {\bibfield
   {journal} {\bibinfo  {journal} {Classical and Quantum Gravity}\ }\textbf
  {\bibinfo {volume} {27}},\ \bibinfo {eid} {084006} (\bibinfo {year}
  {2010})}\BibitemShut {NoStop}%
\bibitem [{\citenamefont {Rieth}\ and\ \citenamefont {Schäfer}(1997)}]{RS96}%
  \BibitemOpen
  \bibfield  {author} {\bibinfo {author} {\bibfnamefont {R.}~\bibnamefont
  {Rieth}}\ and\ \bibinfo {author} {\bibfnamefont {G.}~\bibnamefont
  {Schäfer}},\ }\href {\doibase 10.1088/0264-9381/14/8/029} {\bibfield
  {journal} {\bibinfo  {journal} {Classical and Quantum Gravity}\ }\textbf
  {\bibinfo {volume} {14}},\ \bibinfo {pages} {2357} (\bibinfo {year}
  {1997})}\BibitemShut {NoStop}%
\bibitem [{\citenamefont {Mik\'oczi}\ \emph
  {et~al.}(2015{\natexlab{a}})\citenamefont {Mik\'oczi}, \citenamefont
  {Forg\'acs},\ and\ \citenamefont {Vas\'uth}}]{BPM15}%
  \BibitemOpen
  \bibfield  {author} {\bibinfo {author} {\bibfnamefont {B.}~\bibnamefont
  {Mik\'oczi}}, \bibinfo {author} {\bibfnamefont {P.}~\bibnamefont
  {Forg\'acs}}, \ and\ \bibinfo {author} {\bibfnamefont {M.}~\bibnamefont
  {Vas\'uth}},\ }\href {\doibase 10.1103/PhysRevD.92.044038} {\bibfield
  {journal} {\bibinfo  {journal} {Phys. Rev. D}\ }\textbf {\bibinfo {volume}
  {92}},\ \bibinfo {pages} {044038} (\bibinfo {year}
  {2015}{\natexlab{a}})}\BibitemShut {NoStop}%
\bibitem [{\citenamefont {Buonanno}\ \emph {et~al.}(2009)\citenamefont
  {Buonanno}, \citenamefont {Iyer}, \citenamefont {Ochsner}, \citenamefont
  {Pan},\ and\ \citenamefont {Sathyaprakash}}]{Buonanno09}%
  \BibitemOpen
  \bibfield  {author} {\bibinfo {author} {\bibfnamefont {A.}~\bibnamefont
  {Buonanno}}, \bibinfo {author} {\bibfnamefont {B.~R.}\ \bibnamefont {Iyer}},
  \bibinfo {author} {\bibfnamefont {E.}~\bibnamefont {Ochsner}}, \bibinfo
  {author} {\bibfnamefont {Y.}~\bibnamefont {Pan}}, \ and\ \bibinfo {author}
  {\bibfnamefont {B.~S.}\ \bibnamefont {Sathyaprakash}},\ }\href {\doibase
  10.1103/physrevd.80.084043} {\bibfield  {journal} {\bibinfo  {journal}
  {Physical Review D}\ }\textbf {\bibinfo {volume} {80}} (\bibinfo {year}
  {2009}),\ 10.1103/physrevd.80.084043}\BibitemShut {NoStop}%
\bibitem [{\citenamefont {Damour}\ \emph {et~al.}(2004)\citenamefont {Damour},
  \citenamefont {Gopakumar},\ and\ \citenamefont {Iyer}}]{DGI}%
  \BibitemOpen
  \bibfield  {author} {\bibinfo {author} {\bibfnamefont {T.}~\bibnamefont
  {Damour}}, \bibinfo {author} {\bibfnamefont {A.}~\bibnamefont {Gopakumar}}, \
  and\ \bibinfo {author} {\bibfnamefont {B.~R.}\ \bibnamefont {Iyer}},\ }\href
  {\doibase 10.1103/physrevd.70.064028} {\bibfield  {journal} {\bibinfo
  {journal} {Physical Review D}\ }\textbf {\bibinfo {volume} {70}} (\bibinfo
  {year} {2004}),\ 10.1103/physrevd.70.064028}\BibitemShut {NoStop}%
\bibitem [{\citenamefont {Königsdörffer}\ and\ \citenamefont
  {Gopakumar}(2006)}]{KG06}%
  \BibitemOpen
  \bibfield  {author} {\bibinfo {author} {\bibfnamefont {C.}~\bibnamefont
  {Königsdörffer}}\ and\ \bibinfo {author} {\bibfnamefont {A.}~\bibnamefont
  {Gopakumar}},\ }\href {\doibase 10.1103/physrevd.73.124012} {\bibfield
  {journal} {\bibinfo  {journal} {Physical Review D}\ }\textbf {\bibinfo
  {volume} {73}} (\bibinfo {year} {2006}),\
  10.1103/physrevd.73.124012}\BibitemShut {NoStop}%
\bibitem [{\citenamefont {Klein}\ \emph {et~al.}(2018)\citenamefont {Klein},
  \citenamefont {Boetzel}, \citenamefont {Gopakumar}, \citenamefont {Jetzer},\
  and\ \citenamefont {de~Vittori}}]{KBGJ_18}%
  \BibitemOpen
  \bibfield  {author} {\bibinfo {author} {\bibfnamefont {A.}~\bibnamefont
  {Klein}}, \bibinfo {author} {\bibfnamefont {Y.}~\bibnamefont {Boetzel}},
  \bibinfo {author} {\bibfnamefont {A.}~\bibnamefont {Gopakumar}}, \bibinfo
  {author} {\bibfnamefont {P.}~\bibnamefont {Jetzer}}, \ and\ \bibinfo {author}
  {\bibfnamefont {L.}~\bibnamefont {de~Vittori}},\ }\href {\doibase
  10.1103/physrevd.98.104043} {\bibfield  {journal} {\bibinfo  {journal}
  {Physical Review D}\ }\textbf {\bibinfo {volume} {98}} (\bibinfo {year}
  {2018}),\ 10.1103/physrevd.98.104043}\BibitemShut {NoStop}%
\bibitem [{\citenamefont {Boyle}\ \emph {et~al.}(2007)\citenamefont {Boyle},
  \citenamefont {Brown}, \citenamefont {Kidder}, \citenamefont {Mroué},
  \citenamefont {Pfeiffer}, \citenamefont {Scheel}, \citenamefont {Cook},\ and\
  \citenamefont {Teukolsky}}]{Boyle07}%
  \BibitemOpen
  \bibfield  {author} {\bibinfo {author} {\bibfnamefont {M.}~\bibnamefont
  {Boyle}}, \bibinfo {author} {\bibfnamefont {D.~A.}\ \bibnamefont {Brown}},
  \bibinfo {author} {\bibfnamefont {L.~E.}\ \bibnamefont {Kidder}}, \bibinfo
  {author} {\bibfnamefont {A.~H.}\ \bibnamefont {Mroué}}, \bibinfo {author}
  {\bibfnamefont {H.~P.}\ \bibnamefont {Pfeiffer}}, \bibinfo {author}
  {\bibfnamefont {M.~A.}\ \bibnamefont {Scheel}}, \bibinfo {author}
  {\bibfnamefont {G.~B.}\ \bibnamefont {Cook}}, \ and\ \bibinfo {author}
  {\bibfnamefont {S.~A.}\ \bibnamefont {Teukolsky}},\ }\href {\doibase
  10.1103/physrevd.76.124038} {\bibfield  {journal} {\bibinfo  {journal}
  {Physical Review D}\ }\textbf {\bibinfo {volume} {76}} (\bibinfo {year}
  {2007}),\ 10.1103/physrevd.76.124038}\BibitemShut {NoStop}%
\bibitem [{\citenamefont {Arun}\ \emph {et~al.}(2009)\citenamefont {Arun},
  \citenamefont {Blanchet}, \citenamefont {Iyer},\ and\ \citenamefont
  {Sinha}}]{ABIS}%
  \BibitemOpen
  \bibfield  {author} {\bibinfo {author} {\bibfnamefont {K.}~\bibnamefont
  {Arun}}, \bibinfo {author} {\bibfnamefont {L.}~\bibnamefont {Blanchet}},
  \bibinfo {author} {\bibfnamefont {B.}~\bibnamefont {Iyer}}, \ and\ \bibinfo
  {author} {\bibfnamefont {S.}~\bibnamefont {Sinha}},\ }\href {\doibase
  10.1103/physrevd.80.124018} {\bibfield  {journal} {\bibinfo  {journal}
  {Physical Review D}\ }\textbf {\bibinfo {volume} {80}} (\bibinfo {year}
  {2009}),\ 10.1103/physrevd.80.124018}\BibitemShut {NoStop}%
\bibitem [{\citenamefont {Gopakumar}\ \emph {et~al.}(2008)\citenamefont
  {Gopakumar}, \citenamefont {Hannam}, \citenamefont {Husa},\ and\
  \citenamefont {Br\"ugmann}}]{GHHB}%
  \BibitemOpen
  \bibfield  {author} {\bibinfo {author} {\bibfnamefont {A.}~\bibnamefont
  {Gopakumar}}, \bibinfo {author} {\bibfnamefont {M.}~\bibnamefont {Hannam}},
  \bibinfo {author} {\bibfnamefont {S.}~\bibnamefont {Husa}}, \ and\ \bibinfo
  {author} {\bibfnamefont {B.}~\bibnamefont {Br\"ugmann}},\ }\href {\doibase
  10.1103/PhysRevD.78.064026} {\bibfield  {journal} {\bibinfo  {journal} {Phys.
  Rev. D}\ }\textbf {\bibinfo {volume} {78}},\ \bibinfo {pages} {064026}
  (\bibinfo {year} {2008})}\BibitemShut {NoStop}%
\bibitem [{\citenamefont {Boetzel}\ \emph
  {et~al.}(2017{\natexlab{b}})\citenamefont {Boetzel}, \citenamefont
  {Susobhanan}, \citenamefont {Gopakumar}, \citenamefont {Klein},\ and\
  \citenamefont {Jetzer}}]{BSGKJ_17}%
  \BibitemOpen
  \bibfield  {author} {\bibinfo {author} {\bibfnamefont {Y.}~\bibnamefont
  {Boetzel}}, \bibinfo {author} {\bibfnamefont {A.}~\bibnamefont {Susobhanan}},
  \bibinfo {author} {\bibfnamefont {A.}~\bibnamefont {Gopakumar}}, \bibinfo
  {author} {\bibfnamefont {A.}~\bibnamefont {Klein}}, \ and\ \bibinfo {author}
  {\bibfnamefont {P.}~\bibnamefont {Jetzer}},\ }\href {\doibase
  10.1103/physrevd.96.044011} {\bibfield  {journal} {\bibinfo  {journal}
  {Physical Review D}\ }\textbf {\bibinfo {volume} {96}} (\bibinfo {year}
  {2017}{\natexlab{b}}),\ 10.1103/physrevd.96.044011}\BibitemShut {NoStop}%
\bibitem [{\citenamefont {Nitz}\ \emph {et~al.}(2019)\citenamefont {Nitz},
  \citenamefont {Harry}, \citenamefont {Brown}, \citenamefont {Biwer},
  \citenamefont {Willis}, \citenamefont {Canton}, \citenamefont {Pekowsky},
  \citenamefont {Capano}, \citenamefont {Dent}, \citenamefont {Williamson},
  \citenamefont {De}, \citenamefont {Cabero}, \citenamefont {Machenschalk},
  \citenamefont {Kumar}, \citenamefont {Reyes}, \citenamefont {Massinger},
  \citenamefont {Macleod}, \citenamefont {Lenon}, \citenamefont {Fairhurst},
  \citenamefont {Nielsen}, \citenamefont {Khan}, \citenamefont {Kapadia},
  \citenamefont {Pannarale}, \citenamefont {Singer}, \citenamefont {Finstad},
  \citenamefont {{T\'apai}}, \citenamefont {Gabbard}, \citenamefont {Sugar},
  \citenamefont {Couvares},\ and\ \citenamefont
  {Zertuche}}]{alex_nitz_2019_2556644}%
  \BibitemOpen
  \bibfield  {author} {\bibinfo {author} {\bibfnamefont {A.}~\bibnamefont
  {Nitz}}, \bibinfo {author} {\bibfnamefont {I.}~\bibnamefont {Harry}},
  \bibinfo {author} {\bibfnamefont {D.}~\bibnamefont {Brown}}, \bibinfo
  {author} {\bibfnamefont {C.~M.}\ \bibnamefont {Biwer}}, \bibinfo {author}
  {\bibfnamefont {J.}~\bibnamefont {Willis}}, \bibinfo {author} {\bibfnamefont
  {T.~D.}\ \bibnamefont {Canton}}, \bibinfo {author} {\bibfnamefont
  {L.}~\bibnamefont {Pekowsky}}, \bibinfo {author} {\bibfnamefont
  {C.}~\bibnamefont {Capano}}, \bibinfo {author} {\bibfnamefont
  {T.}~\bibnamefont {Dent}}, \bibinfo {author} {\bibfnamefont {A.~R.}\
  \bibnamefont {Williamson}}, \bibinfo {author} {\bibfnamefont
  {S.}~\bibnamefont {De}}, \bibinfo {author} {\bibfnamefont {M.}~\bibnamefont
  {Cabero}}, \bibinfo {author} {\bibfnamefont {B.}~\bibnamefont
  {Machenschalk}}, \bibinfo {author} {\bibfnamefont {P.}~\bibnamefont {Kumar}},
  \bibinfo {author} {\bibfnamefont {S.}~\bibnamefont {Reyes}}, \bibinfo
  {author} {\bibfnamefont {T.}~\bibnamefont {Massinger}}, \bibinfo {author}
  {\bibfnamefont {D.}~\bibnamefont {Macleod}}, \bibinfo {author} {\bibfnamefont
  {A.}~\bibnamefont {Lenon}}, \bibinfo {author} {\bibfnamefont
  {S.}~\bibnamefont {Fairhurst}}, \bibinfo {author} {\bibfnamefont
  {A.}~\bibnamefont {Nielsen}}, \bibinfo {author} {\bibfnamefont
  {S.}~\bibnamefont {Khan}}, \bibinfo {author} {\bibfnamefont {S.~J.}\
  \bibnamefont {Kapadia}}, \bibinfo {author} {\bibfnamefont {F.}~\bibnamefont
  {Pannarale}}, \bibinfo {author} {\bibfnamefont {L.}~\bibnamefont {Singer}},
  \bibinfo {author} {\bibfnamefont {D.}~\bibnamefont {Finstad}}, \bibinfo
  {author} {\bibfnamefont {M.}~\bibnamefont {{T\'apai}}}, \bibinfo {author}
  {\bibfnamefont {H.}~\bibnamefont {Gabbard}}, \bibinfo {author} {\bibfnamefont
  {C.}~\bibnamefont {Sugar}}, \bibinfo {author} {\bibfnamefont
  {P.}~\bibnamefont {Couvares}}, \ and\ \bibinfo {author} {\bibfnamefont
  {L.~M.}\ \bibnamefont {Zertuche}},\ }\href {\doibase 10.5281/zenodo.2556644}
  {\enquote {\bibinfo {title} {gwastro/pycbc: Pre-o3 release v1},}\ } (\bibinfo
  {year} {2019})\BibitemShut {NoStop}%
\bibitem [{\citenamefont {Hunter}(2007)}]{matplotlib}%
  \BibitemOpen
  \bibfield  {author} {\bibinfo {author} {\bibfnamefont {J.~D.}\ \bibnamefont
  {Hunter}},\ }\href {\doibase 10.1109/MCSE.2007.55} {\bibfield  {journal}
  {\bibinfo  {journal} {Computing In Science \& Engineering}\ }\textbf
  {\bibinfo {volume} {9}},\ \bibinfo {pages} {90} (\bibinfo {year}
  {2007})}\BibitemShut {NoStop}%
\bibitem [{\citenamefont {Damour}\ \emph {et~al.}(2000)\citenamefont {Damour},
  \citenamefont {Iyer},\ and\ \citenamefont {Sathyaprakash}}]{DIS00}%
  \BibitemOpen
  \bibfield  {author} {\bibinfo {author} {\bibfnamefont {T.}~\bibnamefont
  {Damour}}, \bibinfo {author} {\bibfnamefont {B.~R.}\ \bibnamefont {Iyer}}, \
  and\ \bibinfo {author} {\bibfnamefont {B.~S.}\ \bibnamefont
  {Sathyaprakash}},\ }\href {\doibase 10.1103/PhysRevD.62.084036} {\bibfield
  {journal} {\bibinfo  {journal} {Phys. Rev. D}\ }\textbf {\bibinfo {volume}
  {62}},\ \bibinfo {pages} {084036} (\bibinfo {year} {2000})}\BibitemShut
  {NoStop}%
\bibitem [{\citenamefont {Pierro}\ \emph {et~al.}(2001)\citenamefont {Pierro},
  \citenamefont {Pinto}, \citenamefont {Spallicci}, \citenamefont {Laserra},\
  and\ \citenamefont {Recano}}]{PP01}%
  \BibitemOpen
  \bibfield  {author} {\bibinfo {author} {\bibfnamefont {V.}~\bibnamefont
  {Pierro}}, \bibinfo {author} {\bibfnamefont {I.}~\bibnamefont {Pinto}},
  \bibinfo {author} {\bibfnamefont {A.}~\bibnamefont {Spallicci}}, \bibinfo
  {author} {\bibfnamefont {E.}~\bibnamefont {Laserra}}, \ and\ \bibinfo
  {author} {\bibfnamefont {F.}~\bibnamefont {Recano}},\ }\href {\doibase
  10.1046/j.1365-8711.2001.04442.x} {\bibfield  {journal} {\bibinfo  {journal}
  {Monthly Notices of the Royal Astronomical Society}\ }\textbf {\bibinfo
  {volume} {325}},\ \bibinfo {pages} {358} (\bibinfo {year} {2001})},\ \Eprint
  {http://arxiv.org/abs/http://oup.prod.sis.lan/mnras/article-pdf/325/1/358/2833488/325-1-358.pdf}
  {http://oup.prod.sis.lan/mnras/article-pdf/325/1/358/2833488/325-1-358.pdf}
  \BibitemShut {NoStop}%
\bibitem [{\citenamefont {{Pierro}}\ \emph {et~al.}(2002)\citenamefont
  {{Pierro}}, \citenamefont {{Pinto}},\ and\ \citenamefont {{Spallicci di
  F.}}}]{PP02}%
  \BibitemOpen
  \bibfield  {author} {\bibinfo {author} {\bibfnamefont {V.}~\bibnamefont
  {{Pierro}}}, \bibinfo {author} {\bibfnamefont {I.~M.}\ \bibnamefont
  {{Pinto}}}, \ and\ \bibinfo {author} {\bibfnamefont {A.~D.~A.~M.}\
  \bibnamefont {{Spallicci di F.}}},\ }\href {\doibase
  10.1046/j.1365-8711.2002.05557.x} {\bibfield  {journal} {\bibinfo  {journal}
  {mnras}\ }\textbf {\bibinfo {volume} {334}},\ \bibinfo {pages} {855}
  (\bibinfo {year} {2002})}\BibitemShut {NoStop}%
\bibitem [{\citenamefont {Santamar\'{\i}a}\ \emph {et~al.}(2010)\citenamefont
  {Santamar\'{\i}a}, \citenamefont {Ohme}, \citenamefont {Ajith}, \citenamefont
  {Br\"ugmann}, \citenamefont {Dorband}, \citenamefont {Hannam}, \citenamefont
  {Husa}, \citenamefont {M\"osta}, \citenamefont {Pollney}, \citenamefont
  {Reisswig}, \citenamefont {Robinson}, \citenamefont {Seiler},\ and\
  \citenamefont {Krishnan}}]{IMRPhenomD}%
  \BibitemOpen
  \bibfield  {author} {\bibinfo {author} {\bibfnamefont {L.}~\bibnamefont
  {Santamar\'{\i}a}}, \bibinfo {author} {\bibfnamefont {F.}~\bibnamefont
  {Ohme}}, \bibinfo {author} {\bibfnamefont {P.}~\bibnamefont {Ajith}},
  \bibinfo {author} {\bibfnamefont {B.}~\bibnamefont {Br\"ugmann}}, \bibinfo
  {author} {\bibfnamefont {N.}~\bibnamefont {Dorband}}, \bibinfo {author}
  {\bibfnamefont {M.}~\bibnamefont {Hannam}}, \bibinfo {author} {\bibfnamefont
  {S.}~\bibnamefont {Husa}}, \bibinfo {author} {\bibfnamefont {P.}~\bibnamefont
  {M\"osta}}, \bibinfo {author} {\bibfnamefont {D.}~\bibnamefont {Pollney}},
  \bibinfo {author} {\bibfnamefont {C.}~\bibnamefont {Reisswig}}, \bibinfo
  {author} {\bibfnamefont {E.~L.}\ \bibnamefont {Robinson}}, \bibinfo {author}
  {\bibfnamefont {J.}~\bibnamefont {Seiler}}, \ and\ \bibinfo {author}
  {\bibfnamefont {B.}~\bibnamefont {Krishnan}},\ }\href {\doibase
  10.1103/PhysRevD.82.064016} {\bibfield  {journal} {\bibinfo  {journal} {Phys.
  Rev. D}\ }\textbf {\bibinfo {volume} {82}},\ \bibinfo {pages} {064016}
  (\bibinfo {year} {2010})}\BibitemShut {NoStop}%
\bibitem [{\citenamefont {Ajith}\ \emph {et~al.}(2011)\citenamefont {Ajith},
  \citenamefont {Hannam}, \citenamefont {Husa}, \citenamefont {Chen},
  \citenamefont {Br\"ugmann}, \citenamefont {Dorband}, \citenamefont
  {M\"uller}, \citenamefont {Ohme}, \citenamefont {Pollney}, \citenamefont
  {Reisswig}, \citenamefont {Santamar\'{\i}a},\ and\ \citenamefont
  {Seiler}}]{Ajith_2011}%
  \BibitemOpen
  \bibfield  {author} {\bibinfo {author} {\bibfnamefont {P.}~\bibnamefont
  {Ajith}}, \bibinfo {author} {\bibfnamefont {M.}~\bibnamefont {Hannam}},
  \bibinfo {author} {\bibfnamefont {S.}~\bibnamefont {Husa}}, \bibinfo {author}
  {\bibfnamefont {Y.}~\bibnamefont {Chen}}, \bibinfo {author} {\bibfnamefont
  {B.}~\bibnamefont {Br\"ugmann}}, \bibinfo {author} {\bibfnamefont
  {N.}~\bibnamefont {Dorband}}, \bibinfo {author} {\bibfnamefont
  {D.}~\bibnamefont {M\"uller}}, \bibinfo {author} {\bibfnamefont
  {F.}~\bibnamefont {Ohme}}, \bibinfo {author} {\bibfnamefont {D.}~\bibnamefont
  {Pollney}}, \bibinfo {author} {\bibfnamefont {C.}~\bibnamefont {Reisswig}},
  \bibinfo {author} {\bibfnamefont {L.}~\bibnamefont {Santamar\'{\i}a}}, \ and\
  \bibinfo {author} {\bibfnamefont {J.}~\bibnamefont {Seiler}},\ }\href
  {\doibase 10.1103/PhysRevLett.106.241101} {\bibfield  {journal} {\bibinfo
  {journal} {Phys. Rev. Lett.}\ }\textbf {\bibinfo {volume} {106}},\ \bibinfo
  {pages} {241101} (\bibinfo {year} {2011})}\BibitemShut {NoStop}%
\bibitem [{\citenamefont {Pan}\ \emph {et~al.}(2014)\citenamefont {Pan},
  \citenamefont {Buonanno}, \citenamefont {Taracchini}, \citenamefont {Kidder},
  \citenamefont {Mrou\'e}, \citenamefont {Pfeiffer}, \citenamefont {Scheel},\
  and\ \citenamefont {Szil\'agyi}}]{Pan_2014}%
  \BibitemOpen
  \bibfield  {author} {\bibinfo {author} {\bibfnamefont {Y.}~\bibnamefont
  {Pan}}, \bibinfo {author} {\bibfnamefont {A.}~\bibnamefont {Buonanno}},
  \bibinfo {author} {\bibfnamefont {A.}~\bibnamefont {Taracchini}}, \bibinfo
  {author} {\bibfnamefont {L.~E.}\ \bibnamefont {Kidder}}, \bibinfo {author}
  {\bibfnamefont {A.~H.}\ \bibnamefont {Mrou\'e}}, \bibinfo {author}
  {\bibfnamefont {H.~P.}\ \bibnamefont {Pfeiffer}}, \bibinfo {author}
  {\bibfnamefont {M.~A.}\ \bibnamefont {Scheel}}, \ and\ \bibinfo {author}
  {\bibfnamefont {B.}~\bibnamefont {Szil\'agyi}},\ }\href {\doibase
  10.1103/PhysRevD.89.084006} {\bibfield  {journal} {\bibinfo  {journal} {Phys.
  Rev. D}\ }\textbf {\bibinfo {volume} {89}},\ \bibinfo {pages} {084006}
  (\bibinfo {year} {2014})}\BibitemShut {NoStop}%
\bibitem [{\citenamefont {Mik\'oczi}\ \emph
  {et~al.}(2015{\natexlab{b}})\citenamefont {Mik\'oczi}, \citenamefont
  {Forg\'acs},\ and\ \citenamefont {Vas\'uth}}]{MFV_2015}%
  \BibitemOpen
  \bibfield  {author} {\bibinfo {author} {\bibfnamefont {B.}~\bibnamefont
  {Mik\'oczi}}, \bibinfo {author} {\bibfnamefont {P.}~\bibnamefont
  {Forg\'acs}}, \ and\ \bibinfo {author} {\bibfnamefont {M.}~\bibnamefont
  {Vas\'uth}},\ }\href {\doibase 10.1103/PhysRevD.92.044038} {\bibfield
  {journal} {\bibinfo  {journal} {Phys. Rev. D}\ }\textbf {\bibinfo {volume}
  {92}},\ \bibinfo {pages} {044038} (\bibinfo {year}
  {2015}{\natexlab{b}})}\BibitemShut {NoStop}%
\end{thebibliography}%
\nocite{*}
\end{document}